\documentclass[a4paper,11pt]{article}

\usepackage{./auxi/jheppub}
\usepackage[utf8]{inputenc}
\usepackage[ngerman, english]{babel}
\usepackage[T1]{fontenc}
\usepackage[titletoc]{appendix}
\usepackage{titletoc}
\usepackage{graphicx}
\usepackage{subcaption} % to use subfigure environment (pics side by side)
\usepackage{xcolor}
\usepackage{amsmath,amssymb,amsthm}
\usepackage{bm}
\usepackage{float}
\usepackage[format=plain,labelfont={bf,it},textfont=it]{caption}
\usepackage{multirow}
\usepackage{slashed} % for Feynman slashed notation
\usepackage{mathtools} % for subequation numbering
\usepackage{tabularx}
\usepackage{bbold}
\usepackage[shortlabels]{enumitem}
\usepackage{tikz}
\usetikzlibrary{positioning,arrows,calc}
\usetikzlibrary{arrows.meta}
\usetikzlibrary{decorations.pathmorphing}
\usetikzlibrary{decorations.markings}
\usetikzlibrary{shapes.geometric}
\usetikzlibrary{patterns}
\tikzset{
  snake it/.style={
    decorate, 
    decoration=snake,
    segment length=3
  }
}
\usepackage{orcidlink}
\usepackage{xparse}
\usepackage{pdfpages} %for inserting pdfs in the file
\usepackage{titlesec} % for changing chapter style
\usepackage{fancyhdr} % for changing header style
\usepackage{emptypage} % removes header and page number in empty pages
\usepackage{chngcntr} % for changing numbering of equations to chapter
\usepackage{etoolbox} % for making chapter title to uppercase in ToC

\usepackage{dsfont} % for displaying the blackboard 1 properly
\usepackage[normalem]{ulem} % for striking through text
\definecolor{DarkBlueGrey}{RGB}{76,94,107}
\definecolor{MediumBlueGrey}{RGB}{110,135,153}
\definecolor{LightBlueGrey}{RGB}{134,163,184}
\definecolor{VeryLightBlueGrey}{RGB}{242,249,255}
\definecolor{WCOrange}{RGB}{242,146,29}
\definecolor{VeryLightOrange}{RGB}{255,245,233}
\definecolor{SCRed}{RGB}{179,48,48}
\definecolor{VeryLightRed}{RGB}{255,239,239}
\definecolor{VertexColor}{RGB}{242,146,29}
\definecolor{GluonColor}{RGB}{255,172,172}
\definecolor{SEColor}{RGB}{134,163,184}

\definecolor{BGBox}{RGB}{255,254,230}
\definecolor{PlaneColor}{RGB}{230,230,230}
\definecolor{BlobColor}{RGB}{190,180,230}

\newcommand{\pd}{\partial}

\newcommand{\Li}{{\normalfont\text{Li}}}
\newcommand{\vev}[1]{\langle\, #1 \, \rangle}

\def\veps{\varepsilon}

\newcommand{\Wl}{\mathcal{W}_\ell}
\newcommand{\Op}{\mathcal{O}}
\newcommand{\Oh}{\hat{\mathcal{O}}}
\newcommand{\Dh}{{\hat{\Delta}}}
\newcommand{\uh}{{\hat{u}}}
\newcommand{\nh}{{\hat{n}}}
\newcommand{\kh}{{\hat{k}}}

\newcommand{\lambdah}{\hat{\lambda}}

\def\Am{{\mathcal{A}}}
\def\Bm{{\mathcal{B}}}
\def\Cm{{\mathcal{C}}}

\def\Gm{{\mathcal{G}}}

\def\Km{{\mathcal{K}}}
\def\Lm{{\mathcal{L}}}

\def\Nm{{\mathcal{N}}}
\def\Om{{\mathcal{O}}}
\def\Pm{{\mathcal{P}}}

\def\Fds{{\mathds{F}}}

\def\Ids{{\mathds{I}}}

\usepackage{bbm}

\def\veps{\varepsilon}

\newcommand\psib{{\bar{\psi}}}
\newcommand\cb{{\bar{c}}}

\newcommand\zb{{\bar{z}}}

\newif\ifstartcompletesineup
\newif\ifendcompletesineup
\pgfkeys{
    /pgf/decoration/.cd,
    start up/.is if=startcompletesineup,
    start up=true,
    start up/.default=true,
    start down/.style={/pgf/decoration/start up=false},
    end up/.is if=endcompletesineup,
    end up=true,
    end up/.default=true,
    end down/.style={/pgf/decoration/end up=false}
}
\pgfdeclaredecoration{complete sines}{initial}
{
    \state{initial}[
        width=+0pt,
        next state=upsine,
        persistent precomputation={
            \ifstartcompletesineup
                \pgfkeys{/pgf/decoration automaton/next state=upsine}
                \ifendcompletesineup
                    \pgfmathsetmacro\matchinglength{
                        0.5*\pgfdecoratedinputsegmentlength / (ceil(0.5* \pgfdecoratedinputsegmentlength / \pgfdecorationsegmentlength) )
                    }
                \else
                    \pgfmathsetmacro\matchinglength{
                        0.5 * \pgfdecoratedinputsegmentlength / (ceil(0.5 * \pgfdecoratedinputsegmentlength / \pgfdecorationsegmentlength ) - 0.499)
                    }
                \fi
            \else
                \pgfkeys{/pgf/decoration automaton/next state=downsine}
                \ifendcompletesineup
                    \pgfmathsetmacro\matchinglength{
                        0.5* \pgfdecoratedinputsegmentlength / (ceil(0.5 * \pgfdecoratedinputsegmentlength / \pgfdecorationsegmentlength ) - 0.4999)
                    }
                \else
                    \pgfmathsetmacro\matchinglength{
                        0.5 * \pgfdecoratedinputsegmentlength / (ceil(0.5 * \pgfdecoratedinputsegmentlength / \pgfdecorationsegmentlength ) )
                    }
                \fi
            \fi
            \setlength{\pgfdecorationsegmentlength}{\matchinglength pt}
        }] {}
    \state{downsine}[width=\pgfdecorationsegmentlength,next state=upsine]{
        \pgfpathsine{\pgfpoint{0.5\pgfdecorationsegmentlength}{0.5\pgfdecorationsegmentamplitude}}
        \pgfpathcosine{\pgfpoint{0.5\pgfdecorationsegmentlength}{-0.5\pgfdecorationsegmentamplitude}}
    }
    \state{upsine}[width=\pgfdecorationsegmentlength,next state=downsine]{
        \pgfpathsine{\pgfpoint{0.5\pgfdecorationsegmentlength}{-0.5\pgfdecorationsegmentamplitude}}
        \pgfpathcosine{\pgfpoint{0.5\pgfdecorationsegmentlength}{0.5\pgfdecorationsegmentamplitude}}
}
    \state{final}{}
}

\tikzset{
corner/.style={line width=1pt,dashed,draw=black,dash pattern=on 6pt off 4pt},
scalar/.style={line width=1pt,draw=black},
gluon/.style={line width=1pt,decorate, draw=GluonColor,
    decoration={complete sines,aspect=0,amplitude=1.25mm,segment length=1.5mm,start up,end up}},
gluontwo/.style={line width=1pt,decorate, draw=GluonColor,
    decoration={complete sines,aspect=0,amplitude=.7mm,segment length=1mm,start up,end up}},
ghost/.style={line width=1pt,loosely dotted,draw=black},
wilson/.style={line width=2pt,draw=black},
multiscalar/.style={line width=3pt,draw=black},
 }

\NewDocumentCommand\semiloop{O{black}mmmO{}O{above}}
{%
\draw[#1] let \p1 = ($(#3)-(#2)$) in (#3) arc (#4:({#4+180}):({0.5*veclen(\x1,\y1)})node[midway, #6] {#5};)
}

\newcommand\x{.75}

\newcommand\vertexsize{.07}

\newcommand{\IllustrationIntro}{\begin{tikzpicture}[baseline={([yshift=-1ex]current bounding box.center)}]
\draw[wilson] (0,2) -- (0,-2);
\draw[fill=white] (0,1) circle (.2);
\draw[fill=white] (0,-1) circle (.2);
\draw[fill=white] (2,0) circle (.2);
\draw[scalar,-Latex] (0,2.2) -- (0,3);
\node[] at (.35,2.5) {$\tau$};
\node[] at (2.6,-.05) {$\Op_1$};
\node[] at (-.65,1) {$\Oh_3$};
\node[] at (-.65,-1) {$\Oh_2$};
\end{tikzpicture}}

\newcommand{\OPE}{\begin{tikzpicture}[baseline={([yshift=-1ex]current bounding box.center)}]
\draw[wilson] (-1.5,0) -- (1.5,0);
\draw[wilson,SCRed] (-.75,0) -- (.75,0);
\draw[fill=white] (-.75,0) circle (.1);
\draw[fill=white] (.75,0) circle (.1);
\draw[fill=white] (0,1.25) circle (.1);
\node[] at (0,1.75) {$\Op_{\Delta_1}$};
\node[] at (-.75+.1,-.5) {$\Oh_{\Dh_2}$};
\node[] at (.75+.1,-.5) {$\Oh_{\Dh_3}$};
\node[scale=1.5] at (3.4,0) {$= \sum\limits_{\Dh} \lambdah_{\Dh_2 \Dh_3 \Dh}$};
\draw[wilson] (-.75+6,0) -- (.75+6,0);
\draw[fill=white] (0+6,0) circle (.1);
\draw[fill=white] (0+6,1.25) circle (.1);
\node[] at (0+6,1.75) {$\Op_{\Delta_1}$};
\node[] at (0+.1+6,-.5) {$\Oh_{\Dh}$};
\end{tikzpicture}}

\newcommand{\HigherTraceDiagrams}{\begin{tikzpicture}[scale = 1, baseline={([yshift=-3ex]current bounding box.center)}]
\draw[wilson] (0,0) -- (1,0);
\draw[fill=black,black] (.5,0) circle (.1);
\draw[fill=white] (.5,.2) circle (.1);
\draw[fill=white] (.5,.4) circle (.1);
\node[] at (.5,.95) {\footnotesize $\vdots$};
\end{tikzpicture}}
\newcommand{\LargeNDiagramOneB}{\begin{tikzpicture}[scale = 1, baseline={([yshift=-2.9ex]current bounding box.center)}]
\node[] at (0,12*.1+.4) {\scalebox{.7}{$\vdots$}};
\node[] at (2,12*.1+.4) {\scalebox{.7}{$\vdots$}};
\draw[wilson] (-.5,0) -- (2.5,0);
\draw[wilson] (0,0) .. controls (.8,-.5) and (1.2,-.5) .. (2,0);
\draw[fill=SEColor,SEColor] (1,-.35) circle (.15);
\draw[fill=black,black] (0,0) circle (.1);
\draw[fill=black,black] (2,0) circle (.1);
\draw[wilson] (0,6*.1) -- (2,6*.1);
\draw[fill=SEColor,SEColor] (1,6*.1) circle (.15);
\draw[fill=white] (0,6*.1) circle (.1);
\draw[fill=white] (2,6*.1) circle (.1);
\draw[wilson] (0,12*.1) -- (2,12*.1);
\draw[fill=SEColor,SEColor] (1,12*.1) circle (.15);
\draw[fill=white] (0,12*.1) circle (.1);
\draw[fill=white] (2,12*.1) circle (.1);
\end{tikzpicture}}
\newcommand{\LargeNDiagramTwoB}{\begin{tikzpicture}[scale = 1, baseline={([yshift=-.8ex]current bounding box.center)}]
\node[] at (1.5,9*.1+.4) {\scalebox{.7}{$\vdots$}};
\draw[wilson] (-.5,0) -- (3.5,0);
\draw[wilson] (0,0) .. controls (.75,-.5) .. (1.5,0);
\draw[fill=SEColor,SEColor] (.75,-.375) circle (.15);
\draw[wilson] (3,0) .. controls (2.25,-.5) .. (1.5,0);
\draw[fill=SEColor,SEColor] (2.25,-.375) circle (.15);
\draw[wilson] (0,0) .. controls (.7,-1.1) and (2.3,-1.1) .. (3,0);
\draw[fill=SEColor,SEColor] (1.5,-.825) circle (.15);
\draw[wilson] (0,0) .. controls (.75,4*.1) .. (1.5,4*.1);
\draw[fill=SEColor,SEColor] (.75,.35) circle (.15);
\draw[wilson] (0,0) .. controls (.65,9*.1) .. (1.5,8*.1);
\draw[fill=SEColor,SEColor] (.75,.8) circle (.15);
\draw[wilson] (3,0) .. controls (2.25,4*.1) .. (1.5,4*.1);
\draw[fill=SEColor,SEColor] (2.25,.35) circle (.15);
\draw[wilson] (3,0) .. controls (2.25,9*.1) .. (1.5,8*.1);
\draw[fill=SEColor,SEColor] (2.25,.8) circle (.15);
\draw[fill=black,black] (0,0) circle (.1);
\draw[fill=black,black] (1.5,0) circle (.1);
\draw[fill=black,black] (3,0) circle (.1);
\draw[fill=white] (1.5,4*.1) circle (.1);
\draw[fill=white] (1.5,8*.1) circle (.1);
\end{tikzpicture}}
\newcommand{\LargeNDiagramThreeA}{\begin{tikzpicture}[scale = 1, baseline={([yshift=-2ex]current bounding box.center)}]
\draw[wilson] (-.5,0) -- (.5,0);
\draw[wilson] (0,0) -- (0,1);
\draw[fill=SEColor,SEColor] (0,.5) circle (.15);
\draw[fill=black,black] (0,0) circle (.1);
\draw[fill=white] (0,1) circle (.1);
\end{tikzpicture}}
\newcommand{\LargeNDiagramThreeB}{\begin{tikzpicture}[scale = 1, baseline={([yshift=-2ex]current bounding box.center)}]
\draw[wilson] (-.5,0) -- (.5,0);
\draw[wilson] (0,0) -- (0,1);
\draw[wilson] (0,0) .. controls (-.6,.25) and (-.6,.75) .. (0,1);
\draw[fill=SEColor,SEColor] (0,.6) circle (.15);
\draw[fill=SEColor,SEColor] (-.45,.5) circle (.15);
\draw[fill=black,black] (0,0) circle (.1);
\draw[fill=white] (0,.2) circle (.1);
\draw[fill=white] (0,1) circle (.1);
\end{tikzpicture}}
\newcommand{\LargeNDiagramThreeC}{\begin{tikzpicture}[scale = 1, baseline={([yshift=-2ex]current bounding box.center)}]
\draw[wilson] (-.5,0) -- (.5,0);
\draw[wilson] (0,0) -- (0,1);
\draw[fill=SEColor,SEColor] (0,.6) circle (.15);
\draw[fill=black,black] (0,0) circle (.1);
\draw[fill=white] (0,.2) circle (.1);
\draw[fill=white] (0,1) circle (.1);
\end{tikzpicture}}

\newcommand{\WittenDiagrams}{\scalebox{2.5}{\begin{tikzpicture}[baseline={([yshift=-.55ex]current bounding box.center)}]
 \pgfmathsetmacro{\x}{.8*sqrt(2)/2}
 \pgfmathsetmacro{\y}{.8*sqrt(2)/2}
 \pgfmathsetmacro{\z}{.8*.5}
 \draw [wilson, SCRed] (-\x,\y) to[out=-60,in=60] (-\x,-\y);
 \draw [thick] (0,0) circle [radius=.8];
 \draw[fill=white] (.8, 0) circle (.075);
 \draw[fill=white] (-\x,\y) circle (.075);
 \draw[fill=white] (-\x,-\y) circle (.075);
 \node[scale=.5] at (.2,0) {AdS$_5$};
 \node[scale=.5,SCRed] at (-.15,-.35) {AdS$_2$};
 \node[scale=.5] at (1,.45) {$\pd$AdS$_5$};
 \node[scale=.5,SCRed] at (-1.6,-\y) {$\pd$AdS$_2$};
 \draw[-Latex,SCRed] (-1.25,-\y)--(-.75,-\y);
 \node[scale=.5] at (1.1,0) {$\Op_{\Delta_1}$};
 \node[scale=.5,SCRed] at (-\x,\y+.3) {$\Oh_{\Dh_2}$};
 \node[scale=.5,SCRed] at (-\x,-\y-.25) {$\Oh_{\Dh_3}$};
\end{tikzpicture}}}

\newcommand{\WittenDiagramOne}{\begin{tikzpicture}[baseline={([yshift=-.55ex]current bounding box.center)}]
 \pgfmathsetmacro{\x}{.8*sqrt(2)/2}
 \pgfmathsetmacro{\y}{.8*sqrt(2)/2}
 \pgfmathsetmacro{\z}{.8*.5}
 \draw [wilson, SCRed] (-\x,\y) to[out=-60,in=60] (-\x,-\y);
 \draw [thick] (0,0) circle [radius=.8];
 \draw [scalar] (-\z, 0) -- (.8, 0);
 \draw [scalar] (-\x,\y) -- (-\x,-\y);
 \draw[fill=WCOrange,WCOrange] (-.4,0) circle (\vertexsize);
 \draw[fill=white] (.8, 0) circle (.075);
 \draw[fill=white] (-\x,\y) circle (.075);
 \draw[fill=white] (-\x,-\y) circle (.075);
\end{tikzpicture}}
\newcommand{\WittenDiagramTwo}{\begin{tikzpicture}[baseline={([yshift=-.55ex]current bounding box.center)}]
 \pgfmathsetmacro{\x}{.8*sqrt(2)/2}
 \pgfmathsetmacro{\y}{.8*sqrt(2)/2}
 \pgfmathsetmacro{\z}{.8*.5}
 \draw [wilson, SCRed] (-\x,\y) to[out=-60,in=60] (-\x,-\y);
 \draw [thick] (0,0) circle [radius=.8];
 \draw [scalar] (-\z, 0) -- (.8, 0);
 \draw [scalar] (-.4,0) .. controls (-.2,.4) and (-.2,.3) .. (-\x,\y);
 \draw [scalar] (-.4,0) .. controls (-.2,-.3) and (-.2,-.4) .. (-\x,-\y);
 \draw[fill=WCOrange,WCOrange] (-.4,0) circle (\vertexsize);
 \draw[fill=white] (.8, 0) circle (.075);
 \draw[fill=white] (-\x,\y) circle (.075);
 \draw[fill=white] (-\x,-\y) circle (.075);
\end{tikzpicture}}

 \newcommand{\BulkDefectDefectLO}{\begin{tikzpicture}[baseline={([yshift=-2ex]current bounding box.center)}]
\draw[wilson] (0,0) -- (1.1,0);
\draw[scalar] (.42,0) -- (.42,1);
\draw[scalar] (.68,0) -- (.68,1);
\draw[fill=white] (.55,1) circle (.18);
\draw[fill=white] (.42,.88) circle (.075);
\draw[fill=white] (.68,.88) circle (.075);
\draw[fill=white] (.42,0) circle (.075);
\draw[fill=white] (.68,0) circle (.075);
\end{tikzpicture}}

\newcommand{\BulkDefectDefectNLOSEOne}{\begin{tikzpicture}[baseline={([yshift=-1ex]current bounding box.center)}]
\draw[wilson] (0,0) -- (1.1,0);
\draw[scalar] (.2,0) -- (.42,.88);
\draw[scalar] (.9,0) -- (.68,.88);
\draw[fill=white] (.55,1) circle (.18);
\draw[fill=white] (.42,.88) circle (.075);
\draw[fill=white] (.68,.88) circle (.075);
\draw[fill=white] (.2,0) circle (.075);
\draw[fill=white] (.9,0) circle (.075);
\draw[fill=SEColor,SEColor] (.315,.44) circle (.15);
\end{tikzpicture}}
\newcommand{\BulkDefectDefectNLOSETwo}{\begin{tikzpicture}[baseline={([yshift=-1ex]current bounding box.center)}]
\draw[wilson] (0,0) -- (1.1,0);
\draw[scalar] (.2,0) -- (.42,.88);
\draw[scalar] (.9,0) -- (.68,.88);
\draw[fill=white] (.55,1) circle (.18);
\draw[fill=white] (.42,.88) circle (.075);
\draw[fill=white] (.68,.88) circle (.075);
\draw[fill=white] (.2,0) circle (.075);
\draw[fill=white] (.9,0) circle (.075);
\draw[fill=SEColor,SEColor] (.785,.44) circle (.15);
\end{tikzpicture}}
\newcommand{\BulkDefectDefectNLOX}{\begin{tikzpicture}[baseline={([yshift=-1ex]current bounding box.center)}]
\draw[wilson] (0,0) -- (1.1,0);
\draw[scalar] (.35,0) -- (.68,.88);
\draw[scalar] (.75,0) -- (.42,.88);
\draw[fill=white] (.55,1) circle (.18);
\draw[fill=white] (.42,.88) circle (.075);
\draw[fill=white] (.68,.88) circle (.075);
\draw[fill=white] (.35,0) circle (.075);
\draw[fill=white] (.75,0) circle (.075);
\draw[fill=VertexColor,VertexColor] (.55,.53) circle (\vertexsize);
\end{tikzpicture}}
\newcommand{\BulkDefectDefectNLOH}{\begin{tikzpicture}[baseline={([yshift=-1ex]current bounding box.center)}]
\draw[wilson] (0,0) -- (1.1,0);
\draw[scalar] (.2,0) -- (.42,.88);
\draw[scalar] (.9,0) -- (.68,.88);
\draw[gluon] (.31,.44) -- (.79,.44);
\draw[fill=white] (.55,1) circle (.18);
\draw[fill=white] (.42,.88) circle (.075);
\draw[fill=white] (.68,.88) circle (.075);
\draw[fill=white] (.2,0) circle (.075);
\draw[fill=white] (.9,0) circle (.075);
\draw[fill=VertexColor,VertexColor] (.31,.44) circle (\vertexsize);
\draw[fill=VertexColor,VertexColor] (.79,.44) circle (\vertexsize);
\end{tikzpicture}}
\newcommand{\BulkDefectDefectNLOYOne}{\begin{tikzpicture}[baseline={([yshift=-1ex]current bounding box.center)}]
\draw[wilson] (0,0) -- (1.1,0);
\draw[scalar] (.35,0) -- (.42,.88);
\draw[scalar] (.75,0) -- (.68,.88);
\draw[gluon] (.385,.44) -- (0,.44);
\draw[fill=white] (.55,1) circle (.18);
\draw[fill=white] (.42,.88) circle (.075);
\draw[fill=white] (.68,.88) circle (.075);
\draw[fill=white] (.35,0) circle (.075);
\draw[fill=white] (.75,0) circle (.075);
\draw[fill=VertexColor,VertexColor] (.385,.44) circle (\vertexsize);
\draw[fill=VertexColor,VertexColor] (.15,0) circle (\vertexsize);
\draw[fill=VertexColor,VertexColor] (.95,0) circle (\vertexsize);
\draw[fill=VertexColor,VertexColor] (.55,0) circle (\vertexsize);
\end{tikzpicture}}
\newcommand{\BulkDefectDefectNLOYTwo}{\begin{tikzpicture}[baseline={([yshift=-1ex]current bounding box.center)}]
\draw[wilson] (0,0) -- (1.1,0);
\draw[scalar] (.35,0) -- (.42,.88);
\draw[scalar] (.75,0) -- (.68,.88);
\draw[gluon] (.715,.44) -- (1.1,.44);
\draw[fill=white] (.55,1) circle (.18);
\draw[fill=white] (.42,.88) circle (.075);
\draw[fill=white] (.68,.88) circle (.075);
\draw[fill=white] (.35,0) circle (.075);
\draw[fill=white] (.75,0) circle (.075);
\draw[fill=VertexColor,VertexColor] (.715,.44) circle (\vertexsize);
\draw[fill=VertexColor,VertexColor] (.15,0) circle (\vertexsize);
\draw[fill=VertexColor,VertexColor] (.95,0) circle (\vertexsize);
\draw[fill=VertexColor,VertexColor] (.55,0) circle (\vertexsize);
\end{tikzpicture}}
\newcommand{\BulkDefectDefectNLOFoneOne}{\begin{tikzpicture}[baseline={([yshift=-2ex]current bounding box.center)}]
\draw[wilson] (-.5,0) -- (1.6,0);
\draw[scalar] (-.12,0) -- (.42,.88);
\draw[scalar] (.95,.44) -- (.68,.88);
\draw[fill=white] (.55,1) circle (.18);
\draw[fill=white] (.42,.88) circle (.075);
\draw[fill=white] (.68,.88) circle (.075);
\draw[fill=white] (.35,0) circle (.075);
\draw[fill=white] (.75,0) circle (.075);
\draw[fill=VertexColor,VertexColor] (-.12,0) circle (\vertexsize);
\draw[fill=VertexColor,VertexColor] (.125,0) circle (\vertexsize);
\draw[fill=VertexColor,VertexColor] (1.2,0) circle (\vertexsize);
\path[clip] (-.4,0) rectangle (1.9,1.2);
\draw[scalar] (.55,0) circle (.2);
\draw[fill=white] (.55,1) circle (.18);
\draw[fill=white] (.42,.88) circle (.075);
\draw[fill=white] (.68,.88) circle (.075);
\draw[fill=white] (.35,0) circle (.075);
\draw[fill=white] (.75,0) circle (.075);
\draw[fill=VertexColor,VertexColor] (-.12,0) circle (\vertexsize);
\draw[fill=VertexColor,VertexColor] (.125,0) circle (\vertexsize);
\draw[fill=VertexColor,VertexColor] (1.2,0) circle (\vertexsize);
\end{tikzpicture}}

\newcommand{\BulkDefectDefectNLOFoneTwo}{\begin{tikzpicture}[baseline={([yshift=-.2ex]current bounding box.center)}]
\draw[wilson] (-.25,0) -- (1.35,0);
\draw[scalar] (.42,0) -- (.42,.88);
\draw[scalar] (.68,.44) -- (.68,.88);
\draw[fill=white] (.55,1) circle (.18);
\draw[fill=white] (.42,.88) circle (.075);
\draw[fill=white] (.68,.88) circle (.075);
\draw[fill=white] (0,0) circle (.075);
\draw[fill=white] (1.1,0) circle (.075);
\draw[fill=VertexColor,VertexColor] (.42,0) circle (\vertexsize);
\draw[fill=VertexColor,VertexColor] (.72,0) circle (\vertexsize);
\path[clip] (-.4,0) rectangle (1.9,-.41);
\draw[scalar] (.55,.2) circle (.6);
\draw[wilson] (-.1,0) -- (1.2,0);
\draw[scalar] (.42,0) -- (.42,.88);
\draw[scalar] (.68,.44) -- (.68,.88);
\draw[fill=white] (.55,1) circle (.18);
\draw[fill=white] (.42,.88) circle (.075);
\draw[fill=white] (.68,.88) circle (.075);
\draw[fill=white] (0,0) circle (.075);
\draw[fill=white] (1.1,0) circle (.075);
\draw[fill=VertexColor,VertexColor] (.42,0) circle (\vertexsize);
\draw[fill=VertexColor,VertexColor] (.72,0) circle (\vertexsize);
\end{tikzpicture}}

\newcommand{\BulkDefectDefectNLOFoneThree}{\begin{tikzpicture}[baseline={([yshift=-1ex]current bounding box.center)}]
\draw[wilson] (-.5,0) -- (1.6,0);
\draw[scalar] (.15,.44) -- (.42,.88);
\draw[scalar] (1.22,0) -- (.68,.88);
\draw[fill=white] (.55,1) circle (.18);
\draw[fill=white] (.42,.88) circle (.075);
\draw[fill=white] (.68,.88) circle (.075);
\draw[fill=white] (.35,0) circle (.075);
\draw[fill=white] (.75,0) circle (.075);
\draw[fill=VertexColor,VertexColor] (1.22,0) circle (\vertexsize);
\draw[fill=VertexColor,VertexColor] (1.45,0) circle (\vertexsize);
\path[clip] (-.4,0) rectangle (1.9,1.2);
\draw[scalar] (.55,0) circle (.2);
\draw[fill=white] (.55,1) circle (.18);
\draw[fill=white] (.42,.88) circle (.075);
\draw[fill=white] (.68,.88) circle (.075);
\draw[fill=white] (.35,0) circle (.075);
\draw[fill=white] (.75,0) circle (.075);
\draw[fill=VertexColor,VertexColor] (1.22,0) circle (\vertexsize);
\draw[fill=VertexColor,VertexColor] (1.45,0) circle (\vertexsize);
\end{tikzpicture}}

\newcommand{\BulkDefectDefectGeneralLeadingOrder}{\begin{tikzpicture}[baseline={([yshift=-1ex]current bounding box.center)}]
\draw[wilson] (-1.2,0) -- (3.2,0);
\draw[multiscalar] (.9,1.5) to[out=50+180,in=75] (0,0);
\draw[multiscalar] (1.1,1.5) to[out=-50,in=15+90] (2,0);
\draw[multiscalar] (0,0) to[out=-75,in=-15-90] (2,0);
\draw[scalar] (-.4,0) to[out=75,in=180] (1,1.55);
\draw[scalar] (-1,0) to[out=90,in=180-25] (1,1.6);
\draw[fill=VertexColor,VertexColor] (-.4,0) circle (\vertexsize);
\draw[fill=VertexColor,VertexColor] (-1,0) circle (\vertexsize);
\node[scale=.7] at (-.5,.65) {\footnotesize $\ldots$};
\draw[scalar] (.5,0) to[out=90,in=-90] (.95,1.5);
\draw[scalar] (1.5,0) to[out=90,in=-90] (1.05,1.5);
\draw[fill=VertexColor,VertexColor] (.5,0) circle (\vertexsize);
\draw[fill=VertexColor,VertexColor] (1.5,0) circle (\vertexsize);
\node[scale=.7] at (1,.5) {\footnotesize $\ldots$};
\draw[scalar] (2.4,0) to[out=15+90,in=0] (1,1.55);
\draw[scalar] (3,0) to[out=90,in=25] (1,1.6);
\draw[fill=VertexColor,VertexColor] (2.4,0) circle (\vertexsize);
\draw[fill=VertexColor,VertexColor] (3,0) circle (\vertexsize);
\node[scale=.7] at (2.5,.65) {\footnotesize $\ldots$};
\node[scale=.7] at (.225,1) {\footnotesize $a_1$};
\node[scale=.7] at (2-.225,1) {\footnotesize $a_2$};
\node[scale=.7] at (1,-.35) {\footnotesize $a_3$};
\node[scale=.7] at (2.7,1.2) {\footnotesize $a_4$};
\draw[fill=white] (1,1.5) circle (.18);
\draw[fill=white] (0,0) circle (.18);
\draw[fill=white] (2,0) circle (.18);
\end{tikzpicture}}

\newcommand{\ScalarPropagator}{\begin{tikzpicture}[baseline={([yshift=-.85ex]current bounding box.center)}]
\draw[scalar] (0,.5) -- (1,.5);
\draw[fill=white] (0,.5) circle (.075);
\draw[fill=white] (1,.5) circle (.075);
\node[] at (0,.1) {\footnotesize \makebox[\widthof{$\smash{\mu,a}$}][c]{$\smash{I,a}$}};
\node[] at (0,.9) {\footnotesize $1$};
\node[] at (1,.1) {\footnotesize \makebox[\widthof{$\smash{\mu,a}$}][c]{$\smash{J,b}$}};
\node[] at (1,.9) {\footnotesize $2$};
\end{tikzpicture}}
\newcommand{\GluonPropagator}{\begin{tikzpicture}[baseline={([yshift=-.85ex]current bounding box.center)}]
\draw[gluon] (0,.5) -- (1,.5);
\draw[fill=white] (0,.5) circle (.075);
\draw[fill=white] (1,.5) circle (.075);
\node[] at (0,.1) {\footnotesize \makebox[\widthof{$\mu,a$}][c]{$\smash{\mu,a}$}};
\node[] at (0,.9) {\footnotesize $1$};
\node[] at (1,.1) {\footnotesize \makebox[\widthof{$\nu,b$}][c]{$\smash{\nu,b}$}};
\node[] at (1,.9) {\footnotesize $2$};
\end{tikzpicture}}
\newcommand{\FermionPropagator}{\begin{tikzpicture}[baseline={([yshift=-.85ex]current bounding box.center)}]
\draw[scalar] (0,.5) -- (1,.5);
\draw[fill=white] (0,.5) circle (.075);
\draw[fill=white] (1,.5) circle (.075);
\draw[-Latex] (.62,.5)--(.63,.5);
\node[] at (0,.1) {\footnotesize \makebox[\widthof{$\smash{\mu,a}$}][c]{$\smash{a}$}};
\node[] at (0,.9) {\footnotesize $1$};
\node[] at (1,.1) {\footnotesize \makebox[\widthof{$\smash{\nu,b}$}][c]{$\smash{b}$}};
\node[] at (1,.9) {\footnotesize $2$};
\end{tikzpicture}}
\newcommand{\GhostPropagator}{\begin{tikzpicture}[baseline={([yshift=-.85ex]current bounding box.center)}]
\draw[ghost] (0,.5) -- (1,.5);
\draw[fill=white] (0,.5) circle (.075);
\draw[fill=white] (1,.5) circle (.075);
\node[] at (0,.1) {\footnotesize \makebox[\widthof{$\smash{\mu,a}$}][c]{$\smash{a}$}};
\node[] at (0,.9) {\footnotesize $1$};
\node[] at (1,.1) {\footnotesize \makebox[\widthof{$\smash{\nu,b}$}][c]{$\smash{b}$}};
\node[] at (1,.9) {\footnotesize $2$};
\end{tikzpicture}}

\newcommand{\VertexScalarScalarGluon}{\begin{tikzpicture}[baseline={([yshift=-.35ex]current bounding box.center)}]
\draw[gluon] (0,.5) -- (.75,.5);
\draw[scalar] (.75,.5) -- (1.25,1);
\draw[scalar] (.75,.5) -- (1.25,0);
\draw[fill=VertexColor,VertexColor] (.75,.5) circle (\vertexsize);
\draw[fill=white] (0,.5) circle (.075);
\draw[fill=white] (1.25,1) circle (.075);
\draw[fill=white] (1.25,0) circle (.075);
\node[anchor=west] at (1.3,1) {\footnotesize $I,a$};
\node[] at (1.25,1.3) {\footnotesize $1$};
\node[anchor=west] at (1.3,0) {\footnotesize $J,b$};
\node[] at (1.25,-.3) {\footnotesize $2$};
\node[] at (0,.1) {\footnotesize $\mu,c$};
\node[] at (0,.9) {\footnotesize $3$};
\end{tikzpicture}}
\newcommand{\VertexFermionFermionScalar}{\begin{tikzpicture}[baseline={([yshift=-.45ex]current bounding box.center)}]
\draw[scalar] (0,.5) -- (.75,.5);
\draw[scalar] (.75,.5) -- (1.25,1);
\draw[-Latex] (1.05,.8)--(.95,.7);
\draw[scalar] (.75,.5) -- (1.25,0);
\draw[-Latex] (1,.25)--(1.1,.15);
\draw[fill=VertexColor,VertexColor] (.75,.5) circle (\vertexsize);
\draw[fill=white] (0,.5) circle (.075);
\draw[fill=white] (1.25,1) circle (.075);
\draw[fill=white] (1.25,0) circle (.075);
\end{tikzpicture}}
\newcommand{\VertexFermionFermionGluon}{\begin{tikzpicture}[baseline={([yshift=-.35ex]current bounding box.center)}]
\draw[gluon] (0,.5) -- (.75,.5);
\draw[scalar] (.75,.5) -- (1.25,1);
\draw[-Latex] (1.05,.8)--(.95,.7);
\draw[scalar] (.75,.5) -- (1.25,0);
\draw[-Latex] (1,.25)--(1.1,.15);
\draw[fill=VertexColor,VertexColor] (.75,.5) circle (\vertexsize);
\draw[fill=white] (0,.5) circle (.075);
\draw[fill=white] (1.25,1) circle (.075);
\draw[fill=white] (1.25,0) circle (.075);
\end{tikzpicture}}
\newcommand{\VertexGhostGhostGluon}{\begin{tikzpicture}[baseline={([yshift=-.35ex]current bounding box.center)}]
\draw[gluon] (0,.5) -- (.75,.5);
\draw[ghost] (.75,.5) -- (1.25,1);
\draw[ghost] (.75,.5) -- (1.25,0);
\draw[fill=VertexColor,VertexColor] (.75,.5) circle (\vertexsize);
\draw[fill=white] (0,.5) circle (.075);
\draw[fill=white] (1.25,1) circle (.075);
\draw[fill=white] (1.25,0) circle (.075);
\end{tikzpicture}}
\newcommand{\VertexGluonGluonGluon}{\begin{tikzpicture}[baseline={([yshift=-.35ex]current bounding box.center)}]
\draw[gluon] (0,.5) -- (.75,.5);
\draw[gluon] (.75,.5) -- (1.25,1);
\draw[gluon] (.75,.5) -- (1.25,0);
\draw[fill=VertexColor,VertexColor] (.75,.5) circle (\vertexsize);
\draw[fill=white] (0,.5) circle (.075);
\draw[fill=white] (1.25,1) circle (.075);
\draw[fill=white] (1.25,0) circle (.075);
\end{tikzpicture}}
\newcommand{\VertexFourScalars}{\begin{tikzpicture}[baseline={([yshift=-.35ex]current bounding box.center)}]
\draw[scalar] (0,0) -- (1.25,1);
\draw[scalar] (0,1) -- (1.25,0);
\draw[fill=VertexColor,VertexColor] (.625,.5) circle (\vertexsize);
\draw[fill=white] (0,0) circle (.075);
\draw[fill=white] (0,1) circle (.075);
\draw[fill=white] (1.25,1) circle (.075);
\draw[fill=white] (1.25,0) circle (.075);
\node[anchor=west] at (1.3,1) {\footnotesize $I,a$};
\node[] at (1.25,1.3) {\footnotesize $1$};
\node[anchor=west] at (1.3,0) {\footnotesize $J,b$};
\node[] at (1.25,-.3) {\footnotesize $2$};
\node[anchor=east] at (-0.05,1) {\footnotesize $K,c$};
\node[] at (0,1.3) {\footnotesize $3$};
\node[anchor=east] at (-0.05,0) {\footnotesize $L,d$};
\node[] at (0,-.3) {\footnotesize $4$};
\end{tikzpicture}}
\newcommand{\VertexScalarScalarGluonGluon}{\begin{tikzpicture}[baseline={([yshift=-.35ex]current bounding box.center)}]
\draw[scalar] (.625,.5) -- (1.25,1);
\draw[scalar] (.625,.5) -- (1.25,0);
\draw[gluon] (0,0) -- (.625,.5);
\draw[gluon] (0,1) -- (.625,.5);
\draw[fill=VertexColor,VertexColor] (.625,.5) circle (\vertexsize);
\draw[fill=white] (0,0) circle (.075);
\draw[fill=white] (0,1) circle (.075);
\draw[fill=white] (1.25,1) circle (.075);
\draw[fill=white] (1.25,0) circle (.075);
\end{tikzpicture}}
\newcommand{\VertexGluonGluonGluonGluon}{\begin{tikzpicture}[baseline={([yshift=-.35ex]current bounding box.center)}]
\draw[gluon] (.625,.5) -- (1.25,1);
\draw[gluon] (.625,.5) -- (1.25,0);
\draw[gluon] (0,0) -- (.625,.5);
\draw[gluon] (0,1) -- (.625,.5);
\draw[fill=VertexColor,VertexColor] (.625,.5) circle (\vertexsize);
\draw[fill=white] (0,0) circle (.075);
\draw[fill=white] (0,1) circle (.075);
\draw[fill=white] (1.25,1) circle (.075);
\draw[fill=white] (1.25,0) circle (.075);
\end{tikzpicture}}

\newcommand{\SelfEnergyNoText}{\begin{tikzpicture}[baseline={([yshift=-.35ex]current bounding box.center)}]
\draw[scalar] (0,.5) -- (1,.5);
\draw[fill=white] (0,.5) circle (.075);
\draw[fill=white] (1,.5) circle (.075);
\draw[fill=SEColor,SEColor] (.5,.5) circle (.15);
\end{tikzpicture}}
\newcommand{\SelfEnergyDiagramOne}{\begin{tikzpicture}[baseline={([yshift=-.4ex]current bounding box.center)}]
\draw[scalar] (0,.5) -- (2,.5);
\draw[gluon] (.5,.5) arc (180:-20:.5);
\draw[fill=white] (0,.5) circle (.075);
\draw[fill=white] (2,.5) circle (.075);
\draw[fill=VertexColor,VertexColor] (.5,.5) circle (\vertexsize);
\draw[fill=VertexColor,VertexColor] (1.5,.5) circle (\vertexsize);
\end{tikzpicture}}
\newcommand{\SelfEnergyDiagramTwo}{\begin{tikzpicture}[baseline={([yshift=-.5ex]current bounding box.center)}]
\draw[scalar] (0,.5) -- (.5,.5);
\draw[scalar] (1.5,.5) -- (2,.5);
\draw[scalar,postaction={decorate}] (.5,.5) arc (180:0:.5);
\draw[-Latex] (1.1,1)--(1.11,1);
\draw[scalar,postaction={decorate}] (.5,.5) arc (-180:0:.5);
\draw[-Latex] (.9,0)--(.89,0);
\draw[fill=white] (0,.5) circle (.075);
\draw[fill=white] (2,.5) circle (.075);
\draw[fill=VertexColor,VertexColor] (.5,.5) circle (\vertexsize);
\draw[fill=VertexColor,VertexColor] (1.5,.5) circle (\vertexsize);
\end{tikzpicture}}
\newcommand{\SelfEnergyDiagramThree}{\begin{tikzpicture}[baseline={([yshift=-.75ex]current bounding box.center)}]
\draw[scalar] (0,.5) -- (2,.5);
\draw[scalar] (1,1) circle (.5);
\draw[fill=white] (0,.5) circle (.075);
\draw[fill=white] (2,.5) circle (.075);
\draw[fill=VertexColor,VertexColor] (1,.5) circle (\vertexsize);
\end{tikzpicture}}
\newcommand{\SelfEnergyDiagramFour}{\begin{tikzpicture}[baseline={([yshift=-1.05ex]current bounding box.center)}]
\draw[scalar] (0,.5) -- (2,.5);
\draw[gluon] (1,1) circle (.5);
\draw[fill=white] (0,.5) circle (.075);
\draw[fill=white] (2,.5) circle (.075);
\draw[fill=VertexColor,VertexColor] (1,.5) circle (\vertexsize);
\end{tikzpicture}}

\newcommand{\DefectVertexOnePointScalar}{\begin{tikzpicture}[baseline={([yshift=-3ex]current bounding box.center)}]
\draw[wilson] (0,0) -- (1.5,0);
\draw[scalar] (.75,0) -- (.75,1);
\draw[fill=white] (.75,1) circle (.075);
\draw[fill=white] (.3,0) circle (.075);
\draw[fill=white] (1.2,0) circle (.075);
\draw[fill=VertexColor,VertexColor] (.75,0) circle (\vertexsize);
\node[anchor=east] at (.7,1) {\footnotesize $I$};
\node[] at (1.05,1) {\footnotesize $1$};
\node[] at (.3,-.3) {\footnotesize $2$};
\node[] at (1.2,-.3) {\footnotesize $3$};
\end{tikzpicture}}
\newcommand{\DefectVertexOnePointGluon}{\begin{tikzpicture}[baseline={([yshift=-1ex]current bounding box.center)}]
\draw[wilson] (0,0) -- (1.5,0);
\draw[gluon] (.75,0) -- (.75,1);
\draw[fill=white] (.75,1) circle (.075);
\draw[fill=white] (.3,0) circle (.075);
\draw[fill=white] (1.2,0) circle (.075);
\draw[fill=VertexColor,VertexColor] (.75,0) circle (\vertexsize);
\node[anchor=east] at (.7,1) {\footnotesize $\mu$};
\node[] at (1.05,1) {\footnotesize $1$};
\node[] at (.3,-.3) {\footnotesize $2$};
\node[] at (1.2,-.3) {\footnotesize $3$};
\end{tikzpicture}}

 \newcommand{\ONDiagramNLO}{\begin{tikzpicture}[baseline={([yshift=-2ex]current bounding box.center)}]
\draw[wilson] (0,0) -- (1.1,0);
\draw[scalar] (.3,0) -- (.55,.5);
\draw[scalar] (.8,0) -- (.55,.5);
\draw[scalar] (.3,0) -- (.55,.5);
\draw[scalar] (.95,0.3) -- (.55,.5);
\draw[scalar] (.55,.5) -- (.55,1);
\draw[fill=white] (.55,1) circle (.1);
\draw[fill=white] (.55,.85) circle (.075);
\draw[fill=white] (.3,0) circle (.075);
\draw[fill=white] (.8,0) circle (.075);
\draw[fill=VertexColor,VertexColor] (.55,.48) circle (\vertexsize);
\draw[fill=VertexColor,VertexColor] (1,0) circle (\vertexsize);
\draw[fill=VertexColor,VertexColor] (.1,0) circle (\vertexsize);
\draw[fill=VertexColor,VertexColor] (.55,0) circle (\vertexsize);
\end{tikzpicture}}

\newcommand{\XIntegral}{\begin{tikzpicture}[baseline={([yshift=-.35ex]current bounding box.center)}]
\draw[scalar] (0,0) -- (1.25,1);
\draw[scalar] (0,1) -- (1.25,0);
\draw[fill=VertexColor,VertexColor] (.625,.5) circle (\vertexsize);
\draw[fill=white] (0,0) circle (.075);
\draw[fill=white] (0,1) circle (.075);
\draw[fill=white] (1.25,1) circle (.075);
\draw[fill=white] (1.25,0) circle (.075);
\end{tikzpicture}}
\newcommand{\YIntegral}{\begin{tikzpicture}[baseline={([yshift=-.45ex]current bounding box.center)}]
\draw[scalar] (0,.5) -- (.75,.5);
\draw[scalar] (.75,.5) -- (1.25,1);
\draw[scalar] (.75,.5) -- (1.25,0);
\draw[fill=VertexColor,VertexColor] (.75,.5) circle (\vertexsize);
\draw[fill=white] (0,.5) circle (.075);
\draw[fill=white] (1.25,1) circle (.075);
\draw[fill=white] (1.25,0) circle (.075);
\end{tikzpicture}}
\newcommand{\IntegralYOneTwoTwo}{\scalebox{\x}{\begin{tikzpicture}[baseline={([yshift=-.55ex]current bounding box.center)}]
\draw[scalar] (0,.5) -- (1,.5);
\draw[scalar] (1.5,.5) circle (.5);
\draw[fill=white] (0,.5) circle (.075);
\draw[fill=white] (2,.5) circle (.075);
\draw[fill=VertexColor,VertexColor] (1,.5) circle (\vertexsize);
\end{tikzpicture}}}
\newcommand{\IntegralXOneTwoThreeThree}{\scalebox{\x}{\begin{tikzpicture}[baseline={([yshift=-.55ex]current bounding box.center)}]
\draw[scalar] (.5,1) -- (1,.5);
\draw[scalar] (.5,0) -- (1,.5);
\draw[scalar] (1.5,.5) circle (.5);
\draw[fill=white] (.5,1) circle (.075);
\draw[fill=white] (.5,0) circle (.075);
\draw[fill=white] (2,.5) circle (.075);
\draw[fill=VertexColor,VertexColor] (1,.5) circle (\vertexsize);
\end{tikzpicture}}}
\newcommand{\IntegralFOneThreeTwoThree}{\scalebox{\x}{\begin{tikzpicture}[baseline={([yshift=-.55ex]current bounding box.center)}]
\draw[scalar] (0,0) -- (.625,0);
\draw[scalar] (0,1) -- (.625,1);
\draw[scalar] (.625,0) arc (270:450:.5);
\draw[scalar] (.625,0) -- (.625,1);
\draw[fill=VertexColor,VertexColor] (.625,0) circle (\vertexsize);
\draw[fill=VertexColor,VertexColor] (.625,1) circle (\vertexsize);
\draw[fill=white] (0,0) circle (.075);
\draw[fill=white] (0,1) circle (.075);
\draw[fill=white] (1.125,.5) circle (.075);
\end{tikzpicture}}}

\usepackage{simplewick}
\usepackage{dsfont}
\usepackage{float}
\usepackage{pgfplots}
\pgfplotsset{compat=1.14}

\let\oldbfseries=\bfseries
\let\oldmdseries=\mdseries
\let\oldnormalfont=\normalfont
\renewcommand{\bfseries}{\oldbfseries\boldmath}
\renewcommand{\mdseries}{\oldmdseries\unboldmath}
\renewcommand{\normalfont}{\oldnormalfont\unboldmath}

\makeatletter
\newlength{\apb@width}
\newcommand{\autoparbox}[2][c]{\settowidth{\apb@width}{#2}\parbox[#1]{\apb@width}{#2}}

\makeatother

\DeclareMathOperator{\tr}{tr}

\def\Am{{\mathcal{A}}}
\def\Bm{{\mathcal{B}}}
\def\Cm{{\mathcal{C}}}

\def\Gm{{\mathcal{G}}}

\def\Km{{\mathcal{K}}}

\def\Nm{{\mathcal{N}}}
\def\Om{{\mathcal{O}}}
\def\Pm{{\mathcal{P}}}

\def\zb{{\bar{z}}}

\def\veps{\varepsilon}

\def\pd{\partial}

\newcommand{\beq}{\begin{equation}}
\newcommand{\eeq}{\end{equation}}

\definecolor{nicegreen}{rgb}{0.1,0.6,0.1}

\newmuskip\pFqskip
\mathchardef\pFcomma=\mathcode`,

\makeatletter
\renewcommand*\env@matrix[1][\arraystretch]{%
  \edef\arraystretch{#1}%
  \hskip -\arraycolsep
  \let\@ifnextchar\new@ifnextchar
  \array{*\c@MaxMatrixCols c}}
\makeatother

\graphicspath{{./figures/}}

\begin{document} 

\thispagestyle{empty}

\vspace*{-.6in}
\begin{flushright}
    HU-EP-24/27-RTG | DESY-24-166
\end{flushright}

\vspace{1cm}

{\large
\begin{center}
    {\Large \bf Perturbative bootstrap of the Wilson-line defect CFT: Bulk-defect-defect correlators
    }\\
\end{center}}

\vspace{0.5cm}

\begin{center}
    Daniele Artico$^{a,}\footnote{daniele.artico@physik.hu-berlin.de}$ \orcidlink{0009-0000-5016-9291},
    Julien Barrat $^{b,}\footnote{julien.barrat@desy.de}$ \orcidlink{0000-0003-3679-8301} and
    Yingxuan Xu $^{a,}\footnote{yingxuan.xu@physik.hu-berlin.de}$ \orcidlink{0000-0001-6135-8864}
    \\[0.5cm] 
    { \small
    $^{a}$ Institut f\"ur Physik und IRIS Adlershof, Humboldt-Universit{\"a}t zu Berlin, Zum Gro{\ss}en Windkanal 2, 12489 Berlin, Germany\\ 
    \vspace{0.3em}
    $^{b}$ Deutsches Elektronen-Synchrotron DESY, Notkestr. 85, 22607 Hamburg, Germany\\
    }
    \vspace{1cm} 

    \bf Abstract
\end{center}

\begin{abstract}
    \noindent We study the correlators of bulk and defect half-BPS operators in $\Nm=4$ Super Yang-Mills theory with a Maldacena-Wilson line defect, focusing on the case involving one bulk and two defect local operators.
    We analyze the non-perturbative constraints on these correlators, which include a topological sector, pinching and splitting limits, and we compute a variety of bulk-defect-defect correlators up to next-to-leading order at weak coupling, surprisingly observing that transcendental terms cancel.
    Additionally we provide results in the strong-coupling regime for the first two leading orders using a mixture of Witten diagrams and non-perturbative constraints.
\end{abstract}

\newpage

\setcounter{tocdepth}{2}
\tableofcontents
\thispagestyle{empty}
%\maketitle

\newpage

\setcounter{page}{1}

\section{Introduction}
\label{sec:Introduction}
\setcounter{footnote}{0} 
Defects are central to a broad range of physical theories, from condensed-matter systems to high-energy physics.
Despite the presence of a defect, the local structure of a theory remains intact, preserving key features that allow for instance the application of non-perturbative techniques.
In critical systems, conformal defects retain a significant portion of the underlying conformal symmetry, enabling the use of modern tools like the conformal bootstrap to formulate constraints on observables.
Although numerical studies face challenges due to the absence of positivity, substantial progress has been made on the analytic side, inspired by the development of Lorentzian inversion formulas and dispersion relations in conformal field theories (CFTs) without defects \cite{Caron-Huot:2017vep,Carmi:2019cub}.
In recent years, a wide variety of defect CFTs has been investigated , with Wilson lines emerging as crucial probes in both the AdS/CFT correspondence \cite{Maldacena:1997re,Maldacena:1998im} and the study of quark confinement \cite{Polyakov:1978vu,Witten:1998zw}.

In this context, the configuration involving a Maldacena-Wilson line in four-dimensional $\Nm=4$ super Yang-Mills (SYM) theory holds a particularly prominent position.
This setup retains many key features of its parent theory, including one-dimensional conformal symmetry, supersymmetry and integrability.
Recent studies have focused on two canonical configurations: the two-point functions of bulk operators in the presence of the Wilson line and multipoint correlators of defect operators.
For the two-point functions, correlators involving half-BPS operators have been studied at weak and strong coupling, employing analytical bootstrap methods \cite{Barrat:2021yvp,Barrat:2022psm,Bianchi:2022ppi,Meneghelli:2022gps,Gimenez-Grau:2023fcy}, perturbative techniques \cite{Barrat:2020vch}, and exact results obtained through localization in a specific kinematic regime known as topological \cite{Drukker:2007yx,Giombi:2009ds,Giombi:2009ek,Buchbinder:2012vr,Beccaria:2020ykg}.\footnote{More generally, two-point functions of bulk operators in the presence of a defect have been studied in \cite{Billo:2016cpy}, with further results on $n$-point correlation functions (including bulk-defect-defect) to be found in \cite{Lauria:2020emq}.}
In the context of multipoint correlators, extensive work has been conducted on four-point functions of defect half-BPS operators using modern approaches that combine numerical conformal bootstrap and integrability \cite{Cavaglia:2021bnz,Cavaglia:2022qpg,Cavaglia:2022yvv,Cavaglia:2023mmu}.
Strong-coupling results have also been derived through direct computations \cite{Giombi:2017cqn,Giombi:2023zte} or with the analytic bootstrap \cite{Liendo:2016ymz,Liendo:2018ukf,Ferrero:2021bsb,Ferrero:2023znz,Ferrero:2023gnu}.
Meanwhile, exact results have been obtained with localization techniques \cite{Giombi:2018qox}.
Additionally, studies of higher-point functions at weak coupling have led to conjectures about superconformal Ward identities \cite{Barrat:2021tpn,Barrat:2022eim}, which have been confirmed and expanded upon \cite{Bliard:2024und,Barrat:2024ta}, and leading to new results for five- and six-point functions \cite{Bliard:2023zpe,Peveri:2023qip,Barrat:2024nod,Barrat:2024ta2,Artico:2024wut}.

\begin{figure}
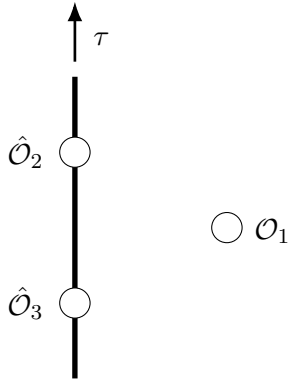

    \centering
    \IllustrationIntro
    \caption{Illustration of a bulk-defect-defect correlator $\vev{\Op_1 \Oh_2 \Oh_3}$. The two \textit{defect} operators $\Oh_{2,3}$ are representations of the one-dimensional CFT preserved by the line, while the \textit{bulk} operator $\Op_1$ on the right lives in the four-dimensional space of $\Nm=4$ SYM.}
    \label{fig:BulkDefectDefect}
\end{figure}

The configurations discussed above are generally considered the simplest correlators with non-trivial kinematics in defect CFTs.
However, there is another canonical setup that has so far received little attention: correlators of one bulk and two defect operators.
This configuration is particularly interesting in the context of the Wilson-line defect CFT, as it depends on just one spacetime cross-ratio and one $R$-symmetry variable for half-BPS operators.
In contrast, two-point functions of bulk operators rely on two spacetime cross-ratios and one $R$-symmetry variable, while four-point functions of defect operators depend on one spacetime cross-ratio and two $R$-symmetry variables.
To date, bulk-defect-defect correlators have been explored primarily in scalar theories with line defects \cite{Lauria:2020emq}, with some insights drawn from locality constraints \cite{Levine:2023ywq,Levine:2024wqn} and conformal block expansions \cite{Buric:2020zea,Okuyama:2024tpg}.\footnote{See also \cite{Gimenez-Grau:2020jvf,Chen:2023oax} for analyses involving boundaries.} The kinematics studied in \cite{Levine:2023ywq,Levine:2024wqn} are analogous to the one of two-point functions in CFT on real projective space studied in \cite{Giombi:2020xah,Zhou:2024ekb}.\footnote{We thank Xinan Zhou for pointing out this correspondence.}
In the context of the Wilson-line defect CFT, the localization machinery has been developed in \cite{Giombi:2018hsx} for calculating these correlators exactly in a special kinematic limit.
The setup is represented in Figure \ref{fig:BulkDefectDefect}.

In this paper we initiate the study of bulk-defect-defect correlators in the Wilson-line defect CFT, focusing on the simplest case where both the bulk and defect operators are half-BPS.
Through a heuristic approach we derive differential constraints (that we interpret as superconformal Ward identities), and demonstrate their equivalence to the existence of a topological sector.
These constraints prove useful in systematically eliminating one $R$-symmetry channel.
We also examine specific limits of these correlators, where they reduce either to bulk-defect two-point functions or to the product of bulk one-point and defect two-point functions. We extend the study of the kinematic limit presented in \cite{Giombi:2018hsx} to NLO, introducing its use in this context as a tool to fix constants and to perform checks. We then study the weak and strong coupling expansions of bulk-defect-defect correlators. At weak coupling, we show that the number of $R$-symmetry channels increases order by order in perturbation theory.
At next-to-leading order, we fully determine the correlators by focusing on the simplest $R$-symmetry channel, which avoids bulk vertices.
Notably, the results contain no transcendental functions, despite their potential presence in individual diagrams.
At strong coupling, we focus on the specific correlator $\vev{2\hat{1}\hat{1}}$ and give explicit results for the correlator up to next-to-leading order using a mixture of Witten diagrams and non-perturbative constraints.

The structure of the paper is as follows.
In Section \ref{sec:Preliminaries}, we establish the foundational elements necessary for the computations in this work.
Section \ref{sec:NonPerturbativeConstraints} compiles the non-perturbative constraints that govern the correlators.
Perturbative results in the weak- and strong-coupling regimes are presented in Section \ref{sec:PerturbativeResults}.
In Section \ref{sec:Conclusions} we summarize the main findings and explore potential future directions. Appendix \ref{app:Integrals} provides the integrals required for the computations discussed in the main text.
In Appendix \ref{app:CheckThroughFeynmanDiagrams}, we perform a check of our weak-coupling results through direct Feynman diagram calculations.
Appendix~\ref{app:WittenIntegrals} presents computational details for the integrals present in the Witten diagrams.

\section{Preliminaries}
\label{sec:Preliminaries}

This section reviews the Wilson-line defect CFT and the correlators of local operators in the presence of a defect.
We begin by presenting the theory both without and with the defect, outlining the Feynman rules and discussing special representations of the symmetry group, known as half-BPS operators.
Next we provide a concise overview of the correlators involving bulk and defect half-BPS operators, first treating them separately and then considering them together.

\subsection{The Wilson-line defect CFT}
\label{subsec:TheWilsonLineDefectCFT}

We start by outlining the key aspects of $\Nm=4$ SYM and the Maldacena-Wilson line.
This includes presenting the action, its symmetries, and the corresponding bulk and defect Feynman rules.
We then review the special role played by half-BPS operators within this defect CFT.

\subsubsection{The action}
\label{subsubsec:TheAction}

We consider as a bulk action $\Nm=4$ SYM in four dimensions.
The theory consists of six scalar fields $\phi^{I=1, \ldots, 6}$, four Weyl fermions and one gauge field.
The action is given by \cite{DAuria:1981zjr}
\begin{equation}
    \begin{split}
    S
    =\ &
    \frac{1}{g^2} \tr \int d^4 x\, \biggl( 
    \frac{1}{2} F_{\mu\nu} F_{\mu\nu}
    + D_\mu \phi^I D_\mu \phi^I
    - \frac{1}{2} [ \phi^I, \phi^J ] [ \phi^I, \phi^J ] \\
    &+
    i\, \psib \slashed{D} \psi
    + \psib \Gamma^I [ \phi^I\,, \psi ]
    + \pd_\mu \cb D_\mu c
    + \xi ( \pd_\mu A_\mu )^2
    \biggr)\,,
    \end{split}
    \label{eq:BulkAction}
\end{equation}
where the Weyl fermions have been compactly written in the form of a single $16$-component Majorana fermion.
The repeated indices are contracted with the flat Euclidean metric $\delta_{\mu\nu}$.
$\Nm=4$ SYM is conformal at the quantum level, and its supersymmetry algebra is $\mathfrak{psu}(2,2|4)$.
We work in the Feynman gauge where the gauge parameter is $\xi = 1$.
In the following we focus on the large $N$ limit of the $\mathfrak{su}(N)$ gauge group and study the perturbative expansions for the coupling
\begin{equation}
    \lambda
    =
    g^2 N\,.
    \label{eq:CouplingConstant}
\end{equation}
At small $\lambda$ $\Nm=4$ SYM is conjectured to be dual to a strongly-coupled string theory in five dimensions \cite{Maldacena:1997re}.

The symmetries of $\Nm=4$ SYM are broken when a line defect is included.
Here, we choose the defect to be the Maldacena-Wilson line, defined as \cite{Maldacena:1998im}
\begin{equation}
    \Wl
    =
    \frac{1}{N}
    \tr \Pm \exp \int_{-\infty}^\infty d\tau\,
    (i \dot{x}_\mu A_\mu (\tau) + |\dot{x}|\, \theta \cdot \phi (\tau))\,,
    \label{eq:WilsonLine}
\end{equation}
with the spacetime orientation determined by
\begin{equation}
    \dot{x}_\mu
    =
    (1,0,0,0)\,,
    \label{eq:LineOrientation}
\end{equation}
and where we choose the $R$-symmetry polarization vector $\theta$ to be
\begin{equation}
    \theta
    =
    (0,0,0,0,0,1)\,.
    \label{eq:theta}
\end{equation}
These choices are arbitrary and do not lead to a loss of generality.
Equation~\eqref{eq:theta} simply defines which scalar fields couple to the line.
The Maldacena-Wilson line is an extended half-BPS operator with the protected expectation value \cite{Erickson:2000af}
\begin{equation}
    \vev{ \Wl } = 1\,.
    \label{eq:Wl_vev}
\end{equation}
This operator breaks the conformal algebra in the following manner:
\begin{equation}
    \mathfrak{so} (5,1) \to \mathfrak{so} (2,1) \times \mathfrak{so} (3)\,,
    \label{eq:SymmetryBreaking}
\end{equation}
where the first term on the right-hand side corresponds to the one-dimensional CFT preserved along the line while the second one refers to rotations around the defect.
We associate to $\mathfrak{so} (2,1)$ the quantum number $\Dh$, which should be understood as the scaling dimension of the defect operators.
For $\mathfrak{so}(3)$ the quantum number is $s$, which is commonly referred to as transverse spin.
On the left-hand side, bulk operators will be determined by their quantum numbers $\Delta$ (the $4d$ scaling dimension) and $\ell$ (the spin).
As we explain in more detail in Section \ref{subsubsec:HalfBPSOperators}, the representations of $\mathfrak{su} (N)$ are distinguished through their trace structure.

The $R$-symmetry algebra is also broken by the presence of the defect:
\begin{equation}
    \mathfrak{so}_R (6) \to \mathfrak{so}_R (5)\,.
    \label{eq:RSymmetryBreaking}
\end{equation}
We associate the quantum numbers $k$ and $\kh$ to $\mathfrak{so} (6)_R$ and to $\mathfrak{so} (5)_R$, respectively.
All in all the supersymmetry algebra $\mathfrak{psu}(2,2|4)$ breaks into the defect algebra $\mathfrak{osp}(4^*|4)$.

\subsubsection{Feynman rules}
\label{subsubsec:FeynmanRules}

In this section we list the Feynman rules of the $\Nm=4$ SYM theory and of the Wilson-line defect CFT generated by the action \eqref{eq:BulkAction} together with the extended operator \eqref{eq:WilsonLine}.

\paragraph{Bulk Feynman rules.}
The free propagators of the $\Nm=4$ SYM fields are given by
\begin{equation}
    \begin{split}
    \text{Scalars:} \qquad 
    & \ScalarPropagator = g^2 \delta^{IJ} \delta^{ab}\, I_{12}\,, \\
    \text{Gluons:} \qquad 
    & \GluonPropagator = g^2 \delta_{\mu\nu} \delta^{ab}\, I_{12}\,, \\
    \text{Fermions:} \qquad 
    & \FermionPropagator = i g^2 \delta^{ab} \slashed{\pd}_1 I_{12}\,, \\
    \text{Ghosts:} \qquad 
    & \GhostPropagator = g^2 \delta^{ab} I_{12}\,,
    \end{split}
    \label{eq:Propagators}
\end{equation}
where we defined the four-dimensional scalar propagator as
\begin{equation}
    I_{ij} = \frac{1}{4\pi^2 x_{ij}^2}\,.
    \label{eq:PropagatorFunction4d}
\end{equation}
In this work, we make explicit use of the following vertices, whose Feynman rules in the form of insertion rules have been given in \cite{Beisert:2002bb,Drukker:2009sf}:
\begin{align}
    \VertexScalarScalarGluon
    &=
    - g^4 f^{abc} \delta^{IJ} (\pd_1 - \pd_2)_\mu Y_{123}\,, \label{eq:VertexScalarScalarGluon} \\
    \VertexFourScalars
    &=
    - g^6
    \left\lbrace f^{abe}f^{cde} \left( \delta^{IK}\delta^{JL}
    -
    \delta^{IL}\delta^{JK} \right)
    +
    f^{ace}f^{bde} \left( \delta^{IJ}\delta^{KL}
    -
    \delta^{IL}\delta^{JK} \right) \right. \notag \\[-1.5em]
    &\phantom{=\ }
    \left. + f^{ade}f^{bce} \left( \delta^{IJ}\delta^{KL}
    -
    \delta^{IK}\delta^{JL} \right) \right\rbrace X_{1234}\,,
    \label{eq:VertexFourScalars}
\end{align}
while, for completion, we simply mention the other ones:
\begin{equation}
    \VertexFermionFermionScalar \quad
    \VertexGluonGluonGluon \quad
    \VertexFermionFermionGluon \quad
    \VertexGhostGhostGluon\ \quad
    \VertexScalarScalarGluonGluon \quad
    \VertexGluonGluonGluonGluon\ .
    \label{eq:MoreVertices}
\end{equation}
In our case, these are only relevant for self-energy calculations, which we give here in the form of an insertion rule.
For instance the scalar self-energy is given by \cite{Erickson:2000af}
\begin{equation}
    \begin{split}
    \SelfEnergyNoText &=
    \SelfEnergyDiagramOne
    +
    \SelfEnergyDiagramTwo
    +
    \SelfEnergyDiagramThree
    +
    \SelfEnergyDiagramFour \\
    &=
    - 2 g^4 N \delta^{ab} \delta^{IJ}\, Y_{112}\,.
    \end{split}
    \label{eq:SelfEnergy}
\end{equation}
Further insertion rules can be found, e.g., in the appendix of \cite{Barrat:2021tpn}.

\paragraph{Defect Feynman rules.}
The Maldacena-Wilson line introduces new Feynman rules in the theory, in particular in the form of integrated vertices coupling the line to the allowed fields.
These rules are given by
\begin{align}
    \DefectVertexOnePointScalar
    &\sim
    \delta^{I6} \int_{\tau_2}^{\tau_3} d\tau_4\, I_{14}\,,
    \label{eq:DefectVertexOnePointScalar} \\
    \DefectVertexOnePointGluon
    &\sim
    i \delta^{\mu 0} \int_{\tau_2}^{\tau_3} d\tau_4\, I_{14}\,,
    \label{eq:DefectVertexOnePointGluon}
\end{align}
where the contribution from the gauge group depends on the number of insertions.

\subsubsection{Half-BPS operators}
\label{subsubsec:HalfBPSOperators}

We now consider a special case of representations of $\Nm=4$ SYM with and without the defect.
These operators preserve half of the supersymmetry and are called half-BPS operators.

\paragraph{Bulk operators.}
We consider in this work correlation functions that involve scalar \textit{single-trace} half-BPS operators, which can be defined as
\begin{equation}
    \Op_{\Delta} (u,x)
    =
    \frac{1}{\sqrt{n_{\Delta}}} \tr \left( u \cdot \phi (x) \right)^\Delta\,.
    \label{eq:SingleTraceHalfBPSOperators_Bulk}
\end{equation}
where the representation is made symmetric and traceless through the condition
\begin{equation}
    u^2
    =
    0\,.
    \label{eq:u_Bulk}
\end{equation}
These operators are protected and have integer scaling dimensions $\Delta = k$.
The normalization constants can be evaluated to be \cite{Semenoff:2002kk}
\begin{equation}
    n_{\Delta}
    =
    \frac{\Delta \lambda^\Delta}{2^{3\Delta} \pi^{2\Delta}}\,.
    \label{eq:NormalizationConstantHalfBPS_Bulk}
\end{equation}
In this work we also consider the AdS dual of the operator $\Om_2 (u,x)$.
Following the notation of~\cite{Gimenez-Grau:2023fcy} we express the dual operator as
\begin{equation}
    \Om_2 (u,x) = u^I u^J \Cm_{IJ}^{\tilde{I}} s_2^{\tilde{I}} (x)\,,
    \label{eq:AdSDualTo20}
\end{equation}
with $\tilde{I} = 1, \ldots, 20$.
The tensors $\Cm_{IJ}^{\tilde{I}}$ form a symmetric traceless basis, which is chosen such that
\begin{equation}
    \Cm_{IJ}^{\tilde{I}} \Cm_{IJ}^{\tilde{J}} = \delta^{\tilde{I}\tilde{J}}\,,
    \label{eq:CmCompletenessRelation}
\end{equation}
and related to spherical harmonics via
\begin{equation}
    Y_2^{\tilde{I}} = \Cm_{IJ}^{\tilde{I}} \theta^I \theta^J\,.
    \label{eq:RelationYToCm}
\end{equation}
Note also that the $s_2$ operator is chosen such that $\Om_2$ is unit-normalized.
Moreover, the restriction to single-trace operators is here purely technical.
All the techniques presented in this paper can be applied to higher-trace operators as well, the only limitation being that the CFT data is in general possibly not known.

\paragraph{Defect operators.}
We can define defect half-BPS operators in a similar manner.
The analogue of the single-trace operators \eqref{eq:SingleTraceHalfBPSOperators_Bulk} take the following explicit form:
\begin{equation}
    \Oh_\Dh (\uh,\tau)
    =
    \frac{1}{\sqrt{\nh_\Dh}}
    \Wl [ (\uh \cdot \phi(\tau))^\Dh ]\,,
    \label{eq:HalfBPSOperators_Defect}
\end{equation}
with the half-BPS conditions
\begin{equation}
    \uh^2
    =
    0
    \quad \text{and} \quad
    \uh \cdot \theta
    =
    0\,.
    \label{eq:DefectHalfBPSConditions}
\end{equation}
Here we defined insertions on the Wilson line $\Wl[\ldots]$ as
\begin{equation}
    \Wl [ \Oh ]
    =
    \frac{1}{N} \tr
    \Pm [\, \Oh\, \exp \int d\tau\,
    (i A_0 (\tau) + \phi^6 (\tau)) ]\,.
    \label{eq:Insertion}
\end{equation}
We call these operators \textit{open-trace}.
The normalization constants can be evaluated explicitly through localization techniques \cite{Giombi:2018qox}.
To the best of our knowledge, there exists no general closed form, but they can be found case by case.
For instance, we have
\begin{align}
    \nh_1
    &=
    \frac{\sqrt{\lambda}}{2\pi^2} \frac{\Ids_1}{\Ids_2}\,, \label{eq:n1} \\
    \nh_2
    &=
    \frac{1}{4\pi^4} (3 \lambda - (\Ids_1 - 2)(\Ids_1 + 10))\,, \label{eq:n2} \\
    \nh_3
    &=
    \frac{3}{8 \pi^6} \biggl(
    \frac{(5 \lambda + 72) \Ids_1 \Ids_2}{\sqrt{\lambda}}
    - \frac{\lambda (26 \Ids_1 + 3\lambda - 32) + 288(\Ids_1 - 1)}{\Ids_1 -2}
    \biggr)\,,
    \label{eq:n3}
\end{align}
where we have defined the help function
\begin{equation}
    \Ids_a
    =
    \frac{\sqrt{\lambda}\, I_0 (\sqrt{\lambda})}{I_a (\sqrt{\lambda})}\,.
    \label{eq:Ids}
\end{equation}
In this paper we also consider the AdS dual of the operator $\Oh_1 (\uh,\tau)$, which can be expressed in terms of $S^5$ fluctuations as~\cite{Giombi:2017cqn}
\begin{equation}
    \Oh_1 (\uh,\tau) = \frac{1}{\sqrt{\nh_1}} \uh^i y^i (\tau)\,.
    \label{eq:AdSDualToOh1}
\end{equation}

We can construct operators with a higher number of traces.
In fact, they are expected to appear in the bulk-to-defect OPE.\footnote{See \cite{Giombi:2018hsx} as well as \cite{Barrat:2021yvp} for a detailed explanation.}
For $n$ traces on top of the one of the Wilson line they take the form
\begin{equation}
    \hat{\mathrm{O}}_{(h | \mathbf{k})} (\uh, \tau)
    =
    \Wl [ (\uh \cdot \phi)^h ]
    \tr (\uh \cdot \phi)^{k_1} \ldots \tr (\uh \cdot \phi)^{k_n}\,,
    \label{eq:HigherTraceBPSOperators}
\end{equation}
with $\mathbf{k} = (k_1\,, \ldots\,, k_n)$ and $\Dh_{(h | \mathbf{k})} = h + \sum k_j$.
We use the notation $\hat{\mathrm{O}}$ instead of $\Oh$ to emphasize that, as they stand, these operators do not form a good basis of operators since they are not orthogonal to the open-trace ones.
This can be achieved by defining new operators following a Gram-Schmidt procedure.
The OPE coefficients that appear in a block expansion give correspond to such an orthonormal basis for which we use the notation $\Oh$.

\subsection{Correlators of half-BPS operators}
\label{subsec:CorrelatorsOfHalfBPSOperators}

We now consider correlation functions of the half-BPS operators introduced in Section \ref{subsubsec:HalfBPSOperators}.
Since we consider mixed correlators of bulk and defect scalar operators, we use for the rest of this paper the shorthand notation
\begin{equation}
    \vev{\Delta_1 \ldots \Delta_m \Dh_1 \ldots \Dh_n}
    =
    \vev{\Op_{\Delta_1} (u_1, x_1) \ldots \Op_{\Delta_m} (u_m, x_m) \Wl [\Oh_{\Dh_1} (\uh_1, \tau_1) \ldots \Oh_{\Dh_n} (\uh_n, \tau_n)] }\,.
    \label{eq:Correlators_ShorthandNotation}
\end{equation}

\subsubsection{Bulk correlators}
\label{subsubsec:BulkCorrelators}

We begin by discussing the correlators of bulk operators, i.e., without the line defect.

\paragraph{Two-point functions.}
Two-point functions are fixed by the $4d$ conformal symmetry to be
\begin{equation}
    \vev{\Delta_1 \Delta_2}
    =
    \delta_{\Om_{\Delta_1} \Om_{\Delta_2}} (12)^{\Delta_1}\,,
    \label{eq:BulkTwoPointFunctions}
\end{equation}
where we use the shorthand notation
\begin{equation}
    (ij)
    =
    \frac{u_i \cdot u_j}{x_{ij}^2}\,.
    \label{eq:4dSuperPropagator}
\end{equation}

\paragraph{Three-point functions.}
Three-point functions are also fixed kinematically, and for scalar operators they read
\begin{equation}
    \vev{\Delta_1 \Delta_2 \Delta_3}
    =
    \lambda_{\Delta_1 \Delta_2 \Delta_3}
    (12)^{\Delta_{123}} (23)^{\Delta_{231}} (31)^{\Delta_{312}}\,,
    \label{eq:BulkThreePointFunctions}
\end{equation}
where we have defined
\begin{equation}
    \Delta_{ijk}
    =
    \frac{1}{2}(\Delta_i + \Delta_j - \Delta_k )\,.
    \label{eq:ShorthandThreePoint}
\end{equation}
In the large $N$ limit, three-point functions of single-trace operators are known to be exactly given by \cite{Lee:1998bxa}
\begin{equation}
    \lambda_{\Delta_1 \Delta_2 \Delta_3}
    =
    \frac{\sqrt{\Delta_1 \Delta_2 \Delta_3}}{N}\,.
    \label{eq:ThreePointBulk_Exact}
\end{equation}
The higher-point functions are not fixed kinematically anymore.
They depend on $n(n-3)/2$ spacetime cross-ratios.
They have been studied at weak and strong-coupling using a variety of techniques \cite{Drukker:2008pi,Drukker:2009sf,Bercini:2024pya}.

\subsubsection{Defect correlators}
\label{subsubsec:DefectCorrelators}

We now consider the case of correlators that involve defect operators only.

\paragraph{Two-point functions.}
Similarly to \eqref{eq:BulkTwoPointFunctions}, defect two-point functions are fixed by the one-dimensional conformal symmetry to be
\begin{equation}
    \vev{\Dh_1 \Dh_2}
    =
    \delta_{\Oh_{\Dh_1} \Oh_{\Dh_2}} (\hat{1}\hat{2})^{\Dh_1}\,,
    \label{eq:DefectTwoPoint}
\end{equation}
where we used the notation
\begin{equation}
    (\hat{i}\hat{j})
    =
    \frac{\uh_i \cdot \uh_j}{\tau_{ij}^2}\,.
    \label{eq:1dSuperPropagator}
\end{equation}

\paragraph{Three-point functions.}
The three-point functions of defect scalar operators are once again fixed by conformal symmetry and they are given by
\begin{equation}
    \vev{\Dh_1 \Dh_2 \Dh_3}
    =
    \lambdah_{\Dh_1 \Dh_2 \Dh_3}
    (\hat{1}\hat{2})^{\Dh_{123}} (\hat{2}\hat{3})^{\Dh_{231}} (\hat{3}\hat{1})^{\Dh_{312}}\,,
    \label{eq:DefectThreePointFunctions}
\end{equation}
with $\Dh_{ijk}$ defined in the same way as in \eqref{eq:ShorthandThreePoint}.

As opposed to the bulk case, the OPE coefficients $\lambdah_{\Dh_1 \Dh_2 \Dh_3}$ are not currently known in an analytical form, even in the large $N$ limit.
They can however be calculated case by case using the integrability techniques of \cite{Giombi:2018qox}.
We give explicitly here the expression for $\lambdah_{\hat{1}\hat{1} \hat{2}}$, as it is useful for our calculations:
\begin{equation}
    \lambdah_{\hat{1}\hat{1}\hat{2}}
    =
    \frac{-2 \sqrt{\lambda}
    \Ids_1 (7 + \Ids_1) \Ids_2 \Ids_3 - (-32 + 14 \Ids_1 + \Ids_1^2) \Ids_2^2 \Ids_3 + \lambda (3 \Ids_1 \Ids_2^2 - \Ids_1^2 \Ids_3 + 9 \Ids_2^2 \Ids_3)}{4 \Ids_1 \sqrt{\lambda (3 \lambda - (-2 + \Ids_1) (10 + \Ids_1))}
    \Ids_2 \Ids_3}\,,
    \label{eq:lambdah112}
\end{equation}
with $\Ids_k$ defined in \eqref{eq:Ids}.

\paragraph{Higher-point functions.}
Higher-point defect correlators take the general form
\begin{equation}
    \vev{\Dh_1 \ldots \Dh_n}
    =
    \Km_{\Dh_1 \ldots \Dh_n} \Am_{\Dh_1 \ldots \Dh_n} (\lbrace x; r,s,t \rbrace)\,,
    \label{eq:HigherPointDefectCorrelators}
\end{equation}
where $\Km_{\Dh_1 \ldots \Dh_n}$ is a (super)conformal prefactor.
In general, these correlators depend on $n-3$ spacetime cross-ratios and $n(n-3)/2$ $R$-symmetry variables.

\subsubsection{Correlators of bulk and defect operators}
\label{subsubsec:CorrelatorsOfBulkAndDefectOperators}

This section is dedicated to correlators that involve both bulk and defect operators.

\paragraph{Bulk-defect two-point functions.}
Two-point bulk-defect correlators are fixed kinematically by conformal symmetry to be
\begin{equation}
    \vev{\Delta_1 \Dh_2}
    =
    b_{\Delta_1 \Dh_2} (1 \hat{2})^{\Dh_2} (1 \theta)^{\Delta_1 - \Dh_2},
    \label{eq:BulkDefect_KinematicStructure}
\end{equation}
with
\begin{equation}
    (i\hat{j})
    =
    \frac{u_i \cdot \hat{u}_j}{x_{ij}^2}\,.
    \qquad
    (i\theta)
    =
    \frac{u_i \cdot \theta}{|\vec{x}_i|}\,.
    \label{eq:MixedSuperPropagators}
\end{equation}
The normalization of the operators is now fixed and it does not leave the possibility of fixing the coefficients $b_{\Delta \Dh}$ to unity.
The bulk-defect coefficients are indeed an additional set of conformal data and at weak coupling they can be expressed as~\cite{Giombi:2018hsx}
\begin{equation}
    b_{\Delta \Dh} = \frac{\sqrt{\Delta}}{(\Delta - \Dh)!} \frac{\lambda^{(\Delta - \Dh)/2}}{2^{3(\Delta - \Dh)/2} N}
    \biggl(
    1
    + \frac{\lambda}{48} \frac{2 - 5 (\Delta - \Dh)}{2 + \Delta - \Dh}
    + \Om(\lambda^2)
    \biggr)\,.
    \label{eq:BulkDefect_WeakCoupling}
\end{equation}
Note that we consider $\Delta \geq \Dh$ in this formula,
since correlators with $\Delta < \Dh$ vanish.

In principle these OPE coefficients can be calculated exactly using either localization techniques \cite{Giombi:2018hsx} or microbootstrap.
The latter was used in \cite{Barrat:2021un,Barrat:2024nod} to efficiently produce a large number of expressions, and to find a closed form for relevant special cases:
\begin{align}
    b_{\Delta \hat{1}}^2
    &=
    \frac{\sqrt{\lambda} \Delta^3}{2^{\Delta + 1} N^2} \frac{\Ids_1 \Ids_2}{\Ids^2_\Delta}\,, \label{eq:BulkDefect_Exact_Delta1} \\
    b_{\Delta \hat{2}}^2
    &=
    -
    \frac{\sqrt{\lambda} \Delta}{2^{\Delta} N^2}
    \frac{\Ids_1 \Ids_2 \Ids_5
    \left(\sqrt{\lambda}\, \Ids_{\Delta-1} (\Ids_3 \Ids_\Delta-\Ids_1
    \Ids_{\Delta+2})+2 (\Delta -1) \Ids_3 \Ids_\Delta \Ids_{\Delta+2}\right)^2}{\Ids_{\Delta-1}^2 \Ids_\Delta^2 \Ids_{\Delta+2}^2
    \left(\sqrt{\lambda }\, \Ids_2 \left(\Ids_1 \Ids_5-\Ids_3^2\right)-4
    \Ids_3^2 \Ids_5\right)}\,.
    \label{eq:BulkDefect_Exact_Delta2}
\end{align}
When the defect operator is the identity, the two-point coefficients become the one-point coefficients of the bulk operators:
\begin{equation}
    b_{\Delta \mathds{1}}
    =
    a_{\Delta}
    =
    \frac{\sqrt{\lambda} \sqrt{\Delta}}{2^{\Delta/2+1} N} \frac{\Ids_1}{\Ids_{\Delta}}\,.
    \label{eq:OnePoint}
\end{equation}
\paragraph{Bulk-defect-defect three-point functions.}
We now move to the observables that are the central focal point of this work: correlators with one bulk and two defect half-BPS operators.
They can be expressed as
\begin{equation}
    \vev{ \Delta_1 \Dh_2 \Dh_3 }
    =
    \Km_{\Delta_1 \Dh_2 \Dh_3} \Am_{\Delta_1 \Dh_2 \Dh_3} (\zeta ; x)\,,
    \label{eq:BDD_Kinematics}
\end{equation}
where $\Am_{\Delta_1 \Dh_2 \Dh_3} (\zeta ; x)$ is a reduced correlator, and $\Km_{\Delta_1 \Dh_2 \Dh_3}$ is a superconformal prefactor.
The prefactor is chosen such that the leading order at weak coupling comes without a factor $\zeta$.
For instance, in the case $\Delta_1 \geq \Dh_2 + \Dh_3$, this results in
\begin{equation}
    \Km_{\Delta_1 \Dh_2 \Dh_3}
    =
    \frac{(1\hat{2})^{\Dh_2} (1\hat{3})^{\Dh_3}}{(1 \theta)^{2\Delta_{231}}}\,.
    \label{eq:BDD_PrefactorExample}
\end{equation}
The conformal cross-ratios are defined as
\begin{equation}
    x
    =
    \frac{|\vec{x}|_{1}^2 \tau_{23}^2}{(\vec{x}_{1}^2+\tau_{12}^2)(\vec{x}_{1}^2+\tau_{13}^2)}\,,
    \label{eq:SpacetimeCrossRatio}
\end{equation}
for the spacetime variables, and 
\begin{equation}
    \zeta
    =
    - \frac{1}{2} \frac{(u_1 \cdot \theta)^2 (\uh_2 \cdot \uh_3)}{(u_1 \cdot \uh_2)(u_1 \cdot \uh_3)}\,,
    \label{eq:RSymmetryCrossRatio}
\end{equation}
for the $R$-symmetry ones.
The $-1/2$ factor in our definition of $\zeta$ is chosen such that the kinematic limit studied in \cite{Giombi:2018hsx}, which is known as topological sector, corresponds to the sum of all the $R$ symmetry channels defined below.

In our conventions the reduced correlator is a polynomial in the $R$-symmetry variable $\zeta$.
Concretely,
\begin{equation}
    \Am_{\Delta_1 \Dh_2 \Dh_3} (\zeta ; x)
    =
    \sum_{j=1}^r
    \left(\frac{\zeta}{x} \right)^{j-1} F_j (x)\,.
    \label{eq:BulkDefectDefect_RSymmetryChannels}
\end{equation}
We call the functions $F_j$ $R$-symmetry channels.
The dependency on $x$ in the prefactors is chosen such that they become $1$ in the topological limit of \cite{Giombi:2018hsx}.
The number of channels $r = r(\Delta_1, \Dh_2, \Dh_3)$ for a specific configuration is given by
\begin{equation}
    r
    =
    \left\lfloor \Delta_{1\hat{3}\hat{2}} \right\rfloor + 1 \,,
    \label{eq:NumberOfRSymmetryChannels1}
\end{equation}
for $\Delta_1 < \Dh_2+\Dh_3$ and 
\begin{equation}
    r
    =
    \Dh_3 + 1 \,,
    \label{eq:NumberOfRSymmetryChannels2}
\end{equation}
for $\Delta_1 \geq \Dh_2+\Dh_3$.\footnote{We always assume $\Dh_2 \geq \Dh_3$ without loss of generality.}

\begin{figure}
    \centering
    \OPE
    \caption{Illustration of the operator product expansion for the bulk-defect-defect correlators.
    The result is an expansion in conformal blocks, with the OPE coefficients being given by the product of defect three-point and bulk-defect two-point coefficients.}
    \label{fig:OPE}
\end{figure}

The OPE can be used to expand the correlator into conformal blocks.
One can either start with an OPE of the two defect operators $\Oh_{\Dh_2} \times \Oh_{\Dh_3}$, or with the OPE of the bulk operator with the line defect, $\Op_{\Delta_1} \times \Wl$.
In fact, these two operations lead to the same conformal block expansion,
\begin{equation}
    F_j (x)
    =
    x^{-\Delta_1}
    \sum_{\Dh\: \mathrm{prim.}}
    b_{\Delta_1 \Dh}^{(j)} \lambdah_{\Dh_2 \Dh_3 \Dh}^{(j)}
    g_{\Dh} (x)\,,
    \label{eq:ConformalBlockExpansion}
\end{equation}
where the sum runs over all the primaries, including superdescendants. Because of the equivalence of the OPE expansions there exists no crossing relations from this point of view.
Although not specified explicitly here to avoid cluttering the notation, the blocks $g_{\Dh} (x)$ depend on the scaling dimensions $\Dh_2$ and $\Dh_3$ (but not $\Delta_1$).
They have been determined in \cite{Buric:2020zea,Okuyama:2024tpg} to be
\begin{equation}
    g_{\Dh} (x)
    =
    x^{\Dh/2}\,
    _2F_1
    \left(
    (\Dh +\Dh_2-\Dh_3)/2,
    (\Dh -\Dh_2+\Dh_3)/2;
    \Dh+1/2;
    x
    \right)\,.
    \label{eq:ConformalBlocks}
\end{equation}
The OPE is illustrated in Figure \ref{fig:OPE}.
In order for the sum in \eqref{eq:ConformalBlockExpansion} to be over superprimaries instead of primaries, we need to use superconformal blocks.
We provide the superblock expansion for the case $\vev{2 \hat{1} \hat{1}}$ in Section \ref{subsec:SuperblockExpansion}.
Finally, note that an alternative expansion in terms of local blocks was derived in \cite{Levine:2023ywq,Levine:2024wqn}.

\paragraph{Large $N$ counting and multi-trace operators.}
We now perform a large $N$ counting of Feynman diagrams to understand which higher-trace operators are allowed to contribute.
Our analysis is based on the well-known genus formula for counting the $N$ factors in Feynman diagrams; for each disconnected part of a diagram, the $N$ counting is~\cite{tHooft:1973alw}
\begin{equation}
    N^{2-2G-T}\,,
    \label{eq:GenusFormula}
\end{equation}
with $G$ the genus of the diagram and $T$ the number of traces.
Since we are only interested in planar diagrams we set $G=0$.
One should also remember that the trace of the Wilson line comes with a normalization $1/N$.
Therefore the rule for counting the $N$ factors can be summarized as follows:
\begin{enumerate}
    \item Each fully connected planar graph not connected to the Wilson line contributes $N^{2-T}$, with $T$ the number of traces;
    \item Each fully connected planar graph connected to the Wilson line contributes $N^{1-T}$.
\end{enumerate}
The immediate conclusion is that dominant diagrams are the \textit{diagrams which are maximally disconnected}.

We now want to apply the genus formula to the study of higher-trace operators, i.e., defect operators of the schematic form
\begin{equation}
    \Oh_m (\tau) = \frac{1}{\sqrt{\nh_{\Oh_m}}} \Wl [\hat{O}_1 (\tau)]\, \tr (\hat{O}_2(\tau)) \ldots \tr (\hat{O}_{m+1}(\tau))\,.
    \label{eq:DefinitionHigherTraceOperators}
\end{equation}
Here $m$ counts the number of extra traces on top of the trace of the Wilson line.
In diagrams we represent these operators as
\begin{equation}
    \Oh_m (\tau) = \HigherTraceDiagrams\,.
    \label{eq:RepresentationHigherTraceOperators}
\end{equation}
The black dot refers to the operator inserted inside the trace of the Wilson line while the white dots carry their own traces and mirror the definition in~\eqref{eq:DefinitionHigherTraceOperators}.

We start by considering two-point functions of identical operators.
These correlators reach maximal counting when the traces are as disconnected as possible, i.e.,\footnote{The vertical distance between the dots is only there to make the diagram more readable; all operators are located on the Wilson line at $\tau_1$ and $\tau_2$.
Moreover, the fat lines going from each trace can contain multiple propagators; the dominant diagrams remaining the ones disconnected in the way displayed.}
\begin{equation}
    \vev{\Oh_m (\tau_1) \Oh_m (\tau_2)} \sim\ \frac{1}{n_{\Oh_m}} \LargeNDiagramOneB\ \sim \frac{N^0}{n_{\Oh_m}}\,,
    \label{eq:NCountingTwoPointFunctions}
\end{equation}
from which we infer that
\begin{equation}
    n_{\Oh_m} \sim N^0\,.
    \label{eq:NCountingNormalizationConstant}
\end{equation}
Moreover one notices from~\eqref{eq:NCountingTwoPointFunctions} that the two-point functions of higher-trace operators can be seen as \textit{products of two-point functions of open- and single-trace operators}.
These single-trace operators can be seen as projections of bulk operators on the defect that \textit{do not interact} with the defect.
This is of course a large $N$ effect that does not persist at finite $N$.

One can perform the same large $N$ counting for three-point functions of defect operators.
We focus on the case of two open-trace and one higher-trace operators since it is the situation that we encounter in this work.
One-point functions vanish in CFTs and thus only a fully connected diagram is allowed, making the counting go as\footnote{For readability we avoid including the fact that there can be lines connecting to arbitrary integrated points along the defect.
This does not affect the $N$ counting.}
\begin{equation}
    \vev{\Oh_0 \Oh_m \Oh_0} \sim\ \LargeNDiagramTwoB\ \sim \frac{1}{N^m}\,.
    \label{eq:NCountingThreePointFunctions}
\end{equation}
This reflects the famous statement that the open-trace sector forms a closed sector: higher-trace operators are always suppressed, as opposed to the bulk theory in which double-trace operators can mix.

Correlators of bulk and defect operators however \textit{enhance} higher-trace operators.
This can be seen in bulk-defect two-point functions in the following way.
First it is easy to see that open-trace operators scale as
\begin{equation}
    \vev{\Om_0 \Oh_0} \sim \LargeNDiagramThreeA\ \sim \frac{1}{N}\,,
\end{equation}
with $\Om_0$ denoting a single-trace bulk operator.
Then double-trace operators can exhibit two kinds of behavior.
In the case in which everything is connected, they scale as $1/N^2$:
\begin{equation}
    \vev{\Om_0 \Oh_{m=1}} \sim \LargeNDiagramThreeB\ \sim \frac{1}{N^2}\,.
\end{equation}
However it can also happen that the number of fundamental fields in the additional trace matches the number of fundamental fields in the bulk operator.
In this case there can be a disconnected diagram and the scaling becomes\footnote{Another way to convince oneself that double-trace operators have to be included is through the bulk-defect OPE.
See~\cite{Giombi:2018hsx} for topological examples.}
\begin{equation}
    \vev{\Om_0 \Oh_{m=1}} \sim \LargeNDiagramThreeC\ \sim N^0\,.
    \label{eq:NCountingSpecial}
\end{equation}
Notice that the open-trace piece cannot have a disconnected part on its own since one-point functions of defect operators vanish.
In other words this $N$ scaling can only happen when the open-trace part is empty.
Using the same counting one can understand that triple- and higher-trace operators are suppressed in our setup, in the sense that the bulk-defect two-point functions can never compensate the $N$ scaling of the defect three-point functions if the external bulk operator is single-trace.

\bigskip

To be concrete and for future purposes, we now apply the discussion above to determine the spectrum of operators having classical scaling dimension two and appearing in the specific correlator of half-BPS operators $\vev{2\hat{1}\hat{1}}$.

First there are two known open-trace operators, which can be defined as~\cite{Correa:2018fgz}
\begin{align}
    \Oh_\pm (\tau) = \frac{1}{\sqrt{\nh_\pm}} \Wl[ \phi^i\phi^i (\tau) \pm \sqrt{5}\, \phi^6\phi^6 (\tau)]\,.
    \label{eq:DefinitionOhpm}
\end{align}
Although these operators are degenerate at the end points of the coupling ($\Delta_\pm(\lambda=0)=2$ and $\Delta_\pm(\lambda=\infty)=4$), they have different scaling dimensions at finite coupling.
On top of that there are only two gauge-invariant double-trace building blocks in $\mathfrak{su}(N)$:
\begin{equation}
    \Wl[\, ]\, \tr (\phi^i\phi^i)\,, \quad \Wl[\, ]\, \tr(\phi^6\phi^6)\,,
    \label{eq:GaugeInvariantBuildingBlocks}
\end{equation}
where $\Wl[\, ]$ indicates that the defect is present but that the operator does not include fields inserted inside the trace of the Wilson line.
We therefore expect a maximum of two double-trace operators on top of the open-trace operators $\Oh_\pm$ discussed in~\eqref{eq:DefinitionOhpm}.
These operators must also be orthogonal to $\Oh_\pm$ in the large $N$ limit.
Equivalently to~\eqref{eq:GaugeInvariantBuildingBlocks}, one can use the basis
\begin{equation}
    \Wl[\, ]\, \Om_{20'}^{66} (\tau)\,, \quad \Wl[\, ]\, \Om_{\Km} (\tau)\,,
    \label{eq:BasisOfBulkOperators}
\end{equation}
which has the advantage to allow us to use the definition of \textit{bulk operators projected onto the defect}:
\begin{align}
    \Oh_{20'}^{IJ} (\tau) &\sim \tr (\phi^I \phi^J) - \frac{\delta^{IJ}}{6} \tr (\phi^K \phi^K)\,, \label{eq:DefinitionO20} \\
    \Oh_{\Km} (\tau) &\sim \tr (\phi^I \phi^I)\,, \label{eq:DefinitionK}
\end{align}
with $I=1,\ldots,6$.
One can convert from one basis to the other through
\begin{align}
    \tr (\phi^6\phi^6) &\sim \Oh_{20'}^{66} + \frac{1}{6} \Oh_{\Km}\,, \label{eq:ConvertBasis1} \\
    \tr (\phi^i\phi^i) &\sim -\Oh_{20'}^{66} + \frac{5}{6} \Oh_{\Km}\,,
    \label{eq:ConvertBasis2}
\end{align}
up to multiplicative constants.
Using the new basis and requiring that the operators have to be eigenstates of the dilatation operator, we immediately conclude that the good operators are
\begin{align}
    \Oh_{(20')} (\tau) &= \frac{1}{\sqrt{\nh_{(20')}}} \Wl[\, ] \left( \tr (\phi^6 \phi^6) - \frac{1}{6} \tr (\phi^I \phi^I) \right) \propto \Wl[\, ] \left(\tr (\phi^6 \phi^6) - \frac{1}{5} \tr (\phi^i \phi^i) \right)\,, \label{eq:DefinitionDT20} \\
    \Oh_{(\Km)} (\tau) &= \frac{1}{\sqrt{\nh_{(\Km)}}} \Wl[\, ]\, \tr (\phi^I \phi^I) \propto \Wl[\, ] \left( \tr (\phi^6\phi^6) + \tr (\phi^i \phi^i) \right)\,, \label{eq:DefinitionDTK}
\end{align}
where the subscripts $\lbrace (20'), (\Km) \rbrace$ are there to indicate that the operators are projections of bulk operators onto the defect.
It should be emphasized that these operators are eigenstates at large $N$ but we do not expect that to hold at finite $N$.
This can be seen from the fact that they are orthogonal to $\Oh_\pm$ in the large $N$ limit but not at finite $N$:\footnote{They are also orthogonal to one another, as follows trivially from the fact that they are projections of bulk operators.}
\begin{equation}
    \vev{\Oh_\pm \Oh_{(20')}} \sim \frac{1}{N}\,, \qquad \vev{\Oh_\pm \Oh_{(\Km)}} \sim \frac{1}{N}\,.
\end{equation}
The scaling dimensions of these double-trace operators can therefore be expressed as
\begin{align}
    \Dh_{(20')} &= 2 + \mathrm{O}(1/N^2)\,, \\
    \Dh_{(\Km)} &= 2 + \gamma_\Km + \mathrm{O}(1/N^2)\,.
\end{align}
with $\gamma_\Km$ the anomalous dimension of the Konishi operator in the bulk $\Nm=4$ SYM theory.
Notice however that we do not expect $\Oh_{(\Km)}$ to be relevant in $\vev{2\hat{1}\hat{1}}$.
This is because this operator is by definition orthogonal to $\Om_2$, and thus we have
\begin{equation}
    \vev{\Om_2 \Oh_{(\Km)}} = 0\,.
\end{equation}
A simple check of the existence of the operator $\Oh_{(20')}$ can be made by looking at the two-point function $\vev{\Om_2 \Om_2}$ in presence of the line defect, for which the defect channel consists of the coefficients $b_{2\Oh}^2$.
In this setup the leading disconnected term is of order $\Om(N^0)$ while the connected terms are $\Om(1/N^2)$~\cite{Giombi:2018hsx,Barrat:2020vch,Barrat:2021yvp,Barrat:2022psm,Bianchi:2022ppi}.
The disconnected term necessarily involves double-trace operators, and the data is protected in the large $N$ limit.
We can anticipate that the OPE data corresponding to $b_{2(20')}$ calculated in this paper (see \eqref{eq:b2dt20}) matches the data in (4.5c) of~\cite{Barrat:2020vch} (up to a factor $4$ explained in footnote \ref{footnote:Factor4}).

\section{Non-perturbative constraints}
\label{sec:NonPerturbativeConstraints}

We now discuss the non-perturbative constraints of the bulk-defect-defect correlators of half-BPS operators.
We start by showing how superconformal symmetry can be used to trade a function of the spacetime cross-ratio for a constant.
This constant, called topological sector, can be evaluated using localization techniques.
We also present the superconformal block expansion to study the case $\vev{2 \hat{1} \hat{1}}$ at strong coupling, which is computed in Section~\ref{sec:PerturbativeResults}.
Finally, we discuss pinching and splitting limits, which relate parts of the bulk-defect-defect correlators to lower-point functions.

\subsection{Superconformal symmetry}
\label{subsec:SuperconformalSymmetry}

Superconformal symmetry can typically be encoded in the form of superconformal Ward identities (SCWI). 
They have been shown to be powerful tools for constraining correlators of half-BPS operators, both in $\Nm=4$ SYM \cite{Dolan:2001tt,Dolan:2004iy,Dolan:2004mu} and in the Wilson-line defect CFT \cite{Liendo:2016ymz,Liendo:2018ukf,Barrat:2021tpn,Bliard:2024und,Barrat:2024ta}.
To the best of our knowledge, SCWI are not known for the setup \eqref{eq:BDD_Kinematics}.
We conjecture in this section that they are fully encoded by the topological sector.

\subsubsection{Topological sector and superconformal Ward identity}
\label{subsubsec:TopologicalSectorAndSuperconformalWardIdentity}

Our starting point is the topological sector calculated in \cite{Giombi:2018hsx}.
In our conventions, it can be expressed as the sum of the $R$-symmetry channels of \eqref{eq:BulkDefectDefect_RSymmetryChannels}:
\begin{equation}
    \sum_{j=1}^{r} F_j (x)
    =
    \Fds_{\Delta_1 \Dh_2 \Dh_3}\,.
    \label{eq:TopologicalSector}
\end{equation}
The right-hand side $\Fds_{\Delta_1 \Dh_2 \Dh_3}$ is a constant, in the sense that it depends neither on spacetime nor $R$-symmetry variables.
It is however a function of the coupling $\lambda$.
Equation \eqref{eq:TopologicalSector} can readily be used to eliminate one $R$-symmetry channel:
\begin{equation}
    F_r (x)
    =
    \Fds_{\Delta_1 \Dh_2 \Dh_3}
    -
    \sum_{j=1}^{r-1} F_j (x)\,.
    \label{eq:SolutionWI}
\end{equation}
In fact, we formulated an Ansatz for the Ward identities based on known examples and used the results of Section \ref{sec:PerturbativeResults} to constrain the numerical coefficients.
We found the differential constraint
\begin{equation}
    \left.
    \bigl(
    \pd_x + \pd_\zeta
    \bigr)
    \Am_{\Delta_1 \Dh_2 \Dh_3} (\zeta; x)
    \right|_{\zeta = x}
    =
    0\,.
    \label{eq:SCWI}
\end{equation}
Unfortunately this equation can be shown not to encode more constraints than \eqref{eq:TopologicalSector}.
It is plausible that the fact that the correlator depends on only one spacetime and one $R$-symmetry cross-ratio does not leave much room for supersymmetry to constrain its form any further.
It is also possible that differential constraints not taking the form \eqref{eq:SCWI} are being missed by this strategy.
For instance, in five-point functions of half-BPS operators in $\Nm=4$ SYM, this Ansatz approach proved unsuccessful, although it is known that differential constraints exist \cite{Meneghelli:2024ta}.\footnote{We are thankful to Carlo Meneghelli and Sophie M{\"u}ller for discussions.}
For the time being we use the constraint \eqref{eq:SolutionWI} to study the correlators at leading and next-to-leading orders in Section \ref{sec:PerturbativeResults}.

\subsubsection{Localization results}
\label{subsubsec:LocalizationResults}

\begin{figure}
\centering
\begin{subfigure}{.5\textwidth}
  \includegraphics[width=.9\linewidth]{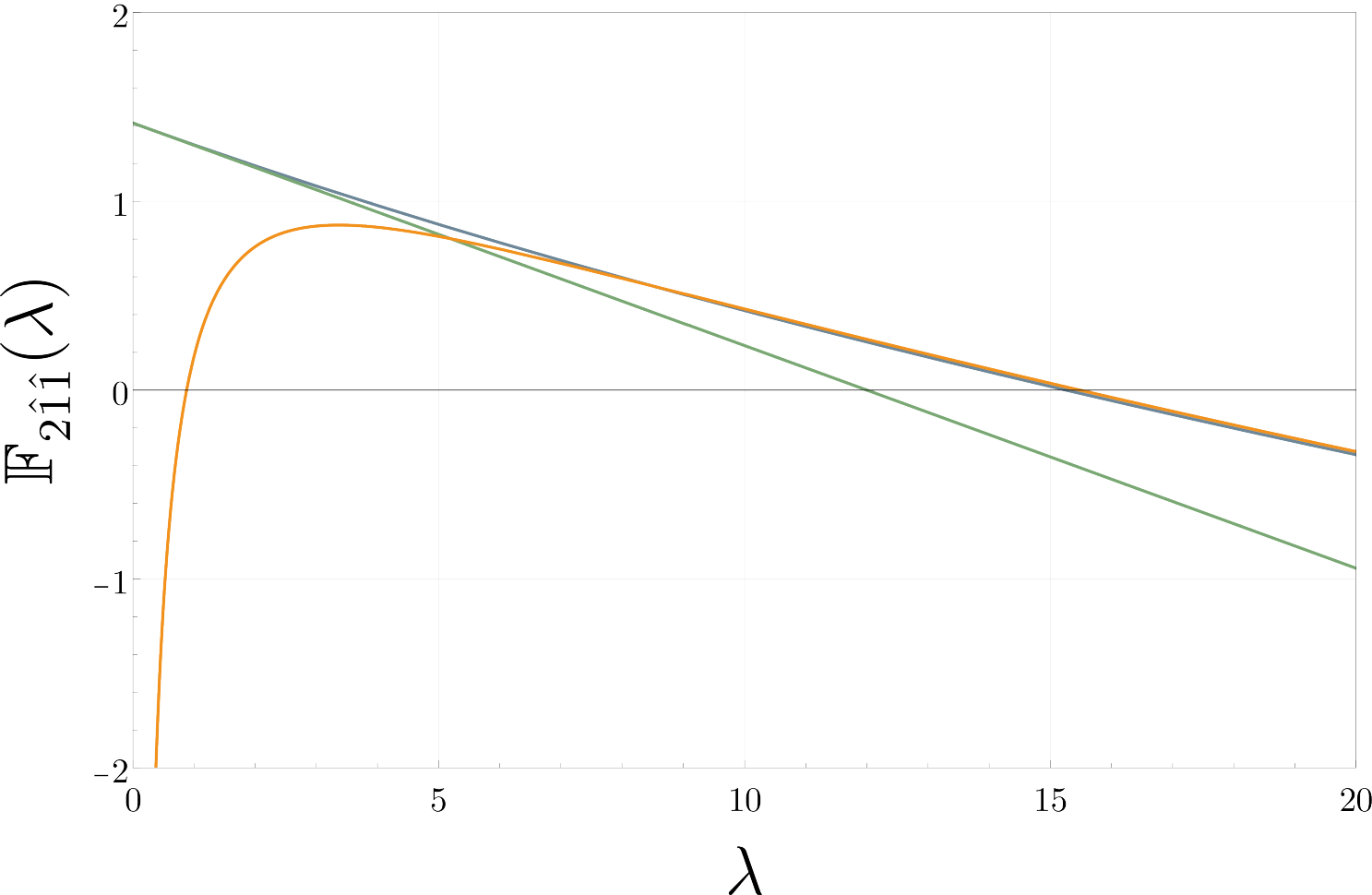}
\end{subfigure}%
\begin{subfigure}{.5\textwidth}
  \includegraphics[width=.9\linewidth]{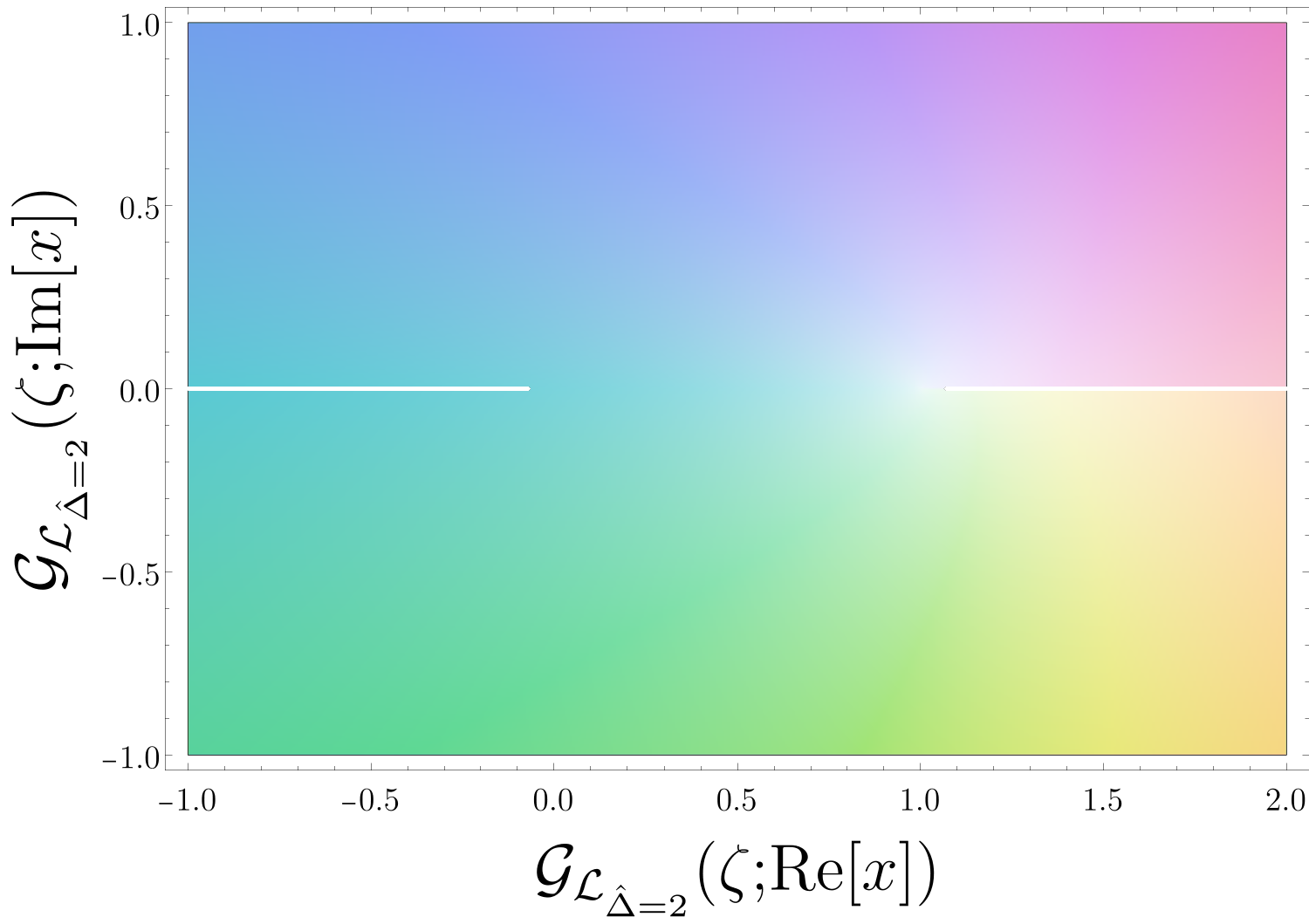}
\end{subfigure}
\caption{The left figure displays the topological sector for $\vev{2 \hat{1} \hat{1}}$.
The blue line is the exact expression, presented in \eqref{eq:Fds211_Exact}.
The green and orange lines correspond, respectively, to the weak- and strong-coupling expansions up to next-to-leading order, derived in \cite{Giombi:2018hsx} and reproduced here in \eqref{eq:Fds_weak}, \eqref{eq:Fds1_weak} and \eqref{eq:Fds211_strong}.
On the right, the analytic structure of the superblock $\Gm_{\Dh =2} (\zeta;x)$ is presented.
We see a discontinuity at $x \geq 1$, which is incompatible with the locality requirement discussed in \cite{Levine:2023ywq,Levine:2024wqn}.
Note that the superblocks $\Gm_{\Dh > 2} (\zeta;x)$ exhibit the same behavior.
}
\label{fig:TopologicalSectorAndContourSuperblocks}
\end{figure}

The topological sector can be evaluated for arbitrary $\Delta_1$, $\Dh_2$, $\Dh_3$ using the localization techniques of \cite{Giombi:2018hsx}.
Here we extend their results to next-to-leading order using the same techniques.

The topological sector follows the perturbative structure
\begin{equation}
    \Fds_{\Delta_1 \Dh_2 \Dh_3}
    =
    \frac{\lambda^{a/2}}{N}
    \biggl(
    \Fds_{\Delta_1 \Dh_2 \Dh_3}^{(0)}
    +
    \lambda
    \Fds_{\Delta_1 \Dh_2 \Dh_3}^{(1)}
    + \ldots
    \biggr)\,,
    \label{eq:Fds_PerturbativeStructure}
\end{equation}
where $a = \text{min} (a_4)$ is introduced in \eqref{eq:GeneralFeynmanDiagrams} and corresponds to the minimum number of propagators connecting the operator $\Op_{\Delta_1}$ and the Wilson line.

\paragraph{Weak coupling.}
For a given configuration $\vev{\Delta_1 \Dh_2 \Dh_3}$ the leading order is given by
\begin{equation}
    \begin{split}
   \Fds_{\Delta_1 \Dh_2 \Dh_3}^{(0)}
   &=
   \frac{2^{\frac{1}{2} (2\Delta_{\hat{2}\hat{3}1}-2)} i^{a-2\Delta_{\hat{2}\hat{3}1}} \sqrt{\Delta_1} }{\Gamma (a+1)} 
   \left(
   (2-2\Delta_{\hat{2}\hat{3}1}) \Theta (1-a)
   +
   2^{1-a} \Theta (a)\right)\,,
   \end{split}
   \label{eq:Fds_weak}
\end{equation}
where we use the following conventions for the Heaviside step function:
\begin{equation}
    \Theta (x)
    =
    \begin{cases}
        0\,, &\text{ if } x \leq 0\,, \\
        +1\,, &\text{ if } x > 0\,.
    \end{cases}
    \label{eq:Heaviside}
\end{equation}
This formula is the compact version of the results presented in Appendix B.1 of \cite{Giombi:2018hsx}.

The next-to-leading order contribution to $\Fds_{\Delta_1 \Dh_2 \Dh_3}$ is 0 for $\Delta_1 \leq \Dh_2-\Dh_3$ and $\Delta_1 < \Dh_2 + \Dh_3$ with $ \Dh_2 + \Dh_3 - \Delta_1$ odd. For the other cases ($\Delta_1 \geq \Dh_2 + \Dh_3$ or $\Delta_1 < \Dh_2 + \Dh_3$ with $ \Dh_2 + \Dh_3 - \Delta_1$ even), we find
\begin{equation}
    \begin{split}
   \Fds_{\Delta_1 \Dh_2 \Dh_3}^{(1)}
   =&
   -\frac{2^{-1+3\Delta_{\hat{2}\hat{3}1}} \sqrt{\Delta_1} \Delta_{\hat{2}\hat{3}1} \lambda^{a/2} }{N}
   \bigg[
   \frac{(-1)^{\Delta_{\hat{2}\hat{3}1}}}{2^{2\Delta_{\hat{2}\hat{3}1}}\Gamma(4+a)}
   \Theta (\Delta_{\hat{2}\hat{3}1})
     \\ 
   &+\frac{1}{\Gamma(3-2\Delta_{\hat{2}\hat{3}1})}
   \left(1-\Theta (\Delta_{\hat{2}\hat{3}1})\right)
   \bigg] -\frac{1}{12}\Fds_{\Delta_1 \Dh_2 \Dh_3}^{(0)}\,,
   \end{split}
   \label{eq:Fds1_weak}
\end{equation}
where the Heaviside step function is defined as in \eqref{eq:Heaviside}.
\paragraph{Strong coupling.}
The strong-coupling limit of the topological sector can be evaluated using the already mentioned tools of \cite{Giombi:2018hsx}.
We focus ourselves here on the case $\vev{2\hat{1}\hat{1}}$, since it is the only one that we consider at strong coupling.
It reads
\begin{equation}
    \Fds_{2\hat{1}\hat{1}}
    =
    \frac{1}{N}
    \biggl(
    \frac{9}{2\sqrt{2}}
    -
    \frac{\sqrt{\lambda}}{\sqrt{2}}
    +
    ...
    \biggr)
    +
    \ldots\,.
    \label{eq:Fds211_strong}
\end{equation}
The topological sector at weak and strong coupling is plotted in Figure \ref{fig:TopologicalSectorAndContourSuperblocks} for $\vev{2 \hat{1} \hat{1}}$.
Note that an exact expression is also provided in \eqref{eq:Fds211_Exact}.

\subsection{Superblock expansion}
\label{subsec:SuperblockExpansion}

The constraints from superconformal symmetry discussed above can be used for deriving an expansion in superblocks for the bulk-defect-defect correlators.\footnote{We thank Gabriel Bliard and Philine van Vliet for sharing preliminary results and for useful conversations around this topic.}
Generically the expansion in superblocks takes the form
\begin{equation}
    \Am_{\Delta_1 \Dh_2 \Dh_3} (\zeta ; x)
    =
    \sum_{\Oh} b_{\Delta_1 \Oh} \lambdah_{\Dh_1 \Dh_2 \Oh} \Gm_{\Oh} (\zeta ; x)\,,
    \label{eq:GeneralCase_SuperblockExpansion}
\end{equation}
where $\Oh$ designates the superprimary operators appearing in the OPE $\Oh_{\Dh_2} \times \Oh_{\Dh_3}$.
For the case of the Wilson-line CFT this OPE is given, e.g., in \cite{Ferrero:2023znz} and reads
\begin{align}
    \Oh_{\Dh_2} \times \Oh_{\Dh_3}
    \longrightarrow
    \sum_{m = \Dh_{23} \text{ step }2}^{\Dh_2 + \Dh_3} \Bm_m
    +
    \sum_{i=0}^{\Dh_2 - 1} \sum_{j=0}^i \sum_{\Dh^{(0)} > 2i + \Dh_{23} + 1} \Lm_{[2i -2j, 2j + \Dh_{23}], 0}^{\Dh}\,.
    \label{eq:OPE}
\end{align}
In the case $\Dh_2 = \Dh_3$, the leading contribution corresponds to the identity through the identification $\Bm_0 = \hat{\mathds{1}}$.
The operators $\Bm_m$ are half-BPS, while the $\Lm_{[a,b],0}^{\Dh}$ have unprotected scaling dimensions.
The lowest-lying operator (both at weak and strong coupling) is $\phi^6$, which has dimension \cite{Alday:2007hr,Grabner:2020nis,Cavaglia:2021bnz}
\begin{align}
    \Delta_{\phi^6}
    &\overset{\lambda \sim 0}{=}
    1 + \frac{\lambda}{4 \pi^2} + \ldots\,, \label{eq:Delta_phi6_weak} \\
    \Delta_{\phi^6}
    &\overset{\lambda \gg 1}{=}
    2 - \frac{5}{\sqrt{\lambda}} + \ldots\,. \label{eq:Delta_phi6_strong}
\end{align}
For $\vev{2 \hat{1} \hat{1}}$, the content of the OPE is
\begin{equation}
    \Oh_1 \times \Oh_1 \longrightarrow \hat{\mathds{1}} + \Bm_2 + \sum_{\Dh^{(0)}\geq 1} \Lm_{[0,0],0}^\Dh\,.
    \label{eq:OPE_Oh1}
\end{equation}
In a perturbative setting, operators can become degenerate.
The degeneracy grows with the tree-level scaling dimension $\Delta^{(0)}$ and is lifted gradually as one includes quantum corrections.
The large $N$ limit helps to reduce the degeneracy, but there are still many operators with the same tree-level scaling dimensions.\footnote{A counting of the number of operators can be found in \cite{Ferrero:2023znz}.}

From \eqref{eq:OPE_Oh1}, we can read the expansion of $\vev{2 \hat{1} \hat{1}}$ in superblocks:
\begin{equation}
    \Am_{2\hat{1}\hat{1}} (\zeta; x)
    =
    a_2 \Gm_{\hat{\mathds{1}}} (\zeta ; x) + b_{2\hat{2}} \lambdah_{\hat{1}\hat{1}\hat{2}} \Gm_{\hat{2}} (\zeta ; x) + \sum_{\Dh} b_{2 \Dh} \lambdah_{\hat{1} \hat{1} \Dh} \Gm_{\Dh} (\zeta ; x)\,,
    \label{eq:211_SuperblockExpansion}
\end{equation}
where $\Dh$ in the last sum refers to the operators of $\Lm_{[0,0],0}^{\Dh}$.

We now discuss how to derive the explicit form of the superblocks.
They can be decomposed into a $R$-symmetry and a spacetime part through the following sum over the Dynkyn labels of the $R$-symmetry and the scaling dimensions, here labelled as $r$:
\begin{equation}
    \Gm_{\Oh} (\zeta ; x)
    =
    \sum_{a,b,r}
    \alpha_{[a,b],r} h_{[a,b]}(\zeta) g_{r} (x)\,,
    \label{eq:Superblock_Decomposition}
\end{equation}
where $g_{\Dh} (x)$ refers to the bosonic blocks already introduced in \eqref{eq:ConformalBlocks}.
The range of the sums is determined by considering the content of the supermultiplets.
This analysis was done in \cite{Ferrero:2023znz} for the case of the Wilson-line defect CFT.
We then apply the SCWI \eqref{eq:SCWI} on the blocks in order to fix the open coefficients $\alpha_{[a,b],r}$.
This amounts to requiring that individual blocks satisfy the symmetries of the correlator.

The identity block is simply given by
\begin{equation}
    \Gm_{\hat{\mathds{1}}} (\zeta ; x)
    =
    - \frac{2 \zeta}{x}\,,
    \label{eq:Superblock_Identity}
\end{equation}
where the factor $2$ is there to account for the one present in the definition \eqref{eq:RSymmetryCrossRatio}.

For the operator $\Oh_2$, we see in (B.4) of \cite{Ferrero:2023znz} that the block should take the form
\begin{equation}
    \Gm_{\hat{2}} (\zeta; x)
    =
    \alpha_{[0,2],2} h_{[0,2]} (\zeta) x^{-1} g_2 (x)
    +
    \alpha_{[2,0],3} h_{[2,0]} (\zeta) x^{-1} g_3 (x)
    +
    \alpha_{[0,0],4} h_{[0,0]} (\zeta) x^{-1} g_4 (x)\,.
    \label{eq:Superblock_B2_Form}
\end{equation}
Note that we have selected scalar contributions in the supermultiplet.
Generically, $R$-symmetry blocks depend on the external operators, and for $\vev{2\hat{1}\hat{1}}$ they are expected to take the form
\begin{equation}
    h_{[a,b]} (\zeta)
    =
    \beta_{[a,b]}^{(0)}
    +
    \beta_{[a,b]}^{(1)} \zeta\,,
    \label{eq:RSymmetryBlock_Ansatz}
\end{equation}
where the coefficients $\beta_{[a,b]}^{(0)}$ are so far unfixed.
Applying the SCWI \eqref{eq:SCWI} and choosing the normalization according to the highest-weight channel we obtain
\begin{equation}
    \Gm_{\hat{2}} (\zeta; x)
    =
    \biggl( 1 - \frac{2}{5} \zeta \biggr) x^{-1} g_2 (x) - \frac{12}{175} \frac{\zeta}{x} g_4 (x)\,,
    \label{eq:Superblock_B2}
\end{equation}
The same method can be applied for the superblocks of the long operators, yielding
\begin{equation}
    \begin{split}
        \Gm_{\Dh} (\zeta ; x)
        &=
        -\frac{\zeta}{x} g_\Dh (x)
        +
        \biggl( 1-\frac{2 \left(\Delta ^2+3 \Delta +1\right) \zeta }{4 \Delta ^2+12 \Delta +5} \biggr) x^{-1} g_{\Dh+2} (x) \\
        &\phantom{=\ }
        -
        \frac{(\Delta +2)^2 (\Delta +3)^2 }{(2 \Delta +3) (2 \Delta +5)^2 (2 \Delta +7)} \frac{\zeta}{x} g_{\Dh+4} (x)\,.
    \end{split}
    \label{eq:SuperblockLong}
\end{equation}

The expansion in superblocks can be expanded in the coupling constant, through
\begin{equation}
    \Delta
    =
    \Delta^{(0)} + \lambda \gamma^{(1)}_{\Delta^{(0)}} + \ldots\,,
    \label{eq:Delta_Expansion}
\end{equation}
and similarly for OPE coefficients.
This allows a perturbative analysis, which will play an important role in Section \ref{sec:PerturbativeResults}.
It is important however to notice that, as mentioned above, OPE coefficients and scaling dimensions become degenerate in a perturbative setting.
Instead of individual coefficients, we are therefore forced to consider their \textit{average} only.
For instance,
\begin{equation}
    \vev{ b_{2 \Dh}^{(\ell)} \lambdah_{\hat{1} \hat{1} \Dh}^{(\ell)} }
    =
    \sum_{\text{deg}} b_{2 \Dh}^{(\ell)} \lambdah_{\hat{1} \hat{1} \Dh}^{(\ell)}\,,
    \label{eq:AverageOPE_Definition}
\end{equation}
where the sum is over all the operators that have the same tree-level scaling dimension $\Delta^{(0)}$.
Note that we are using here the notation of \cite{Ferrero:2023znz}.
Moreover the same analysis can be performed at strong coupling, and it is known that in this regime the degeneracy is being lifted slower than at weak coupling \cite{Liendo:2018ukf,Ferrero:2021bsb}.

The superblock expansion can be studied in the topological sector as well.
It is well-known that only half-BPS contributions survive in this kinematic limit.
For the case of $\vev{2 \hat{1} \hat{1}}$ we have
\begin{equation}
    \Fds_{2 \hat{1} \hat{1}}
    =
    -2 a_2 + \vev{b_{2\hat{2}} \lambdah_{\hat{1} \hat{1} \hat{2}}}\,.
    \label{eq:Fds211_AlmostExact}
\end{equation}
This provides an exact formula for the topological sector discussed in Section \ref{subsubsec:LocalizationResults}.
The average above corresponds to the contributions from the two operators $\Oh_{2}$ and $\Oh_{(0|2)}$.
This degeneracy is in fact absent, as can be seen in the following way.
From the Gram-Schmidt procedure to build an orthogonal basis of operators mentioned above, the good operator $\Oh_{(0|2)}$ takes the form
\begin{equation}
    \Oh_{(0|2)}
    \sim
    \bigl(
    \hat{\mathrm{O}}_{(0|2)}
    -
    \vev{\Oh_2 \hat{\mathrm{O}}_{(0|2)}} \Oh_2
    \bigr)\,.
    \label{eq:DoubleTrace_GramSchmidt}
\end{equation}
The corresponding OPE coefficient $b_{2(0|2)}$ is of order $\Om(N^0)$ following the $N$ counting of~\eqref{eq:NCountingSpecial}.
However the three-point function $\lambdah_{\hat{1} \hat{1} (0|\hat{2})}$ vanishes as a consequence of \eqref{eq:DoubleTrace_GramSchmidt}.
The reason is that the coefficient $\lambdah_{\hat{1} \hat{1} (0|\hat{2})}$ of the defect three-point function $\vev{\Oh_{1} \Oh_{1} \Oh_{(0|2)}}$ can be extracted from $\vev{\Oh_{2} \Oh_{(0|2)}}$, corresponding to the pinching of the two open-trace operators.
Taking into account orthogonality between operators we can write
\begin{equation}
\frac{\nh_1}{\sqrt{\nh_2}}\lambdah_{\hat{1} \hat{1} (0|\hat{2})} =
  \delta_{\Oh_{2} \Oh_{(0|2)}} = 0\,,
\end{equation}
precisely because of the Gram-Schmidt procedure.
Equivalently, we can state that the defect operator $\Oh_{(0|2)}$ does not appear in the defect OPE $\Oh_1 \times \Oh_1$ and that it only contains the operator $\Oh_2$ at protected dimension two.
The topological sector is therefore simply given by
\begin{equation}
    \Fds_{2 \hat{1} \hat{1}}
    =
    -2 a_2 + b_{2 \hat{2}} \lambdah_{\hat{1} \hat{1} \hat{2}}\,.
    \label{eq:Fds211_Exact}
\end{equation}
This expression is plotted in Figure \ref{fig:TopologicalSectorAndContourSuperblocks} from weak to strong coupling using the analytic results \eqref{eq:OnePoint}, \eqref{eq:BulkDefect_Exact_Delta2} and \eqref{eq:lambdah112}, and is shown to agree well with the perturbative results of \cite{Giombi:2018hsx}.

To conclude this section, let us make some remarks on the analytic structure of the conformal blocks.
As noted in \cite{Levine:2023ywq,Levine:2024wqn}, conformal blocks are discontinuous at $x \geq 1$, which means that they do not satisfy locality on their own.
This is shown in Figure \ref{fig:TopologicalSectorAndContourSuperblocks}.
As a consequence, the (infinite) sum of superblocks should produce a function that is free of this discontinuity.
This imposes strong constraints on the CFT data formulated as sum rules.
Although we do not pursue this line of thoughts further in this paper, it would be interesting to understand how these constraints combine with the other constraints presented in this section.

\subsection{Pinching and splitting limits}
\label{subsec:PinchingAndSplittingLimits}

In this section we present two kinematic limits corresponding to the reduction of some bulk-defect-defect correlators to simple functions, in particular to two-point bulk-defect functions (pinching limit) and products of one-point bulk correlators and two-point defect correlators (splitting limit).
In our framework lower-point functions serve as further non-perturbative information constraining the correlators under study.

\subsubsection{Pinching limit}
\label{subsubsec:PinchingLimits}

The pinching of the two defect operators into one having scaling dimension being the sum of the original two, i.e., the limit of the correlator $\tau_3 \rightarrow \tau_2$ and  $\uh_3 \rightarrow \uh_2$,  results in the OPE coefficient $b_{\Delta_1 \Dh}$ after accounting for the different normalization constants:
\begin{align}
    \vev{ \Delta_1 \Dh }
    =
    \frac{\sqrt{\nh_{\Dh_2} \nh_{\Dh_3}}}{\sqrt{ \nh_{\Dh}} }
    \lim\limits_{3 \to 2}\,
    \vev{ \Delta_1 \Dh_2 \Dh_3 }
    =
    b_{\Delta_1 \Dh}\,
     (12)^{\Dh} (1 \theta)^{\Delta_1 - \Dh} \,,
    \label{eq:PinchingBulkDefectDefect}
\end{align}
with $\Dh = \Dh_2 + \Dh_3$.
This provides a valuable limit of bulk-defect-defect correlators since the coefficients $b_{\Delta_1 \Dh}$ can be evaluated (see \eqref{eq:BulkDefect_WeakCoupling}).
If one inserts the expansion in $R$-symmetry channels \eqref{eq:BulkDefectDefect_RSymmetryChannels} in \eqref{eq:PinchingBulkDefectDefect}, only the channel $F_1$ computed at $x = 0$ survives:
\begin{align}
   \vev{ \Delta_1 \Dh }
    &=  (12)^{\Dh_2 + \Dh_3} (1\theta)^{-2\Delta_{\hat{2}\hat{3}1}}
    \biggl(
    F_1^{(0)} (0)
    + \lambda
    \biggl(
    F_1^{(1)} (0) - \frac{F_1^{(0)} (0)}{48}
    \biggr)
    + \Om(\lambda^2)
    \biggr)\,,
    \label{eq:BulkDefectInF0}
\end{align}
from which it is possible to extract the pinching constraints at leading and next-to-leading orders:
\begin{align}
    F_1^{(0)} (0) &=
    \frac{\sqrt{\Delta_1}}{(-2\Delta_{\hat{2}\hat{3}1})!} \frac{\lambda^{-\Delta_{\hat{2}\hat{3}1}}}{2^{-3\Delta_{\hat{2}\hat{3}1}} N}\,, \label{eq:PinchingConstraint_LO} \\
    F_1^{(1)} (0) &=
    \frac{\sqrt{\Delta_1}}{(-2\Delta_{\hat{2}\hat{3}1})!} \frac{\lambda^{-\Delta_{\hat{2}\hat{3}1}}}{2^{(-3\Delta_{\hat{2}\hat{3}1}+2)} N} 
    \frac{(1+2\Delta_{\hat{2}\hat{3}1})}{3 (2 - 2\Delta_{\hat{2}\hat{3}1})}\,.
    \label{eq:PinchingConstraint_NLO}
\end{align}

It might be tempting to consider another pinching limit, which consists in bringing the bulk operator in a third position on the line in order to generate a double-trace defect operator.
This operator would however be of the form \eqref{eq:HigherTraceBPSOperators} and not orthonormal to the open-trace operators, as discussed in Section \ref{subsubsec:LocalizationResults}.
Therefore, the fact that this pinching limit gives a non-vanishing result for $\vev{2 \hat{1} \hat{1}}$ is not an inconsistency with respect to \eqref{eq:DoubleTrace_GramSchmidt}.

\subsubsection{Splitting limit}
\label{subsubsec:SplittingLimits}

A different limit comes from considering a large separation between the bulk operator and the defect ones, i.e., $|\vec{x}_1| \to \infty$.
In the correlator expressed in $R-$symmetry channels \eqref{eq:BulkDefectDefect_RSymmetryChannels}, this corresponds to the limit $x \rightarrow 0$ of the whole correlator without setting $\zeta = 0$ (which would instead correspond to the previous pinching).
In this case the bulk-defect-defect correlator factorizes into a product of a two-point defect correlator and a one-point function of a bulk operator: 
\begin{equation}
    \lim\limits_{\chi \to 0}\,
    \vev{ \Delta_1 \Dh_2 \Dh_3 } = \vev{\Delta_1} \vev{\Dh_2 \Dh_3}\,,
    \label{eq:SplittingADelta}
\end{equation}
from which \eqref{eq:BulkDefectDefect_RSymmetryChannels} simplifies to 
\begin{equation}
        \lim\limits_{x \to 0}\,
    \vev{ \Delta_1 \Dh_2 \Dh_3 } =  a_{\Delta_1} \Km_{\Delta_1\Dh_2\Dh_3} \left(\frac{\zeta}{x}\right)^{\Dh_2} \delta_{\Dh_2\Dh_3}\,,
    \label{eq:Splitting}
\end{equation}
where the coefficient $a_{\Delta_1}$ is given in \eqref{eq:OnePoint}.
This can serve as a constraint or consistency check for the case $\Dh_2 = \Dh_3$: Equation \eqref{eq:Splitting} combined with \eqref{eq:BulkDefectDefect_RSymmetryChannels} gives
\begin{equation}
        F_{\Dh_2 +1}(0) =  \left( - 2\right)^{\Dh_2} a_{\Delta_1} \delta_{\Dh_2\Dh_3}\,,
    \label{eq:SplittingCheck}
\end{equation}
which is valid non-perturbatively.

\section{Perturbative results}
\label{sec:PerturbativeResults}

In this section we present explicit results for the bulk-defect-defect correlators introduced in Section \ref{subsubsec:CorrelatorsOfBulkAndDefectOperators} and for which we listed non-perturbative constraints in Section \ref{sec:NonPerturbativeConstraints}.
We begin by calculating correlators up to next-to-leading order at weak coupling.
The results are obtained by combining the constraints mentioned above with simple calculations of Feynman diagrams that do not involve bulk vertices.
For the case of $\vev{2 \hat{1} \hat{1}}$, we determine the CFT data using the superblocks presented in Section \ref{subsec:SuperblockExpansion}.
We observe that transcendental terms are systematically absent at this order, which results into new relations for the OPE coefficients.
We then consider the strong-coupling limit using a mixture of Witten diagrams and non-perturbative constraints.
We determine the contributions to the correlator $\vev{2 \hat{1} \hat{1}}$ up to next-to-leading order and extract the corresponding CFT data.

\subsection{Weak coupling}
\label{subsec:WeakCoupling}

\subsubsection{Perturbative structure}
\label{subsubsec:PerturbativeStructure_weak}

At weak coupling, the bulk-defect-defect correlator $\vev{2 \hat{1} \hat{1}}$ has the following perturbative structure at large $N$:
\begin{equation}
    \Am_{2\hat{1}\hat{1}} (\zeta;x)
    =
    \frac{1}{N}
    \bigl(
    \Am_{2\hat{1}\hat{1}}^{(0)} (\zeta;x)
    +
    \lambda\, \Am_{2\hat{1}\hat{1}}^{(1)} (\zeta;x)
    +
    \ldots
    \bigr)
    + \ldots\,,
    \label{eq:PerturbativeStructure_211_weak}
\end{equation}
where the $\ldots$ inside the brackets refer to higher powers of $\lambda$, while the $\ldots$ at the end of the expression refer to corrections in $N$ and $\lambda$.
We focus mainly on this correlator in this section for ease of readability, before generalizing the expressions for arbitrary external dimensions in Section \ref{subsubsection:GeneralizationToArbitraryDelta1Dh2Dh3}.
As explained in \eqref{eq:BulkDefectDefect_RSymmetryChannels} $\vev{2\hat{1}\hat{1}}$ consists of \textit{two} $R$-symmetry channels:
\begin{equation}
    \Am_{2\hat{1}\hat{1}}^{(\ell)} (\zeta ; x)
    =
    F_1^{(\ell)} (x) + \frac{\zeta}{x} F_2^{(\ell)} (x)\,.
    \label{eq:RSymmetryChannels_211}
\end{equation}

\subsubsection{Leading order}
\label{subsubsec:LeadingOrder_weak}

\paragraph{Correlator.}
As a warm-up, let us determine the leading order for the case $\vev{2\hat{1}\hat{1}}$.
This correlator consists of a single Feynman diagram, which can be represented as
\begin{equation}
    \Am_{2 \hat{1} \hat{1}}^{(0)} (\zeta ; x)
    =
    \BulkDefectDefectLO\,,
    \label{eq:LO_FeynmanDiagram_211}
\end{equation}
and only consists of the free propagators defined in \eqref{eq:Propagators}.
It is easy to see that the $R$-symmetry channels take the form
\begin{equation}
    F_1^{(0)} (x)
    =
    c_0\,,
    \qquad
    F_2^{(0)} (x)
    =
    0\,,
    \label{eq:LO_Fixed_211}
\end{equation}
where $c_0$ is a constant that can be fixed either through direct calculation, from the pinching constraint of \eqref{eq:PinchingConstraint_LO} or using the topological sector \eqref{eq:Fds_weak}.
We find
\begin{equation}
    c_0
    =
    b_{2\hat{2}}^{(0)} \lambdah_{\hat{1}\hat{1}\hat{2}}^{(0)}
    =
    \sqrt{2}\,.
    \label{eq:LO_Constant_211}
\end{equation}
Note that the identity does not contribute at this order, i.e., we have
\begin{equation}
    a_2^{(0)}
    =
    0\,.
    \label{eq:LO_Identity_211}
\end{equation}

\paragraph{CFT data.}
It is a trivial task to extract the CFT data for these results using the superblock expansion \eqref{eq:211_SuperblockExpansion}.
For the long operators labelled $\Lm_{[0,0],0}^\Dh$ we find the following closed form:
\begin{equation}
    \vev{ b_{2\Dh}^{(0)} \lambdah_{\hat{1}\hat{1}\Dh}^{(0)} }
    =
    \begin{cases}
        0\,, & \text{ if } \Dh^{(0)} \text{ odd}\,, \\
        \frac{(-1)^{\Dh/2} \sqrt{\pi} \Gamma (\Dh+2)}{2^{\Dh+1/2} \Gamma (\Dh+3/2)}\,, & \text{ if } \Dh^{(0)} \text{ even}\,.
    \end{cases}
    \label{eq:LO_CFTData_211}
\end{equation}
This is the supersymmetric version of the formula given in \cite{Levine:2023ywq,Levine:2024wqn}, which corresponds to an expansion in bosonic blocks.

\subsubsection{Next-to-leading order}
\label{subsubsec:NextToLeadingOrder_weak}

\begin{table}
    \centering
    \caption{Feynman diagrams contributing to the correlator $\vev{2\hat{1}\hat{1}}$ at next-to-leading order.}
    \begin{tabular}{lc}
        \hline \\[1pt]
        $F_1^{(1)} (x)$ & 
        \BulkDefectDefectNLOSEOne \;\;
        \BulkDefectDefectNLOSETwo \;\;
        \BulkDefectDefectNLOX \;\;
        \BulkDefectDefectNLOH \;\;
        \BulkDefectDefectNLOYOne \;\;
        \BulkDefectDefectNLOYTwo  \\[5ex]
        \hline \\[1pt]
        $F_2^{(1)} (x)$ & 
        \BulkDefectDefectNLOFoneOne \;\;
        \BulkDefectDefectNLOFoneTwo \;\;
        \BulkDefectDefectNLOFoneThree\\[5ex]
        \hline
    \end{tabular}
    \label{table:Diagrams211NLO}
\end{table}

We now consider the next-to-leading order.

\paragraph{Correlator.}
The Feynman diagrams corresponding to the correlator are listed in Table \ref{table:Diagrams211NLO}.
Thanks to the non-perturbative constraints, we do not have to calculate all the diagrams.
In particular, we can focus on the diagrams that do not contain bulk vertices.
They are easy to compute using the integrals \eqref{eq:MasterIntegralDefet1} and \eqref{eq:MasterIntegralDefet2}.
We obtain
\begin{align}
    \BulkDefectDefectNLOFoneOne
    &= - \mathcal{K}_{2\hat{1}\hat{1}} \frac{1}{16 \sqrt{2} \pi^2} \frac{\zeta}{x} \biggl( \pi + 2 \text{arctan} \biggl( \sqrt{\frac{1-x}{x}} \biggr) \biggr)^2\,, \label{eq:NLO_FeynmanDiagrams_F2_211_1} \\[1.5ex]
    \BulkDefectDefectNLOFoneTwo
    &= - \mathcal{K}_{2\hat{1}\hat{1}} \frac{1}{16 \sqrt{2} \pi^2} \frac{\zeta}{x} \biggl( \pi - 2 \text{arctan} \biggl( \sqrt{\frac{1-x}{x}} \biggr) \biggr)^2\,,
    \label{eq:NLO_FeynmanDiagrams_F2_211_2}
\end{align}
where the prefactors result from symmetry factors and trace contributions.
Note that each open line in \eqref{eq:NLO_FeynmanDiagrams_F2_211_1}-\eqref{eq:NLO_FeynmanDiagrams_F2_211_2} can end on the orange dots placed along the defect.
Putting the two diagrams together, we obtain for the $R$-symmetry channel $F_2$ the result
\begin{align}
    F_2^{(1)} (x)
    &=
    - \frac{1}{8 \sqrt{2} \pi^2}
    (\pi^2 + 4 \text{arccos}^2 \left(\sqrt{x}\right))\,.
    \label{eq:NLO_ResultF2_211}
\end{align}
As a consistency test we can check that this expression is compatible with the splitting constraint of Section \ref{subsec:PinchingAndSplittingLimits}:
\begin{equation}
    F_2^{(1)} (0)
    =
    -2 a_2^{(1)}
    =
    - \frac{1}{4 \sqrt{2}}\,.
    \label{eq:NLO_Identity_211}
\end{equation}

\begin{figure}
\centering
\begin{subfigure}{.5\textwidth}
  \includegraphics[width=.9\linewidth]{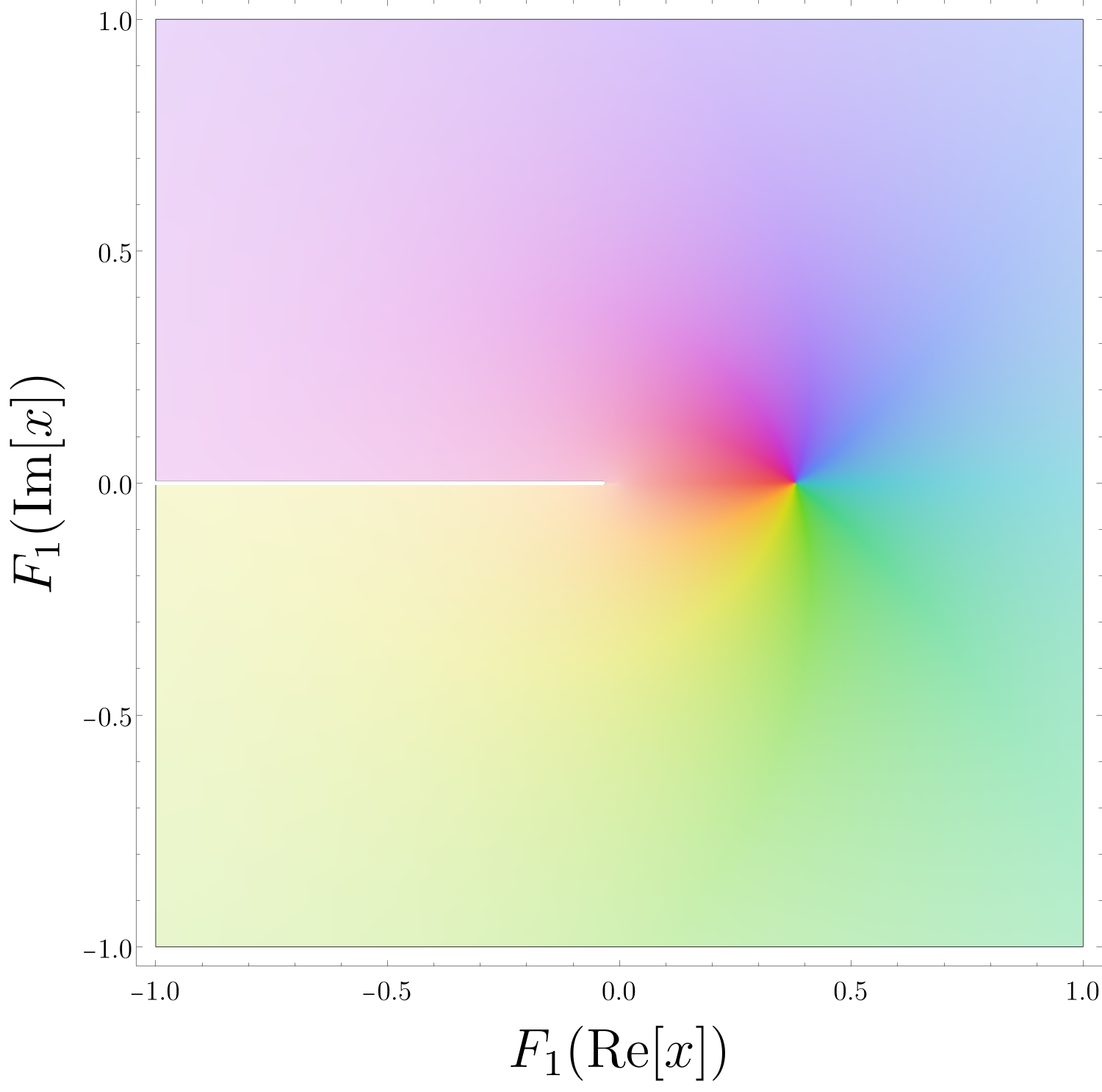}
\end{subfigure}%
\begin{subfigure}{.5\textwidth}
  \includegraphics[width=.9\linewidth]{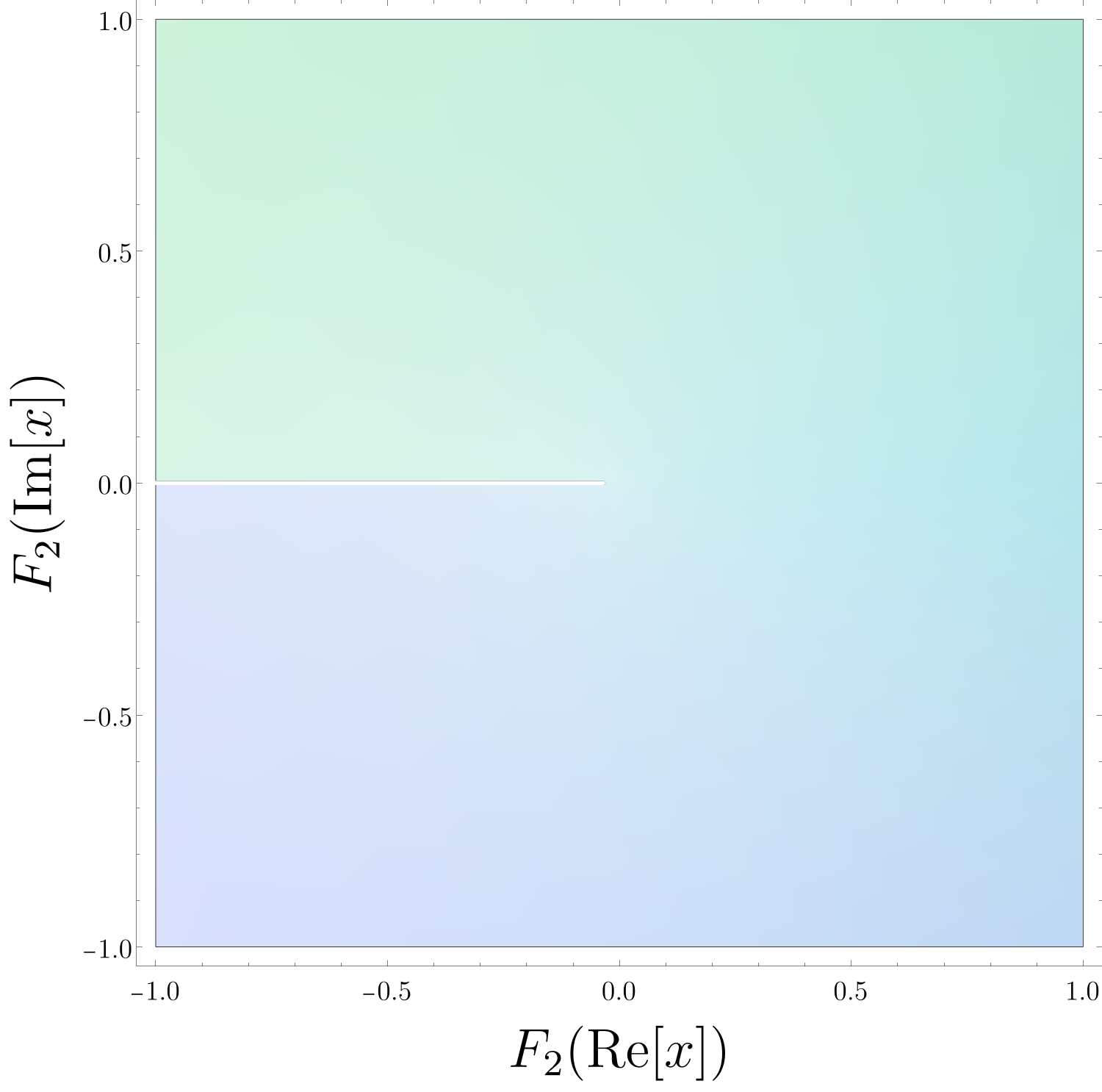}
\end{subfigure}
\caption{Analytic structure of the correlator $\Am_{2 \hat{1} \hat{1}}$ at next-to-leading order at weak coupling.
The left figure shows the $R$-symmetry channel $F_1(x)$, while the second one corresponds to $F_2(x)$.
We observe that the correlator is discontinuous at $x \leq 0$, while it is continuous at $x=1$.
This matches perfectly the analysis of \cite{Levine:2023ywq,Levine:2024wqn}.
}
\label{fig:ContourPlots_NLO}
\end{figure}

The channel $F_1$ can be obtained through the supersymmetry constraints of Section \ref{subsec:SuperconformalSymmetry}.
Using the solution \eqref{eq:SolutionWI} to the Ward identity, we obtain
\begin{equation}
    F_1^{(1)} (x)
    =
    \Fds_{2\hat{1}\hat{1}}^{(1)}
    -
    F_2^{(1)} (x)\,.
    \label{eq:NLO_AfterWI_211}
\end{equation}
Using the results of \cite{Giombi:2018hsx} (summarized in Section \ref{subsubsec:LocalizationResults}) we can calculate the topological sector and get
\begin{equation}
    \Fds_{2\hat{1}\hat{1}}^{(1)}
    =
    - \frac{1}{6 \sqrt{2}}\,.
    \label{eq:NLO_Topological_211}
\end{equation}
The correlator is now completely fixed.
Notice that this result was obtained without considering the second line of Table \ref{table:Diagrams211NLO}, where the Feynman diagrams consist of bulk vertices and are significantly harder than the first line.
We nevertheless calculated these diagrams as a check of our results and found perfect agreement.
This computation is reported in Appendix \ref{app:CheckThroughFeynmanDiagrams}.

\begin{figure}
\centering
\begin{subfigure}{.65\textwidth}
  \includegraphics[width=1\linewidth]{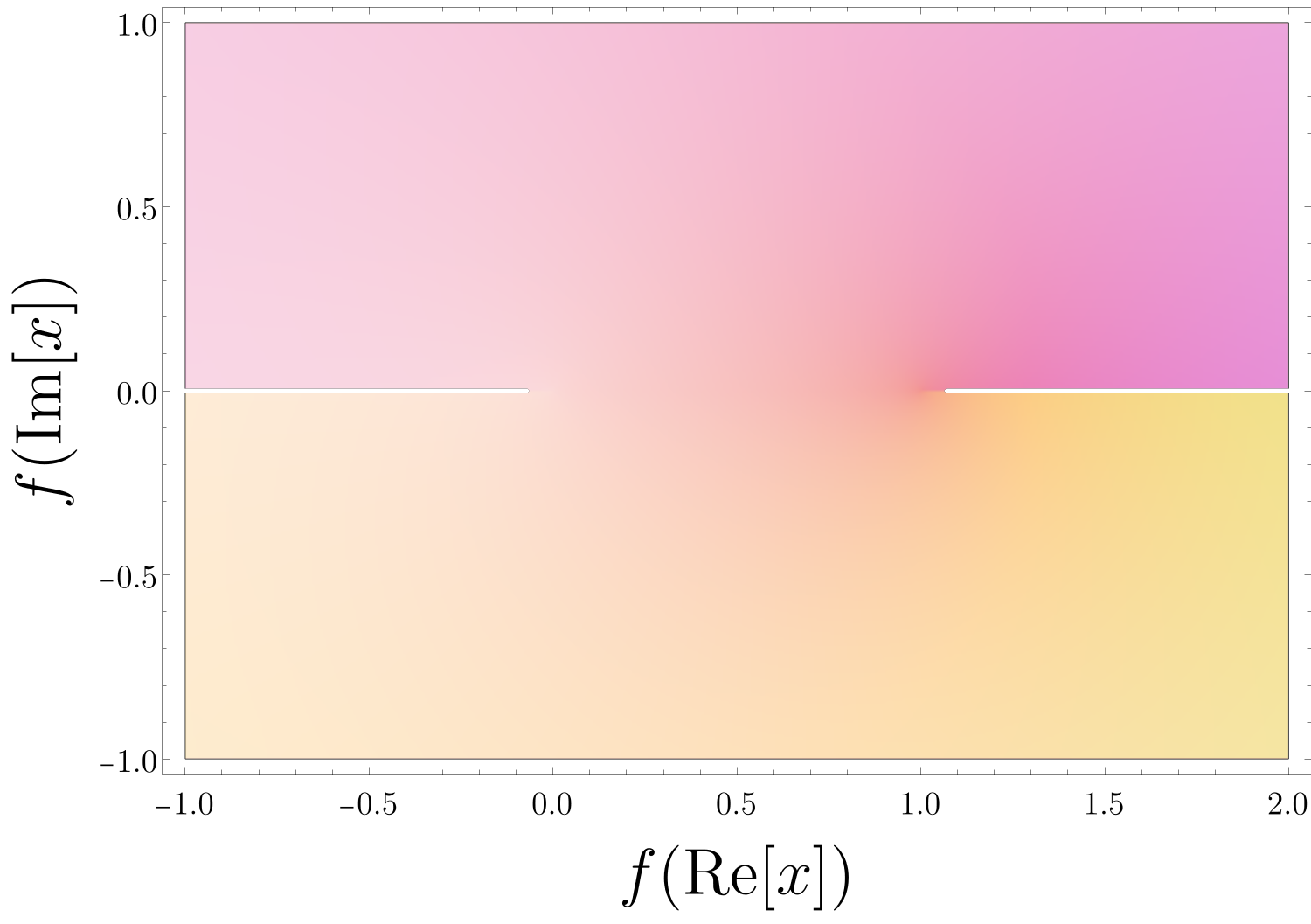}
\end{subfigure}%
\caption{Complex plot of the Feynman diagram given in Equation \eqref{eq:NLO_FeynmanDiagrams_F2_211_1}
Here, we defined $f(x) = \bigl(\pi + \arctan (\sqrt{(1-x)/x}) \bigr)^2$.
We observe a discontinuity at $x \geq 0$, incompatible with the locality constraint.
}
\label{fig:ComplexPlot_ArcTan}
\end{figure}

It is instructive to look at the analytic structure of our results.
In order to satisfy locality constraints, bulk-defect-defect correlators are expected to have a branch cut at $x \leq 0$ only.
The $R$-symmetry channels $F_1$ and $F_2$ are plotted in Figure \ref{fig:ContourPlots_NLO}, where they are seen to satisfy this property.
Notice also that the individual diagrams of \eqref{eq:NLO_FeynmanDiagrams_F2_211_1}-\eqref{eq:NLO_FeynmanDiagrams_F2_211_2} do not satisfy the locality constraints, as can be seen in Figure \ref{fig:ComplexPlot_ArcTan}.

\paragraph{CFT data.}
The results above can be used for determining the CFT data at next-to-leading order.
Notice that the expressions \eqref{eq:NLO_ResultF2_211} and \eqref{eq:NLO_AfterWI_211} are entirely \textit{rational}, i.e., there are no transcendental functions appearing at this order.
This is rather surprising since we know that the operators exchanged in the OPE generally have anomalous dimensions at one loop, while we have seen in \eqref{eq:LO_CFTData_211} that operators with even $\Dh$ have a tree-level OPE coefficient.
This observation can be summarized in the following equation:
\begin{equation}
    \vev{ b_{2 \Dh}^{(0)} \lambdah_{\hat{1}\hat{1}\Dh}^{(0)} \gamma_\Dh^{(1)} }
    =
    0\,,
    \label{eq:NLO_NoLog_211}
\end{equation}
where the brackets $\vev{ \ldots }$ should be understood in the average sense explained in \eqref{eq:AverageOPE_Definition} and below.
We comment on the absence of logarithmic terms in Section \ref{subsubsec:OnTheAbsenceOfTranscendentalFunctions}.

In \eqref{eq:LO_CFTData_211} we saw that operators with odd tree-level scaling dimension have vanishing OPE coefficients at order $\Om(\lambda^0/N)$.
They start to contribute at one loop, and we find the following closed form for their average:
\begin{equation}
    \vev{ b_{2 \Dh} \lambdah_{\hat{1}\hat{1}\Dh}}^{(1)}_{\Dh \text{ odd}}
    =
    \frac{(-1)^{\Dh/2} i\, \Gamma(\Dh+2)}{2^{\Dh+5/2} \sqrt{\pi} \Gamma(\Dh+3/2)} H_{\Dh+1}\,,
    \label{eq:NLO_OPECoeffsOdd_211}
\end{equation}
with $H_n$ the harmonic numbers defined through
\begin{equation}
    H_n
    =
    \sum_{k=1}^n \frac{1}{k}\,.
    \label{eq:HarmonicNumbers}
\end{equation}
The lowest operator described by \eqref{eq:NLO_OPECoeffsOdd_211} corresponds to $\phi^6$ and is not degenerate.
Its corresponding OPE coefficient is given by
\begin{equation}
    (b_{2 {\phi^6}} \lambdah_{\hat{1}\hat{1}\phi^6})^{(1)}
    =
    - \frac{1}{2\sqrt{2} \pi}\,,
    \label{eq:NLO_OPECoeffsphi^6_211}
\end{equation}
from which we read, after reinstating the powers of $\lambda$ and $N$,
\begin{equation}
    b_{2 {\phi^6}} = \frac{\sqrt{\lambda}}{N} + \Om(\lambda^{3/2})\,,
    \label{eq:b2phi6Leading}
\end{equation}
after using~\cite{Barrat:2022eim}
\begin{equation}
    \lambdah_{\hat{1}\hat{1}\phi^6} = - \frac{\sqrt{\lambda}}{2\sqrt{2}\pi} + \Om(\lambda^{3/2})\,.
    \label{eq:lambda11phi6Leading}
\end{equation}
Equation~\eqref{eq:b2phi6Leading} is also in agreement with Equation~(4.7b) of~\cite{Barrat:2020vch}.\footnote{\label{footnote:Factor4}Note that there is a factor $4$ difference between the definition of the CFT data in this paper and in~\cite{Barrat:2020vch}, which can be tracked back to the definition of the $R$-symmetry channels.}

Long operators with an even tree-level scaling dimension receive corrections to their tree-level value given in \eqref{eq:LO_CFTData_211}.
It is difficult to find an analytic closed form, however they can be expressed in the form of a recursion relation:
\begin{align}
    \vev{ b_{2 \Dh} \lambdah_{\hat{1}\hat{1}\Dh} }^{(1)}_{\Dh \text{ even}}
    &=
    - \frac{(1)_{\Dh/2}}{12 \sqrt{2} (5/2)_{\Dh/2}}
    \biggl( 1 - \frac{4}{\pi^2} \frac{(\Dh+1)(\Dh+3)}{\Dh^2} \biggr) \notag \\
    &\phantom{=\ }
    - \frac{\Gamma(\Dh/2+1)}{\Gamma(-\Dh/2)}
    \sum_{h=0 \text{ step } 2}^{\Dh-2}
    \frac{i^h \Gamma (h+5/2) \Gamma((h-\Dh)/2)}{\Gamma^2(h/2+1) \Gamma((h+\Dh+5)/2)}
    \vev{ b_{2 h} \lambdah_{\hat{1}\hat{1}h}}^{(1)}\,,
    \label{eq:NLO_OPECoeffsEven_211}
\end{align}
with initial value
\begin{equation}
    \vev{ b_{2 0} \lambdah_{\hat{1}\hat{1}0}}^{(1)}
    =
    (b_{2 \mathds{1}} \lambdah_{\hat{1}\hat{1}\mathds{1}})^{(1)}
    =
    0\,.
    \label{eq:NLO_CFTData_InitialValue}
\end{equation}
Although this is not an analytic expression, note that recursion relations are extremely efficient, even for high $\Dh$.
In the formula above we used the Pochhammer symbols, defined through
\begin{equation}
    (a)_n
    =
    \frac{a!}{(a-n)!}\,.
    \label{eq:Pochhammer}
\end{equation}

\subsubsection{Generalization to arbitrary $\Delta_1$, $\Dh_2$, $\Dh_3$}
\label{subsubsection:GeneralizationToArbitraryDelta1Dh2Dh3}

We now repeat the analysis of the correlators presented above for the general correlator $\vev{\Delta_1 \Dh_2 \Dh_3}$. We do not extract the CFT data although we note that it is in principle possible to do it case by case in the same way as in Sections \ref{subsubsec:LeadingOrder_weak} and \ref{subsubsec:NextToLeadingOrder_weak}.

\paragraph{General Feynman diagrams.}
We have seen in \eqref{eq:LO_FeynmanDiagram_211} and \eqref{eq:NLO_FeynmanDiagrams_F2_211_1}-\eqref{eq:NLO_FeynmanDiagrams_F2_211_2} that the correlator $\vev{2 \hat{1} \hat{1}}$ is fixed up to next-to-leading order by Feynman diagrams that do not contain bulk vertices.
This observation holds for the general case $\vev{\Delta_1 \Dh_2 \Dh_3}$.
We are therefore interested in diagrams of the form
\begin{equation}
    \BulkDefectDefectGeneralLeadingOrder\,.
    \label{eq:GeneralFeynmanDiagrams}
\end{equation}
Here, the thick solid lines designate $a_k$ free propagators.
The coefficients $a_{k=1, \ldots, 4}$ refer to the possible allowed contractions without involving a bulk vertex.
They satisfy the consistency relations
\begin{equation}
    \begin{split}
        a_1 + a_2 + a_4 &= \Delta_1\,, \\
        a_1 + a_3 &= \Dh_2\,, \\
        a_2 + a_3 &= \Dh_3\,.
    \end{split}
    \label{eq:ConsistencyConditions_GeneralCase}
\end{equation}

\paragraph{Perturbative structure.}
The Feynman diagrams \eqref{eq:GeneralFeynmanDiagrams} can be used to understand the perturbative structure of the bulk-defect-defect correlators.
It is not hard to see that general correlators have to take the following form:
\begin{equation}
    \Am_{\Delta_1 \Dh_2 \Dh_3} (\zeta ; x)
    =
    \frac{\lambda^{a/2}}{N} \biggl(
    \Am_{\Delta_1 \Dh_2 \Dh_3}^{(0)} (\zeta ; x)
    +
    \lambda \Am_{\Delta_1 \Dh_2 \Dh_3}^{(1)} (\zeta ; x)
    + \ldots
    \biggr)
    + \ldots\,,
    \label{eq:PerturbativeStructure_weak_General}
\end{equation}
where we defined $a=\text{min}(a_4)$, i.e., the lowest number of propagators that can connect the bulk operators and the line defect for a given configuration characterized by $\Delta_1$, $\Dh_2$, $\Dh_3$, while still satisfying \eqref{eq:ConsistencyConditions_GeneralCase}.
The number of $R$-symmetry channels can grow arbitrarily high, as can be seen in \eqref{eq:NumberOfRSymmetryChannels1}.
However, at low orders, most of the $R$-symmetry channels are suppressed for a given configuration.
Concretely,
\begin{align}
    \Am_{\Delta_1 \Dh_2 \Dh_3}^{(0)}
    &=
    F_1^{(0)} (x)\,, \\
    \Am_{\Delta_1 \Dh_2 \Dh_3}^{(1)}
    &=
    F_1^{(1)} (x) + \frac{\zeta}{x} F_2^{(1)} (x)\,, \\
    &\phantom{||}\vdots
    \label{eq:RSymmetryChannels_PerturbativeStructure}
\end{align}
A new channel appears at each order until the full number of channels has been reached.
In the following we study the correlators up to next-to-leading order, in which only \textit{two} $R$-symmetry channels contribute.

\paragraph{Leading order.}
The leading order is given by the topological sector since only one channel contributes and it is constant.
To see that this is the case, consider the Feynman diagrams \eqref{eq:GeneralFeynmanDiagrams} for the distinct cases $\Dh_2 - \Dh_3 \leq \Delta_1 < \Dh_2 + \Dh_3$ and $\Delta_1 \geq \Dh_2 + \Dh_3$.
In the first case, there is no propagator connecting to the defect when minimizing $a_4$, and thus the leading $R$-symmetry channel is constant.
The second configuration leads to considering \textit{nested} defect integrals.
Using the identity \eqref{eq:IntegratePhi6OnTheLine} we can rewrite them in terms of a single master defect integral:
\begin{equation}
    \vev{\Delta_1 \Dh_2 \Dh_3}^{(0)}
    =
    c_0 (\Delta_1, \Dh_2, \Dh_3)
    \biggl(
    \int_{-\infty}^{\infty} d\tau_4\, I_{14}
    \biggr)^{-2\Delta_{\hat{2}\hat{3}1}}\,.
    \label{eq:LO_Integrals_Case2}
\end{equation}
This integral is elementary and its solution is given by the sum of \eqref{eq:MasterIntegralDefet1} and \eqref{eq:MasterIntegralDefet2}.
The coefficient $c_0 (\Delta_1, \Dh_2, \Dh_3)$ encodes the symmetry factors and the contributions from the traces.
It is given by the topological sector \eqref{eq:Fds_weak} (or the pinching limit \eqref{eq:PinchingConstraint_NLO} when applicable).

\paragraph{Next-to-leading order.}
At next-to-leading order we use the supersymmetry constraint \eqref{eq:SolutionWI} and focus on calculating the channel $F_2$ only:
\begin{equation}
    F_1^{(1)} (x)
    =
    \Fds_{\Delta_1 \Dh_2 \Dh_3}^{(1)}
    -
    F_2^{(1)} (x)\,.
    \label{eq:211_SolutionWI}
\end{equation}
We only need to solve the generalized version of \eqref{eq:NLO_FeynmanDiagrams_F2_211_1}-\eqref{eq:NLO_FeynmanDiagrams_F2_211_2} in order to obtain all the correlators. We can group the correlators into two different groups:
\begin{itemize}
    \item $\Dh_2-\Dh_3 \leq \Delta_1 \leq \Dh_2 - \Dh_3 + 1$:
    These configurations are the ones where only one $R$-symmetry channel exists.
    In this case, the correlator is topological and $\Am_{\Delta_1 \Dh_2 \Dh_3}$ is equal to the topological sector $\Fds_{\Delta_1 \Dh_2 \Dh_3}$;
    \item $\Dh_2-\Dh_3 + 1  < \Delta_1$:
    In the conformal frame $\tau_3 \to \infty$, the only surviving diagrams are the ones depicted in \eqref{eq:GeneralFeynmanDiagrams}, for which the value of $a_4$ is now raised by $2$.
    The correlator can then be determined from the function
    \begin{equation}
        F_2^{(1)} (x)
        =
        c_1 (\Delta_1, \Dh_2, \Dh_3)
        \sum_{\pm} \biggl(
        \pi \pm 2 \text{arctan} \biggl( \sqrt{\frac{1-x}{x}} \biggr)
        \biggr)^{a+2}\,,
        \label{eq:F2_NLO_Generalization}
    \end{equation}
    which is the generalization of \eqref{eq:NLO_FeynmanDiagrams_F2_211_1}-\eqref{eq:NLO_FeynmanDiagrams_F2_211_2}. The sum over the signs should be understood as the presence of two summands with opposite relative signs.
    The constant $c_1$ is then fixed by the topological sector and pinching/splitting limits of Section \ref{subsec:PinchingAndSplittingLimits}.
\end{itemize}

\subsubsection{On the absence of transcendental functions}
\label{subsubsec:OnTheAbsenceOfTranscendentalFunctions}

We now comment on the absence of transcendental functions observed at next-to-leading order.
In this setup the combination of Feynman diagrams gathered in Appendix~\ref{app:CheckThroughFeynmanDiagrams} forms a rational function and the superconformal Ward identity forces the other channel to be rational as well.
It is unclear at present whether we should expect this property to appear in other theories.
We provide in this section an explicit example of this cancellation.

\paragraph{Example: $\Dh^{(0)}=2$ for $\vev{2 \hat{1} \hat{1}}$.}
Let us illustrate Equation \eqref{eq:NLO_NoLog_211} for the OPE coefficient corresponding to $\Dh^{(0)}=2$ to convince ourselves that it is not a trivial relation.
From the correlator $\vev{2 \hat{1} \hat{1}}$, we expect through \eqref{eq:LO_CFTData_211} that
\begin{equation}
    \vev{b_{2 \Dh}^{(0)} \lambdah_{\hat{1}\hat{1}\Dh}^{(0)} \gamma_\Dh^{(1)}}_{\Dh^{(0)} = 2} = 0\,.
    \label{eq:211_CFTDataCancellation_2}
\end{equation}
We now need to identify the operators that contribute to this expression.
As discussed in Section~\ref{subsubsec:CorrelatorsOfBulkAndDefectOperators}, for $\Dh^{(0)} = 2$ there are three operators contributing in the OPE $\Oh_1 \times \Oh_1$: the two open-trace operators $\Oh_\pm$ and the double-trace one $\Oh_{(20')}$.
The one-loop anomalous dimensions of $\Oh_\pm$ are given by~\cite{Correa:2018fgz}
\begin{equation}
    \gamma_{\pm}^{(1)}
    =
    \frac{5 \pm \sqrt{5}}{16 \pi^2}\,,
    \label{eq:gammaOpm}
\end{equation}
while $\Oh_{(20')}$ has protected dimension two.
We therefore have the following relations:
\begin{align}
    0 &= b_{2 \Oh_+}^{(0)} \lambdah_{\hat{1}\hat{1}\Oh_+}^{(0)} \gamma_{\Oh_+}^{(1)} + b_{2 \Oh_-}^{(0)} \lambdah_{\hat{1}\hat{1}\Oh_-}^{(0)} \gamma_{\Oh_-}^{(1)}\,, \label{eq:Relation1} \\
    -\frac{2\sqrt{2}}{5 N} &= b_{2 \Oh_+}^{(0)} \lambdah_{\hat{1}\hat{1}\Oh_+}^{(0)} + b_{2 \Oh_-}^{(0)} \lambdah_{\hat{1}\hat{1}\Oh_-}^{(0)} + b_{2 \Oh_{(20')}}^{(0)} \lambdah_{\hat{1}\hat{1}\Oh_{(20')}}^{(0)}\,.\label{eq:Relation2}
\end{align}
It is easy to calculate explicitly the tree-level OPE data for these operators and we find
\begin{align}
    \lambdah_{\hat{1}\hat{1}\Oh_\pm}^{(0)} &= \frac{1}{\sqrt{10}}\,, \label{eq:lambdah11pm} \\
    \lambdah_{\hat{1}\hat{1}\Oh_{(20')}}^{(0)} &= - \frac{1}{\sqrt{15} N}\,, \label{eq:lambdah11dt20} \\
    b_{2 \Oh_\pm}^{(0)}  &= \frac{- \sqrt{5} \pm 5}{5 N}\,, \label{eq:b2pm} \\
    b_{2 \Oh_{(20')}}^{(0)}  &= \sqrt{\frac{6}{5}}\,.
    \label{eq:b2dt20}
\end{align}
Putting everything together we find that the relations~\eqref{eq:Relation1}-\eqref{eq:Relation2} are satisfied.
Moreover, the $N$ scaling matches the analysis performed in~Section~\ref{subsubsec:CorrelatorsOfBulkAndDefectOperators}.
The fact that everything appears to be consistent provides a nice check of the basis of double-trace operators.

\subsection{Strong coupling}
\label{subsec:StrongCoupling}

\begin{figure}
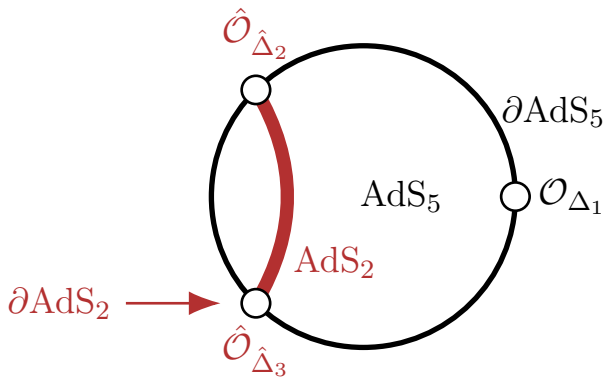

    \centering
    \WittenDiagrams
    \caption{Illustration of the Witten diagrams for the bulk-defect-defect correlators at strong coupling.
    As labelled on the figure, the inside of the circle represents the AdS$_5$ spacetime, while its boundary corresponds to the dual CFT$_4$ (in this case, $\Nm=4$ SYM).
    The bold red line represents the string worldsheet that is dual holographically to the Wilson line.
    The worldsheet and the boundary CFT coincide on a line, which is the Maldacena-Wilson line \eqref{eq:WilsonLine}.
    The bulk operator $\Op_{\Delta_1}$ lives in the CFT$_4$, while the operators $\Oh_{\Dh_2}$ and $\Oh_{\Dh_3}$ are representations of the CFT$_1$ dual to the AdS$_2$ surface.
    Note that the fact that the worldsheet and the CFT$_4$ coincide on two points only on the figure is an artefact of the representation.
    }
    \label{fig:WittenDiagrams}
\end{figure}

We now consider the bulk-defect-defect correlators in the strong-coupling regime.\footnote{We are thankful to Ziwen Kong and Jake Belton (see also~\cite{Belton:2025ief}) for pointing out mistakes in an earlier version of this section.}
We provide the leading and next-to-leading orders with the corresponding CFT data for $\vev{2\hat{1}\hat{1}}$, applying non-perturbative constraints on the expected contributions from vertices in the effective AdS$_2$ action.

\subsubsection{Perturbative structure}
\label{subsubsec:PerturbativeStructure_strong}

We study the strong-coupling regime through a perturbative expansion at large $N$ of the form
\begin{equation}
    \Am_{2 \hat{1} \hat{1}} (\zeta;x)
    =
    \frac{\sqrt{\lambda}}{N}
    \biggl(
    \Am_{2 \hat{1} \hat{1}}^{(0)} (\zeta;x)
    +
    \frac{1}{\sqrt{\lambda}} \Am_{2 \hat{1} \hat{1}}^{(1)} (\zeta;x)
    +
    \ldots
    \biggr)
    + \ldots\,.
    \label{eq:StrongCoupling_PerturbativeStructure}
\end{equation}
For simplicity, we focus on $\vev{2 \hat{1} \hat{1}}$.
As in \eqref{eq:PerturbativeStructure_211_weak} the $\ldots$ refer to corrections in $\lambda$ inside the brackets and in $N$ outside the brackets.

\subsubsection{Leading order}
\label{subsubsec:LeadingOrder_strong}

At leading order there is only one disconnected Witten diagram, which corresponds to the contribution of the identity operator:
\begin{equation}
    \left. \vev{2\hat{1}\hat{1}} \right|_{\Om(\sqrt{\lambda})} 
    =\ 
    \WittenDiagramOne\ .
\end{equation}
We have seen in Section \ref{subsec:SuperblockExpansion} that the identity only appears in $F_2$ (in the form of a constant), and thus we have
\begin{equation}
    F_2^{(0)} (x)
    =
    - 2 a_2^{(0)}
    =
    - \frac{1}{\sqrt{2}}\,,
    \qquad
    F_1^{(0)}
    =
    0\,.
    \label{eq:StrongCoupling_Results_LO}
\end{equation}
The CFT data for the long operators is thus
\begin{equation}
    \vev{ b_{2 \Dh}^{(0)} \lambdah_{\hat{1}\hat{1}\Dh}^{(0)} }
    =
    0\,.
    \label{eq:StrongCoupling_CFTData_LO}
\end{equation}

\subsubsection{Next-to-leading order}
\label{subsubsec:NextToLeadingOrder_strong}

We now consider the correlator at next-to-leading order.
Our approach is to analyze which terms are allowed to appear from the effective AdS$_2$ action and use our constraints to fix the remaining coefficients.

\paragraph{Effective action.}
Our starting point is the effective action in AdS$_2$ induced by the presence of the minimal worldsheet surface.
In the conventions of~\cite{Gimenez-Grau:2023fcy} it is given by
\begin{equation}
    S_{F1} = \frac{\sqrt{\lambda}}{2\pi} \int \text{d}^2\hat{z}\, \sqrt{\text{det} (\gamma_{\alpha\beta})}\,,
    \label{eq:ActionF1}
\end{equation}
where $\alpha, \beta = 1,2$ are the AdS$_2$ indices and $\gamma_{\alpha\beta} = G_{MN} (X) \pd_\alpha X^M \pd_\beta X^N$ is the induced worldsheet metric.
The AdS duals of the CFT operators are defined in~\eqref{eq:AdSDualTo20} and~\eqref{eq:AdSDualToOh1}.
For our purposes we only need to track the vertices of the type $yys$ and therefore we can set
\begin{equation}
    X^0 = \hat{z}^0\,, \quad X^\mu = \hat{z}^1 \delta^{\mu 1}\,, \quad X^i = y^i\,,
    \label{eq:XToy}
\end{equation}
effectively ignoring the interactions involving the displacement operator.
The idea is now to expand $\gamma_{\alpha\beta}$ around the static $y^i=0$ to obtain the effective Lagrangian $\Lm_{F1}^{(2,1)}$, with the superscripts referring to how many defect fields (two) and bulk fields (one) are coupled, using again the conventions of~\cite{Gimenez-Grau:2023fcy}.
We identify five possible vertices:
\begin{align}
    &s_2^{\tilde{I}} y_i y_j \pd^i \pd^j Y_2^{\tilde{I}}\,, \quad
    \nabla_\alpha \nabla^{\alpha} s_2^{\tilde{I}} y_i y_j \pd^i \pd^j Y_2^{\tilde{I}}\,, \label{eq:Vertices1} \\
    &\nabla_\alpha \nabla^{\alpha} s_2^{\tilde{I}} \pd^\beta y_i \pd_\beta y^i Y_2^{\tilde{I}}\,, \quad
    \nabla_\alpha \nabla^\beta s_2^{\tilde{I}} \pd_\beta y_i \pd_\alpha y^i Y_2^{\tilde{I}}\,, \quad
    s_2^{\tilde{I}} \pd^\alpha y_i \pd_\alpha y^i Y_2^{\tilde{I}}\,.
    \label{eq:Vertices2}
\end{align}
In the following we bootstrap the coefficients in front of each of these vertices in our correlator of interest instead of working out their prefactors generally.

\paragraph{Witten diagram.}
At the order $\Om(\lambda^0)$ there is only one Witten diagram contributing to the bulk-defect-defect correlator:
\begin{equation}
    \left. \vev{2\hat{1}\hat{1}} \right|_{\Om(\lambda^0)} 
    =\
    \WittenDiagramTwo\,,
    \label{eq:WittenDiagram2}
\end{equation}
which is formed by the five vertices given in \eqref{eq:Vertices1} and \eqref{eq:Vertices2}.
We therefore formulate an Ansatz of the following form for the two channels:
\begin{align}
    F_1^{(1)} (x) &= \sum_{i=1}^5 \alpha_i V_i (x)\,, \label{eq:F1Ansatz} \\
    F_2^{(1)} (x) &= c + x \sum_{i=1}^5 \beta_i V_i (x)\,,
    \label{eq:F2Ansatz}
\end{align}
with $V_i$ the expressions generated by the vertices once one performs the associated integrals.
They are performed explicitly in Appendix \ref{app:WittenIntegrals} and the expressions associated to each vertex are given by
\begin{align}
    V_1 (x) &= - \frac{\pi}{4} \frac{\log(x)}{1-x}\,, \label{eq:V1} \\
    V_2 (x) &= - \frac{\pi}{2(1-x)} \left(3 + \frac{2x+1}{1-x} \log(x) \right)\,, \label{eq:V2} \\
    V_3 (x) &= - \frac{\pi}{(1-x)^2} \left( 3(1+4x) + \frac{1+19x+10x^2}{2(1-x)} \log(x) \right)\,, \label{eq:V3} \\
    V_4 (x) &= - \frac{\pi}{2 (1-x)^2} \left( \frac{3}{2} (3+7x) + \frac{1+10x+4x^2}{1-x} \log(x) \right)\,, \label{eq:V4} \\
    V_5 (x) &= \frac{V_2(x)}{2}\,. \label{eq:V5}
\end{align}
Note that we included an additional constant $c$ in $F_2^{(1)} (x)$ which corresponds to a potential contribution from the tree-level diagram due to the normalization constant.

\paragraph{Applying the constraints.}
We now fix the coefficients $\alpha_i$ and $\beta_i$.
To begin, note that only the vertices $V_1$ and $V_2$ can contribute to $F_2^{(1)} (x)$ due to the contraction $y_i y^i$ in~\eqref{eq:Vertices2}.
We immediately conclude that
\begin{equation}
    \beta_3 = \beta_4 = \beta_5 = 0\,.
    \label{eq:FixCoefficients1}
\end{equation}
We can then impose that the sum $F_1 + F_2$ is topological, as explained in Section~\ref{subsec:SuperconformalSymmetry}.
This fixes five coefficients:
\begin{align}
    \alpha_1 &= - \frac{4}{3\pi} (c - \Fds_{2\hat{1}\hat{1}})\,, \label{eq:FixCoefficients2} \\
    \alpha_2 &= \frac{2}{3\pi} (c - \Fds_{2\hat{1}\hat{1}}) - \frac{\alpha_4 + \alpha_5}{2}\,, \label{eq:FixCoefficients3} \\
    \alpha_3 &= - \frac{\alpha_4}{2}\,, \label{eq:FixCoefficients4} \\
    \beta_1 &= - \frac{8}{3\pi} (c - \Fds_{2\hat{1}\hat{1}})\,, \label{eq:FixCoefficients5} \\
    \beta_2 &= - \frac{2}{3\pi} (c - \Fds_{2\hat{1}\hat{1}})\,.
    \label{eq:FixCoefficients6}
\end{align}
Moreover the relation $\beta_2/\beta_1 = 4$ imposes that $\alpha_2/\alpha_1 = 4$, fixing
\begin{equation}
    \alpha_5 = \frac{2}{\pi} (c - \Fds_{2\hat{1}\hat{1}}) - \alpha_4\,.
    \label{eq:FixCoefficients7}
\end{equation}
Note that at this point $F_2$ is already fixed up to the constant $c$, while $F_1$ is fixed up to $c$ and $\alpha_4$.
Since the channels are related by the topological sector, it is clear that $\alpha_4$ drops from the correlator and we only need to fix $c$.
To do so we use the splitting property \eqref{eq:SplittingCheck} of $F_2$, which reads
\begin{equation}
    c = F_2^{(1)} (0) = \frac{3}{2\sqrt{2}}\,.
    \label{eq:FixCoefficients8}
\end{equation}
We can fix $\Fds_{2\hat{1}\hat{1}}$ either via the pinching limit of $F_1$ or through the topological sector given in \eqref{eq:Fds211_Exact}.
Both methods yield the same value.
The final result for the correlator reads\footnote{It would be interesting to see whether $\log$ terms persist for non-extremal correlators and/or correlators with odd/even $\Delta_1 + \Dh_2 + \Dh_3$.
A crude analysis based on Witten diagrams suggest that logarithms might be absent in $\vev{3\hat{1}\hat{1}}$ but present in $\vev{4\hat{1}\hat{1}}$.
We reserve a thorough study of these correlators to future work.}
\begin{align}
    F_1^{(1)} (x) &= \frac{3}{\sqrt{2} (1-x)} \left(1 + \frac{x}{1-x} \log(x) \right)\,, \label{eq:F1Final} \\
    F_2^{(1)} (x) &= \frac{3}{\sqrt{2} (1-x)} \left( \frac{1-3x}{2} - \frac{x}{1-x} \log(x) \right)\,.
    \label{eq:F2Final}
\end{align}

\paragraph{CFT data.}
We can now extract the CFT data from the correlator~\eqref{eq:F1Final}-\eqref{eq:F2Final}.
From the $\log$ terms we read that
\begin{equation}
    \vev{ b_{2 \Dh}^{(0)} \lambdah_{\hat{1}\hat{1}\Dh}^{(0)} \gamma_\Dh^{(1)} }
    =
    \frac{\sqrt{2} \Dh (\Dh+1) \Gamma\left(\frac{\Dh + 1}{2}\right) \Gamma\left(\frac{\Dh + 5}{2}\right)}{\sqrt{\pi}\Gamma\left(\Dh + \frac{3}{2}\right)}\,.
    \label{eq:CFTDataStrong1}
\end{equation}
Noticing that $b_{2 (20')}$ is of order $\Om(1)$ and protected, thus given by~\eqref{eq:b2dt20}, one can solve~\eqref{eq:StrongCoupling_CFTData_LO} and \eqref{eq:CFTDataStrong1} with $\Dh=2$ for $b_{2 \phi^6}^{(0)}$.
Using the known data
\begin{equation}
    \lambdah_{\hat{1}\hat{1}\phi^6}^{(0)} = \sqrt{\frac{2}{5}}\,, \qquad \gamma^{(1)}_{\phi^6} = - 5\,,
\end{equation}
we obtain
\begin{align}
    b_{2 \phi^6} = - \frac{3}{\sqrt{5}} \frac{\sqrt{\lambda}}{N} + \ldots\,.
    \label{eq:b2phi6Strong}
\end{align}
A byproduct of our analysis is the three-point function
\begin{equation}
    \lambda_{\hat{1} \hat{1} (20')} = \sqrt{\frac{3}{5}} \frac{\sqrt{\lambda}}{N} + \ldots\,.
\end{equation}
It should be noted that~\eqref{eq:b2phi6Strong} does not appear to agree with Equation (3.36) in~\cite{Barrat:2021yvp}.
The mismatch might be due to corrections in $N$ of the double-trace operator $\Oh_{(20')}$ that would mix with $\phi^6$.
We reserve a better understanding of double-trace operators in the correlator $\vev{\Om_2 \Om_2}$ for future studies.

We have not been able to find a closed form for the corrections to the OPE coefficients which can be read from the rational terms.
We provide here the OPE data for the lowest-lying operators:
\begin{align}
    \vev{ b_{2 \hat{2}} \lambdah_{\hat{1}\hat{1}\hat{2}}}^{(1)} &= \frac{9}{5\sqrt{2}}\,, \label{eq:CFTDataStrong2} \\
    \vev{ b_{2 \hat{4}} \lambdah_{\hat{1}\hat{1}\hat{4}}}^{(1)} &= - \frac{145}{189\sqrt{2}}\,, \label{eq:CFTDataStrong3} \\
    \vev{ b_{2 \hat{6}} \lambdah_{\hat{1}\hat{1}\hat{6}}}^{(1)} &= - \frac{25235}{20449\sqrt{2}}\,, \label{eq:CFTDataStrong4} \\
    &\ \vdots \notag
\end{align}
Based on the lowest $14$ OPE coefficients, we observe the following asymptotic exponential behavior:
\begin{equation}
    \vev{ b_{2 \Dh} \lambdah_{\hat{1}\hat{1}\Dh}}^{(1)}
    \overset{\Dh \to \infty}{\sim}
    350\, e^{-.57\Dh}\,.
    \label{eq:CFTDataStrongAsymptotic}
\end{equation}
It would be interesting to refine this expression and interpret it on the AdS side.

\section{Conclusions}
\label{sec:Conclusions}

\begin{figure}
\centering
\begin{subfigure}{.65\textwidth}
  \includegraphics[width=1\linewidth]{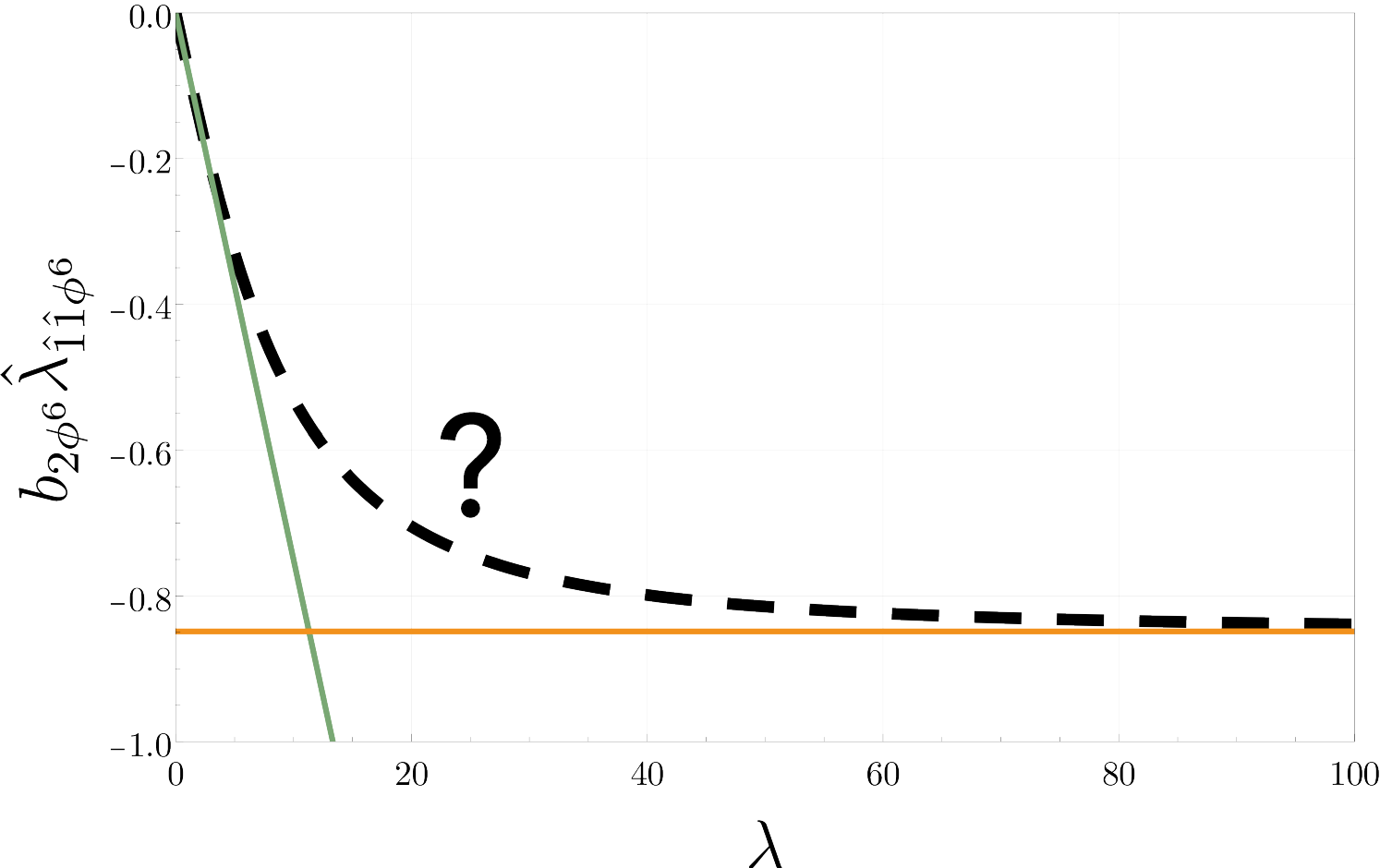}
\end{subfigure}%
\caption{Plot of the OPE coefficient $b_{2\phi^6} \lambdah_{\hat{1}\hat{1}\phi^6}$ as a function of the coupling constant.
The green line corresponds to the weak-coupling value while the orange line represents the strong-coupling regime, both extracted from our results for $\vev{2\hat{1}\hat{1}}$.
The black dashed line is a two-point Pad{\'e} approximation between these two regimes.
An interesting goal would be to use numerical bootstrap methods for studying this intermediate regime.
}
\label{fig:PlotConclusions}
\end{figure}

In this work, we have studied the correlation functions involving one bulk and two defect half-BPS operators in the context of the Wilson-line defect CFT in $\Nm=4$ SYM.
This setup preserves many properties of $\Nm=4$ SYM without defects, such as supersymmetry, integrability and a large subset of conformal symmetry.
Through a combination of non-perturbative constraints and perturbative insights, we derived novel results for external operators with arbitrary scaling dimensions at weak coupling.
At next-to-leading order, we observe a cancellation of transcendental terms.
This is explained through intricate relations between the OPE coefficients and the scaling dimensions, which follows from supersymmetry. It is unclear what the fate of this observation is in other models.
In the strong-coupling regime, we provide results for the correlator $\vev{2 \hat{1} \hat{1}}$ up to next-to-leading order, combining the calculation of Witten diagrams with the known non-perturbative constraints.
In this case the correlator includes logarithmic terms.
Using this result we extract some previously unknown conformal data at strong coupling that also include the contribution of higher-trace operators.

There are many interesting directions that one can take after this work.
We list here some that we find particularly promising:
\begin{itemize}
    \item In the weak-coupling regime little is known about the functional space of these correlators.
    In the case of the Wilson line in $\Nm=4$ SYM, we encounter trigonometric functions of the spacetime cross-ratio $x$, which are rational for the whole range $x \in [0,1]$.
    This is not surprising as $x$ is directly related to the angle formed by the two defect operators while keeping the bulk operator $\Op_{\Delta_1}$ fixed.
    It would be interesting to study the integrals encountered at next-to-next-to-leading order using the variable $\phi = \arccos x$ and see if it leads to simplifications. If the space of functions is understood, a valuable tool to further constrain the result might be locality, in the spirit of the discontinuity analysis presented in \cite{Levine:2023ywq}. As Figure \ref{fig:ComplexPlot_ArcTan} shows, the functions appearing as the result of individual Feynman diagrams can have non-physical branch cuts: If the space of functions is known, applying locality constraints can partially fix the contribution from other diagrams, as non-physical discontinuities must cancel;
    \item We observe a surprising cancellation of the transcendental terms at next-to-leading order.
    Although this can be explained through relations between the OPE coefficients, it would be valuable to gain a deeper insight into the reasons that prevent logarithms to appear, in particular to see if supersymmetry is ultimately responsible for this cancellation.
    One idea is to consider the bulk-defect-defect correlator of the lowest-lying operators in the $O(N)$ model using the $\veps$-expansion at the Wilson-Fisher fixed point (see \cite{Cuomo:2021kfm,Gimenez-Grau:2022czc,Gimenez-Grau:2022ebb,Bianchi:2022sbz,Pannell:2023pwz,Giombi:2023dqs} for related works).
    At next-to-leading order a single diagram contributes:
    \begin{equation}
        \vev{\phi(x_1) \hat{\phi}(\tau_2) \hat{\phi}(\tau_3)} = \ONDiagramNLO \sim \pi^2-4\arccos \left(\sqrt{x}\right)
        \label{eq:ONModel}
    \end{equation}
    We observe in this case as well an absence of transcendental terms, which might hint at a more universal mechanism that does not involve supersymmetry.
    Future studies could investigate this phenomenon for more models (e.g. fishnet field theory in presence of a defect \cite{Wu:2020nis,Gromov:2021ahm} or fermionic defect CFT \cite{Giombi:2022vnz, Barrat:2023ivo,Pannell:2023pwz})  and different external operators;
    \item The sum rules presented in \cite{Levine:2023ywq} are amenable to a numerical study, especially in the Wilson-line defect case where the spectrum is known through integrability across the conformal manifold \cite{Cavaglia:2021bnz,Cavaglia:2022qpg,Cavaglia:2022yvv}.
    Typical bootstrap strategies are however suffering from the fact that the OPE coefficients are not positive, as it is the case for four-point functions of identical operators.\footnote{Note that an additional difficulty lies in the fact that double-trace operators contribute in this correlator. This might be solvable for practical purposes in the same way as done in Section~\ref{subsubsec:CorrelatorsOfBulkAndDefectOperators}.}
    An alternative method would be to truncate the sum \textit{à la} Gliozzi \cite{Gliozzi:2016cmg} and use an adapted version of the Tauberian theorem to estimate the tail, similarly to what was done in thermal cases \cite{Qiao:2017xif,Marchetto:2023xap,Barrat:2024aoa}.
    It would also be interesting to see a full-fledged bootstrap study of a defect CFT that involves the four-point functions of defect operators $\vev{\Dh_1 \Dh_2 \Dh_3 \Dh_4}$, the two-point functions of bulk operators (in presence of the defect) $\vev{\Delta_1 \Delta_2}$, together with the bulk-defect-defect correlators $\vev{\Delta_1 \Dh_2 \Dh_3}$.
    This should result in intertwining relations that might provide interesting relations for lifting degeneracies in perturbative settings.
    A good candidate of an observable to study numerically and that appears in $\vev{2 \hat{1} \hat{1}}$ is the combination $b_{2\phi^6} \lambdah_{\hat{1}\hat{1} \phi^6}$ (see Figure \ref{fig:PlotConclusions}).
    The three-point function $\lambdah_{\hat{1}\hat{1} \phi^6}$ has been studied extensively and is known precisely from weak to strong coupling, while presently very little is known about $b_{2\phi^6}$.
    The numerical approach could also profit from the integrated correlator relations derived in~\cite{Girault:2025kzt,Belton:2025ief}, similar to the ones derived for $\Nm=4$ SYM in \cite{Dorigoni:2022zcr,Alday:2023pet} and for line defects in \cite{Drukker:2022pxk,Cavaglia:2022qpg,Cavaglia:2022yvv,Pufu:2023vwo,Billo:2023ncz,Billo:2024kri,Dempsey:2024vkf}.
\end{itemize}

\acknowledgments

We are particularly grateful to Jake Belton, Lorenzo Bianchi, Gabriel Bliard, Valentina Forini, Julius Julius, Ziwen Kong, Enrico Marchetto, Elli Pomoni, Nika Sokolova, Roman Stemplowski, and Philine Van Vliet for useful discussions.
DA and YX are funded by the Deutsche Forschungsgemeinschaft (DFG, German Research Foundation) -- Projektnummer 417533893/ GK2575 ``Rethinking Quantum Field Theory''.
JB is supported by ERC-2021-CoG - BrokenSymmetries 101044226, and has benefited from the German Research Foundation DFG under Germany’s Excellence Strategy – EXC 2121 Quantum Universe – 390833306.
DA is especially grateful to the organizers of the workshop \textit{Defects, from condensed matter to quantum gravity} for providing the framework for useful discussions.

\appendix

\section{Integrals}
\label{app:Integrals}

\subsection{Bulk integrals}
\label{subsec:BulkIntegrals}

We gather here the bulk integrals used throughout this work.
We often encounter the well-known $X$-integral, for which the definition and the solution are
\begin{align}
    X_{1234}
    =
    \XIntegral\
    =
    \int d^4 x_5\, I_{15} I_{25} I_{35} I_{45}
    =
    \frac{I_{12} I_{34}}{16\pi^2}\ z \zb\, D( z, \zb)\,,
    \label{eq:X1234}
\end{align}
where we have used the Bloch-Wigner function \cite{bloch1978applications}, defined as
\begin{equation}
    D(z, \zb)
    =
    \frac{1}{z - \zb} \left[ 2 \Li_2 (z) - 2 \Li_2 (\zb) + \log (z \zb)\, \log \left(\frac{1-z}{1-\zb}\right) \right]\,.
    \label{eq:BlochWigner}
\end{equation}
Here the cross-ratios $z$ and $\zb$ are related to the coordinates of the external points in the standard way:
\begin{equation}
    z \zb = \frac{x_{12}^2 x_{34}^2}{x_{13}^2 x_{24}^2}\,.
    \qquad
    (1-z)(1-\zb) = \frac{x_{14}^2 x_{23}^2}{x_{13}^2 x_{24}^2}\,.
    \label{eq:4dCrossRatios}
\end{equation}
Note that the Bloch-Wigner function is crossing symmetric:
\begin{equation}
    D(z, \zb) = D(1-z\,, 1-\zb)\,.
    \label{eq:BWIsCrossingSymmetric}
\end{equation}
If two external points are coincident, the $X$-integral becomes divergent.
It can be expressed in the following way:
\begin{equation}
    X_{1233}
    =
    \IntegralXOneTwoThreeThree\
    =
    \frac{1}{2} ( I_{13} Y_{223} + I_{23} Y_{113} ) - \frac{I_{13} I_{23}}{32 \pi^2} \log \left( \frac{x_{12}^4}{x_{13}^2 x_{23}^2} \right)\,,
    \label{eq:X1233}
\end{equation}
where the integral $Y_{iij}$ is given in \eqref{eq:Y112}.

We also encounter the $Y$-integral, which can be obtained through the limit $\tau_4 \to \infty$ of the $X$-integral:
\begin{align}
    Y_{123}
    =
    \YIntegral\
    &=
    \int d^4 x_4\, I_{14} I_{24} I_{34} \notag \\
    &=
    \lim\limits_{\tau_4 \to \infty} I_{34}^{-1} X_{1234}
    =
    \frac{I_{12}}{16\pi^2} z \zb\, D( z, \zb)\,.
    \label{eq:Y123}
\end{align}
The cross-ratios $z, \zb$ are now defined as
\begin{equation}
    z \zb
    =
    \frac{I_{13}}{I_{12}}\,,
    \qquad
    (1-z)(1-\zb)
    =
    \frac{I_{13}}{I_{23}}\,.
    \label{eq:CrossRatiosY}
\end{equation}
The $Y$-integral is log-divergent in the limit where two external points are coincident.
The corresponding expression is given by
\begin{equation}
    Y_{112} = Y_{122} = \IntegralYOneTwoTwo\ = - \frac{I_{12}}{16\pi^2} \left[ \log \left( \frac{\veps^2}{x_{12}^2} \right) - 2 \right]\,.
    \label{eq:Y112}
\end{equation}
Note that we use point-splitting regularization.
The $Y$-integral is often encountered with derivatives.
The following identities are useful for manipulating the integrals:
\begin{equation}
    \begin{split}
    \partial_{1 \mu} Y_{123} &= - (\partial_{2 \mu} + \partial_{3 \mu}) Y_{123}\,, \\[.6em]
    \partial_1^2\, Y_{123} &= - I_{12} I_{13}\,, \\
    \left( \partial_1 \cdot \partial_2 \right) Y_{123} &= \frac{1}{2} \left( I_{12} I_{13} + I_{12} I_{23} - I_{13} I_{23} \right)\,,
    \end{split}
    \label{eq:YIdentities}
\end{equation}
where we have use the scalar Green's equation
\begin{equation}
    \pd_1^2 I_{12}
    =
    - \delta^{(4)} (x_{12})\,.
    \label{eq:GreensEquation}
\end{equation}

It is useful to define the following integral:
\begin{equation}
    F_{12,34}
    =
    \frac{\pd_{12} \cdot \pd_{34}}{I_{12} I_{34}} H_{12,34}\,,
    \label{eq:F1234_Definition}
\end{equation}
where we have introduced the shorthand notation $\pd_{ij}^\mu := \pd_i^\mu - \pd_j^\mu$.
This integral is encountered in Feynman diagrams with two gluon vertices.
It can be expressed in terms of $X$- and $Y$-integrals \cite{Beisert:2002bb}:
\begin{equation}
    \begin{split}
    F_{12,34}
    =\ &
    \frac{X_{1234}}{I_{13}I_{24}} - \frac{X_{1234}}{I_{14}I_{23}} + \left( \frac{1}{I_{14}} - \frac{1}{I_{13}} \right) Y_{134} + \left( \frac{1}{I_{23}} - \frac{1}{I_{24}} \right) Y_{234} \\
    & + \left( \frac{1}{I_{23}} - \frac{1}{I_{13}} \right) Y_{123} + \left( \frac{1}{I_{14}} - \frac{1}{I_{24}} \right) Y_{124}\,.
    \end{split}
    \label{eq:F1234_Result}
\end{equation}
The $F$-integral is also log-divergent in the coincident limit:
\begin{equation}
    \begin{split}
    F_{13,23} =\ \IntegralFOneThreeTwoThree\ =\ &  \frac{1}{2}
    \left(\frac{Y_{113}}{I_{13}} + \frac{Y_{223}}{I_{23}} \right) + Y_{123} \left( \frac{1}{I_{13}} + \frac{1}{I_{23}} - \frac{2}{I_{12}} \right) \\
    &+ \frac{1}{32 \pi^2} \log \left( \frac{x_{12}^4}{x_{13} ^2 x_{23}^2} \right)\,.
    \end{split}
    \label{eq:F1323}
\end{equation}

\subsection{Defect integrals}
\label{subsec:DefectIntegrals}

We present here explicitly the defect integrals necessary to compute the bulk-defect-defect correlators at NLO. The diagrams appearing in the channel $F_2(x)$ do not have bulk vertices, therefore the integrals arise from the coupling of $\phi^6$ with the Wilson line
\begin{equation}
  I(a) =  \int_{-\infty}^a d\tau_1 I_{1\tau_1}  ... \int_{-\infty}^{\tau_{n-1}} d\tau_{\Delta_1-(\Dh_2 + \Dh_3)} I_{1\tau_{\Delta_1-(\Dh_2 + \Dh_3)}} = \frac{1}{n!}\left(\int_{-\infty}^a d\tau I_{1\tau}\right)^{n}\,,
  \label{eq:IntegratePhi6OnTheLine}
\end{equation}
and similar from a point $a$ to $+\infty$.
Equation \eqref{eq:IntegratePhi6OnTheLine} implies that the only integrals determining bulk-defect-defect correlators as next-to-leading order are
\begin{equation}
    \int_{-\infty}^a d\tau\, I_{1\tau} = \frac{\pi + 2 \arctan\left( \frac{a}{|x_\perp|}\right) }{2 |x_\perp|}\,,
    \label{eq:MasterIntegralDefet1}
\end{equation}
and
\begin{equation}
    \int_a^{+\infty} d\tau\, I_{1\tau} = \frac{\pi - 2 \arctan\left( \frac{a}{|x_\perp|}\right) }{2 |x_\perp|}\,.
    \label{eq:MasterIntegralDefet2}
\end{equation}

\section{Checks through Feynman diagrams}
\label{app:CheckThroughFeynmanDiagrams}

We present here the results for the Feynman diagrams of Table \ref{table:Diagrams211NLO} that contribute to the channel $F_1$ of the correlator $\vev{2 \hat{1} \hat{1}}$.
The results are obtained using the Feynman rules of Section \ref{subsubsec:FeynmanRules} and the integrals of Appendix \ref{app:Integrals}.

\subsection{Self-energy diagrams}
\label{subsec:SelfEnergyDiagrams}

We begin by the self-energy diagrams.
Using the insertion rule \eqref{eq:SelfEnergy}, it is easy to find that
\begin{equation}
    \BulkDefectDefectNLOSEOne
    = \frac{\log (x)+2 \log (\veps )-2}{4 \sqrt{2} \pi ^2}\,,
    \label{eq:DiagramSEOne}
\end{equation}
while
\begin{equation}
    \BulkDefectDefectNLOSETwo
    = \frac{-\log \left(\tau _3\right)+\log (\veps )-1}{2 \sqrt{2} \pi ^2}\,.
    \label{eq:DiagramSETwo}
\end{equation}
Note that these results are given in the conformal limit $\tau_3 \sim \infty$.
In other words, each diagram contains corrections in $1/\tau_3$, however since we know that the correlator depends on a single cross-ratio $x$, they are expected to cancel once added up.
As visible in the final result \eqref{eq:NLO_AfterWI_211}, we observe indeed a cancellation of the $\log \tau_3$ terms.

\subsection{Diagrams with bulk vertices}
\label{subsec:DiagramsWithBulkVertices}

There are two diagrams that involve bulk vertices but no integration along the Wilson line.
The first one contains the $X$-integral defined in \eqref{eq:X1234}, and pinched to \eqref{eq:X1233}.
Taking into account all the prefactors, it reads
\begin{equation}
    \BulkDefectDefectNLOX
    = -\frac{\log (x)+2 \log (\veps )-2}{8 \sqrt{2} \pi ^2}\,,
    \label{eq:DiagramX}
\end{equation}
The second diagram involves the F-integral \eqref{eq:F1234_Definition}.
Using \eqref{eq:F1323}, we obtain\footnote{We thank Julius Julius and Philine van Vliet for pointing out corrections to this expression.}
\begin{equation}
    \BulkDefectDefectNLOH
    = -\frac{\log \left(x\right)+2 \log (\veps )}{8 \sqrt{2} \pi ^2}\,.
    \label{eq:DiagramF}
\end{equation}

\subsection{Boundary diagrams}
\label{subsec:BoundaryDiagrams}

Finally there are two diagrams that involve an integral along the Wilson line as well as a bulk vertex.
These integrals are more challenging but can be performed analytically in the conformal frame $\tau_3 \to \infty$.
The first one gives
\begin{equation}
    \BulkDefectDefectNLOYOne
    = -\frac{3 \log (x)-12 \left[\arctan\left(\sqrt{\frac{1-x}{x}}\right)\right]^2+6 \log (\veps )+\pi ^2-6}{24 \sqrt{2} \pi ^2}\,,
    \label{eq:DiagramYOne}
\end{equation}
where it should be understood that the gluon propagator can connect to each orange dot on the line.
The second one yields
\begin{equation}
    \BulkDefectDefectNLOYTwo
    = \frac{3\log \left( x \right) -2\left(-6 \log \left(\tau _3\right)+3 \log (\veps )+\pi ^2-6\right)}{24 \sqrt{2} \pi ^2}\,.
    \label{eq:DiagramYTwo}
\end{equation}

When adding all the diagrams, we find that the divergences cancel as well as the $\log \tau_3$ terms.
The result is in perfect agreement with \eqref{eq:NLO_AfterWI_211}.

\section{The Witten integrals for $\vev{2\hat{1}\hat{1}}$}
\label{app:WittenIntegrals}

We perform here the Witten integrals for the five vertices appearing in \eqref{eq:Vertices1} and \eqref{eq:Vertices2}, whose results are already listed in (\ref{eq:V1}-\ref{eq:V5}).
We begin by defining our notation for the bulk-to-boundary AdS$_5$ propagator, matching the references \cite{Freedman:1998tz,Giombi:2017cqn,Gimenez-Grau:2023fcy}.
The propagator of a scalar field of dimension $\Delta$ connecting the boundary of AdS$_5$ to a point in the bulk is
\begin{equation}
    \mathcal{K}(\Delta,x_{12},z_1) = \left( \frac{z_1}{z_1^2+x_{12}^2}\right)^\Delta\,,
    \label{eq:WittenProp}
\end{equation}
where $x_i$ is a four-dimensional vector parametrizing the boundary of AdS$_5$, $z_i$ is the bulk coordinate and $x_{12} = x_1 - x_2$.
In order to declutter the notation we can write the explicit form of the only three propagators that we are going to need.
We first consider a $\Delta=2$ propagator from the point $(0,t_1,\vec{x}_1)$ on the AdS$_5$ boundary to the point $(z,\tau,\vec{0})$ living on the induced AdS$_2$ dual of the Wilson line:
\begin{equation}
        \mathcal{K}_{1x} = \left( \frac{z}{z^2+x_{1}^2+(\tau-\tau_1)^2}\right)^2\,.
        \label{eq:K1x}
\end{equation}
The other two propagators appearing in the following integrals correspond to a dimension one scalar connecting a point on the Wilson line $(0,\tau_2,\vec{0})$ to the point $(z,\tau,\vec{0})$:
\begin{equation}
        \hat{\mathcal{K}}_{2x} = \frac{z}{z^2+(\tau-\tau_2)^2}\,,
        \label{eq:K2x}
\end{equation}
and similarly for the point $(0,\tau_3,\vec{0})$:
\begin{equation}
        \hat{\mathcal{K}}_{3x} = \frac{z}{z^2+(\tau-\tau_3)^2}\,..
        \label{eq:K3x}
\end{equation}
In the following subsections we list the integrals corresponding to each vertex in $\Lm_{F1}^{(2,1)}$ and describe our way to obtain the results \eqref{eq:V1}-\eqref{eq:V5}.

\subsection{The vertex with no derivatives}

We consider first the vertex $s_2^{\tilde{I}} y_i y_j \pd^i \pd^j Y_2^{\tilde{I}}$ in which no derivative act on the bulk-to-boundary propagators.
Due to the way the $R$-symmetry indices of the $y_i$ fields contract, this vertex can appear in both channels $F_1(x)$ and $F_2(x)$.
We also use this section to introduce the general form of the Witten integrals to consider.
In the absence of derivatives acting on the fields in the vertex, the integral we consider is
\begin{equation}
    \mathcal{V}_1(\vec{x}_1,\tau_1,\tau_2,\tau_3) = \int_0^{+\infty} \frac{dz}{z^2} \int_{-\infty}^{+\infty} d\tau\, \,\mathcal{K}_{1x}  \hat{\mathcal{K}}_{2x} \hat{\mathcal{K}}_{3x}\,,
    \label{eq:WittenIntegral1}
\end{equation}
where the $z^{-2}$ factor is due to the induced AdS$_2$ metric.
In order to perform this integral we go to the conformal frame by taking
\begin{equation}
    V_1(x) = \tau_3^2(\vec{x_1}^2+\tau_1^2) \lim_{\tau_3\to +\infty\,,\,\,\tau_2\to 0} \mathcal{V}_1(\vec{x}_1,\tau_1,\tau_2,\tau_3) |_{\vec{x}_1^2=\frac{\tau_1^2 {x}}{{1-x}}} \,.
    \label{eq:V1_conformal_frame}
\end{equation}
Whenever there are no derivatives acting on $\hat{\mathcal{K}}_{3x}$ the conformal frame can be taken at the integrand level and the integral becomes simpler. The result for the function $V_1(x)$ is
\begin{equation}
    V_1 (x) = - \frac{\pi}{4} \frac{\log(x)}{1-x}\,, \label{eq:resultV1} 
\end{equation}

\subsection{The vertices with two derivatives}

We analyze in this section the two vertices in which each integral has two derivatives acting on the bulk-to-boundary propagators: $\nabla_\alpha \nabla^{\alpha} s_2^{\tilde{I}} y_i y_j \pd^i \pd^j Y_2^{\tilde{I}}$ and $s_2^{\tilde{I}} \pd^\alpha y_i \pd_\alpha y^i Y_2^{\tilde{I}}$.
The second one can only appear in the channel $F_1(x)$.
We associate to the first vertex the integral
\begin{equation}
    \mathcal{V}_2(\vec{x}_1,\tau_1,\tau_2,\tau_3) = \int_0^{+\infty} \frac{dz}{z^2} \int_{-\infty}^{+\infty} d\tau\, z^2 \delta_{\alpha\beta} \pd_\alpha (\pd_\beta \mathcal{K}_{1x} ) \hat{\mathcal{K}}_{2x} \hat{\mathcal{K}}_{3x}\,,
    \label{eq:WittenDiagram2}
\end{equation}
where for the sake of clarity we made explicit that index raising in AdS$_5$ carries a factor $z^2$; therefore each index contraction comes with a factor $z^2$.
The variables with respect to which we differentiate are $z$ and $\tau$.
This integral can be performed in the conformal frame and gives
\begin{equation}
 V_2 (x) = - \frac{\pi}{2(1-x)} \left(3 + \frac{2x+1}{1-x} \log(x) \right)\,.
    \label{eq:V2result}
\end{equation}
The second vertex gives the integral
\begin{equation}
    \mathcal{V}_5(\vec{x}_1,\tau_1,\tau_2,\tau_3) = \int_0^{+\infty} \frac{dz}{z^2} \int_{-\infty}^{+\infty} d\tau\, z^2 \delta_{\alpha\beta} \mathcal{K}_{1x}  \pd_\alpha (\hat{\mathcal{K}}_{2x} )\pd_\beta( \hat{\mathcal{K}}_{3x})\,,
    \label{eq:WittenDiagram2}
\end{equation}
where, because of the derivatives acting on $\hat{\mathcal{K}}_{3x} $, we cannot immediately apply the conformal frame.
The safest way is usually to perform integration by parts to keep the propagator $\hat{\mathcal{K}}_{3x} $ undifferentiated; for this particular integral we can also simply use the identity \cite{Giombi:2017cqn}
\begin{align}
    \pd_\alpha \hat{\mathcal{K}}(\Dh_1,z,\tau_1-\tau)\pd^\alpha \hat{\mathcal{K}}(\Dh_2,z,\tau_2-\tau) =\, & \Dh_1\Dh_2 \left[\hat{\mathcal{K}}(\Dh_1,z,\tau_1-\tau)\hat{\mathcal{K}}(\Dh_2,z,\tau_2-\tau) \right. \nonumber \\  & \left. -2(\tau_{12})^2\hat{\mathcal{K}}(\Dh_1+1,z,\tau_1-\tau)\hat{\mathcal{K}}(\Dh_2+1,z,\tau_2-\tau) \right]\,.
    \label{eq:WittenDerivativesIdentity}
\end{align}
The result we obtain is
\begin{equation}
    V_5 (x) = - \frac{\pi}{4(1-x)} \left(3 + \frac{2x+1}{1-x} \log(x) \right)\,.
    \label{eq:V5result}
\end{equation}

\subsection{The vertices with four derivatives}

Finally we consider the two remaining vertices: $\nabla_\alpha \nabla^{\alpha} s_2^{\tilde{I}} \pd^\beta y_i \pd_\beta y^i Y_2^{\tilde{I}}\,$ and $ \nabla_\alpha \nabla^\beta s_2^{\tilde{I}} \pd_\beta y_i \pd_\alpha y^i Y_2^{\tilde{I}}$.
They only differ for the index contraction of the derivatives; in particular the first one has the same structures of derivatives on $y_i$ of $\mathcal{V}_5(\vec{x}_1,\tau_1,\tau_2,\tau_3)$.
We can use again equation \eqref{eq:WittenDerivativesIdentity} to replace the derivatives $ \pd^\beta y_i \pd_\beta y^i$ in the first vertex and integrate the result in the conformal frame.
We find that the $x$ dependent part of the result is
\begin{align}
    V_3 (x) = - \frac{\pi}{(1-x)^2} \left( 3(1+4x) + \frac{1+19x+10x^2}{2(1-x)} \log(x) \right)\,.\label{eq:V3result}
\end{align}
For the last vertex we did not find any specific identity to simplify the integral
\begin{equation}
    \mathcal{V}_4(\vec{x}_1,\tau_1,\tau_2,\tau_3) = \int_0^{+\infty} \frac{dz}{z^2} \int_{-\infty}^{+\infty} d\tau\, z^4 \delta_{\alpha\beta}\delta_{\gamma\delta} \pd_\alpha (\pd_\beta \mathcal{K}_{1x} )   \pd_\gamma (\hat{\mathcal{K}}_{2x} )\pd_\delta( \hat{\mathcal{K}}_{3x})\,.
    \label{eq:WittenDiagram2}
\end{equation}
We can however use integration by parts to move the derivative acting on $\hat{\mathcal{K}}_{3x}$ on the other factors of the integrand, and then go to the conformal frame to find
\begin{align}
    V_4 (x) = - \frac{\pi}{2 (1-x)^2} \left( \frac{3}{2} (3+7x) + \frac{1+10x+4x^2}{1-x} \log(x) \right)\,. \label{eq:V4result} \\
\end{align}
The calculation of all five vertices is complete and is reported in \eqref{eq:V1}-\eqref{eq:V5} in the main text.

\bibliography{./auxi/biblio.bib}
\bibliographystyle{./auxi/JHEP}

\end{document}